\documentclass[aps,prd,nofootinbib,twocolumn,superscriptaddress,preprintnumbers,balancelastpage,longbibliography]{revtex4-1}
\usepackage{style}

\begin{document}
\title{Projecting the ultimate pulsar timing sensitivity to dark matter substructure in a stochastic gravitational wave background}

\author{Joshua~W.~Foster\,\orcidlink{0000-0002-7399-2608}}
\email{jwfoster@wisc.edu}
\affiliation{Department of Physics, University of Wisconsin-Madison, Madison, WI 53706, USA}

\author{Tanner Trickle\,\orcidlink{0000-0003-1371-4988}}
\affiliation{Department of Physics, University of Illinois Urbana-Champaign, Urbana, IL 61801, USA}
\affiliation{Illinois Center for Advanced Studies of the Universe,
University of Illinois Urbana-Champaign, Urbana, IL 61801, USA}

\author{Fabrizio Vassallo\,\orcidlink{0000-0002-0541-5606}}
\affiliation{Department of Physics, University of Wisconsin-Madison, Madison, WI 53706, USA}

\date{\today}
\begin{abstract}
Pulsar timing arrays (PTAs) are sensitive to the gravitational influence of passing compact substructures, which can produce Doppler timing delays by accelerating pulsars or the Solar System barycenter, and Shapiro timing delays when passing near Earth--pulsar lines of sight.
Projections for the complete PTA sensitivity to compact dark matter (DM) substructures, such as primordial black holes and axion miniclusters, are challenging due to the variety of signal types ranging from rare, nearly static encounters, to dynamic flybys, to the stochastic limit of many substructures.
We address this challenge with a framework that combines Monte Carlo signal modeling and machine-learned surrogate likelihoods, enabling a unified likelihood-level analysis of signals previously treated only in simplified limiting regimes. 
We then use this framework to precisely assess the impact of a stochastic gravitational wave background (SGWB), for which evidence was recently found, on the PTA sensitivity to compact DM substructures.
The SGWB substantially weakens the sensitivity, and we find that in even the most optimistic observing scenario only a Shapiro search retains sensitivity to subdominant DM components when assuming SGWB parameters inferred from current measurements.
\end{abstract}

\maketitle

\section{Introduction}

The abundance and phase-space distribution of dark matter (DM) substructure encode important information about its microscopic nature and structure formation history. In the standard cold DM paradigm, gravitational clustering proceeds hierarchically and produces bound subhalos over a broad range of masses. Departures from this expectation can arise in many well-motivated scenarios, including DM with enhanced primordial power on small scales \cite{Kolb:1993hw, Zurek:2006sy, Graham:2015rva, Buschmann:2019icd, Xiao:2021nkb, Ellis:2022grh, Pierobon:2023ozb}, nontrivial self-interactions~\cite{Tulin:2017ara, Arvanitaki:2019rax}, compact-object components~\cite{Carr:2020gox}, or modified cosmologies with early matter-dominated eras~\cite{Erickcek:2011us, Fan:2014zua, Erickcek:2020wzd}. Observational probes of substructure therefore provide a way to test the behavior of DM on scales far below those directly accessible through galaxy surveys.

Most existing probes of DM substructure rely on luminous, electromagnetic, or dynamical tracers. Examples include satellite-galaxy counts~\cite{DES:2020fxi}, perturbations to stellar streams \cite{Carlberg_2012, Bonaca:2018fek,Adams:2024zhi,Lu:2025qbp}, dynamical heating or disruption of ultra-faint dwarf galaxies and their star clusters~\cite{Brandt:2016aco, Graham:2023unf, Graham:2024hah, Graham:2025opw}, disruption of wide binaries \cite{Yoo:2003fr, Ramirez:2022mys}, and indirect searches for annihilation radiation \cite{Buckley:2010vg, Blanco:2019eij} or conversion flux \cite{Edwards:2020afl}. Lensing searches provide another powerful class of constraints, especially for sufficiently compact substructures~\cite{Barnacka_2012, EROS-2:2006ryy, Niikura:2019kqi, Croon:2020wpr,Croon:2020ouk,Tamta:2026zju}. Pulsar timing arrays (PTAs) offer a complementary gravitational channel. A sufficiently massive substructure passing near a pulsar or Earth can accelerate it and produce a Doppler timing residual, while a substructure passing near an Earth--pulsar line-of-sight can produce a Shapiro time delay. These effects have been studied as possible PTA probes of DM substructure in Refs.~\cite{Siegel:2007fz, Seto:2007kj, Baghram:2011is, Kashiyama_2012, Clark:2015sha, Schutz:2016khr, Kashiyama:2018gsh, Dror:2019twh, Ramani:2020hdo,Lee:2020wfn, Lee:2021zqw, NANOGrav:2023hvm, Berghaus:2025kvn}.

The statistical structure of these substructure signals in PTAs is richer than that of many standard targets. Depending on the expected number of substructures contributing over the observing span, the signal may be dominated by slowly varying distant objects, by one or a few dynamical encounters, or by the incoherent sum of many independent substructures, often referred to as the static, dynamic, and stochastic regimes, respectively~\cite{Dror:2019twh, Ramani:2020hdo}. Previous studies have used these limiting descriptions to estimate the sensitivity of future PTA datasets \cite{Lee:2021zqw} or perform analyses with existing PTA data \cite{NANOGrav:2023hvm}, but a unified likelihood-level treatment across regimes is challenging because the signal distribution is generally non-Gaussian and is defined only implicitly through a stochastic population of substructures.

A second important issue is the treatment of backgrounds. Recent PTA evidence for a nanohertz stochastic gravitational-wave background (SGWB) from the North American Nanohertz Observatory for Gravitational Waves (NANOGrav)~\cite{NANOGrav:2023gor}, the European Pulsar Timing Array (EPTA)~\cite{EPTA:2023fyk}, the Parkes Pulsar Timing Array (PPTA)~\cite{Reardon:2023gzh}, the Chinese Pulsar Timing Array (CPTA) \cite{Xu:2023wog}, and the MeerKAT Pulsar Timing Array (MPTA)~\cite{Miles:2024seg} implies that the SGWB is an important source of red noise that was only included in determining the PTA sensitivity to compact DM substructures in Ref.~\cite{NANOGrav:2023hvm}. 
For the uncorrelated pulsar-term Doppler and Shapiro signals, even perfect reconstruction and subtraction of the correlated SGWB Earth-term realization through the correlated measurements of the local GW via the timing of many pulsars would leave the statistically independent SGWB pulsar terms as an irreducible background \cite{Hellings:1983fr}. This component is especially important because it is uncorrelated across pulsars, so it has the same structure as pulsar-term signals from substructure. For the correlated Doppler signal sourced by the acceleration of the Solar System barycenter \cite{Ramani:2020hdo}, the relevant comparison is instead with the full correlated angular structure of the SGWB.

In this work, we revisit the projected sensitivity of pulsar timing to compact DM substructure with these issues in mind. We construct a Monte Carlo signal model for Doppler and Shapiro timing residuals sourced by a spatially homogeneous population of common-mass substructures with Standard Halo Model velocities~\cite{Freese:2012xd}. The model extends the treatment of Ref.~\cite{Lee:2021zqw} by sampling the finite spatial regions relevant for the pulsar-acceleration and line-of-sight delay geometries, applying timing-model subtraction, and capturing the transition between static, dynamic, and stochastic signal morphologies. This provides a common statistical framework for Doppler and Shapiro signals across the range of substructure abundances relevant for PTA forecasts.

To turn this signal model into projected sensitivities, we develop complementary approximations of the signal likelihood, which we refer to as \textit{surrogate} likelihoods, for the substructure abundance-dependent signal distribution. In the large substructure number limit, we approximate the signal as a Gaussian process with a covariance computed from the underlying substructure population. In the small substructure number limit, timing-model subtraction leaves a leading cubic mode that captures much of the residual signal power; we model the distribution of this cubic amplitude with a normalizing flow~\cite{Durkan:2019, 2019arXiv191202762P}. For our fiducial pulsar-term projections, we construct a Monte Carlo likelihood using signal realizations generated by a conditional diffusion model~\cite{ho2020denoisingdiffusionprobabilisticmodels, nichol2021improveddenoisingdiffusionprobabilistic}. The covariance and cubic-mode likelihoods serve as interpretable limiting cases and cross-checks of the diffusion approach where their assumptions apply. We also extend the cubic-mode construction to a cubic-vector likelihood for the correlated Doppler signal, preserving the angular coherence of the Solar-System-barycenter response.

Using these likelihood constructions, we project the sensitivity of two benchmark PTA observing scenarios: an SKA-like array~\cite{5136190} and a more optimistic future array. For the pulsar-term Doppler and Shapiro searches, we compare idealized white-noise-only datasets to datasets that include the irreducible SGWB pulsar-term red-noise contribution. For the correlated Doppler search, we retain the multi-pulsar angular structure of both the signal and the SGWB covariance. Our treatment is deliberately optimistic: we do not include intrinsic pulsar red noise, and we assume idealized observing scenarios and known SGWB parameters. The resulting forecasts should therefore be interpreted as best-case projections in the presence of an SGWB.

Our main conclusion is that the SGWB substantially weakens the prospects for detecting compact DM substructure through pulsar timing. In the white-noise-only limit, our projections broadly reproduce previous estimates, while inclusion of an SGWB-motivated background can degrade the sensitivity by orders of magnitude. For SGWB parameters motivated by the current NANOGrav 15-year power-law fit~\cite{NANOGrav:2023gor}, neither the pulsar-term nor correlated Earth-term Doppler searches reach fractional abundances \(f_\mathrm{sub}\leq 1\) in either the SKA-like or more optimistic benchmark scenarios considered here. The Shapiro channel is more robust: while the SKA-like scenario is marginal in the presence of the NANOGrav-fiducial SGWB, the more optimistic benchmark retains sensitivity to fractional abundances below unity. Still, the Shapiro reach is significantly weakened relative to the white-noise-only case. Recovering most of the previously anticipated sensitivity would require the relevant SGWB to be substantially smaller than the current PTA-inferred level.

Our projections apply directly when the DM substructure is compact: for the Doppler signal, when its radius is smaller than its point of closest approach to the Earth or pulsar, and for the Shapiro signal, when its radius is smaller than its distance from the Earth--pulsar line of sight.
For the masses considered here, this corresponds roughly to $r_\text{sub} \ll 10^{-3} \, \text{pc}$, where $r_\text{sub}$ is the substructure radius~\cite{Ramani:2020hdo}.
This is also the limit in which PTA sensitivity is expected to be strongest. At fixed mass, extended substructures produce smaller timing signals because smoothing the density profile reduces the spatial gradients of the gravitational potential and suppresses the induced residuals~\cite{Ramani:2020hdo,Lee:2021zqw}. Red noise is expected to strengthen this conclusion: extended substructures generate broader, longer-timescale events that are more easily absorbed by timing-model subtraction and more degenerate with low-frequency stochastic backgrounds. Thus, the strong degradation found here even for pointlike substructures should be interpreted as the impact of the SGWB in the most detectable substructure scenario. 
% At the same time, the framework developed here is not tied to point masses: 
Replacing the signal model and retraining the surrogate likelihood would enable likelihood-level forecasts and searches for extended profiles, mass spectra, or other substructure populations.

The remainder of this paper is organized as follows. In Sec.~\ref{sec:SignalModel}, we define the Doppler and Shapiro timing delays and construct the Monte Carlo signal model for substructure-induced residuals. In Sec.~\ref{sec:mock_data}, we define the benchmark PTA scenarios, the white-noise and SGWB red-noise models, and the formal substructure likelihood obtained by marginalizing over stochastic signal realizations. In Sec.~\ref{sec:likelihoods}, we introduce the covariance, cubic-mode, and diffusion-based surrogate likelihoods used to approximate this marginalization in practice. In Sec.~\ref{sec:projections}, we present projected sensitivities for the uncorrelated pulsar-term Doppler and Shapiro signals, including comparisons with previous white-noise-only estimates and variations of the SGWB. In Sec.~\ref{sec:correlated_doppler}, we extend the cubic-mode flow treatment to the correlated Earth-term Doppler signal sourced by acceleration of the Solar System barycenter, replacing the independent scalar cubic amplitudes with a cubic-vector likelihood that preserves angular coherence across the PTA. We conclude in Sec.~\ref{sec:Discussion}. Additional details of the projection procedure, signal covariance calculations, surrogate training, and likelihood convergence are given in the appendices.

\section{Signals and Statistics of Doppler and Shapiro Time Delays}
\label{sec:SignalModel}

\begin{figure}[!t]
  \centering
  \includegraphics[width=0.4\textwidth]{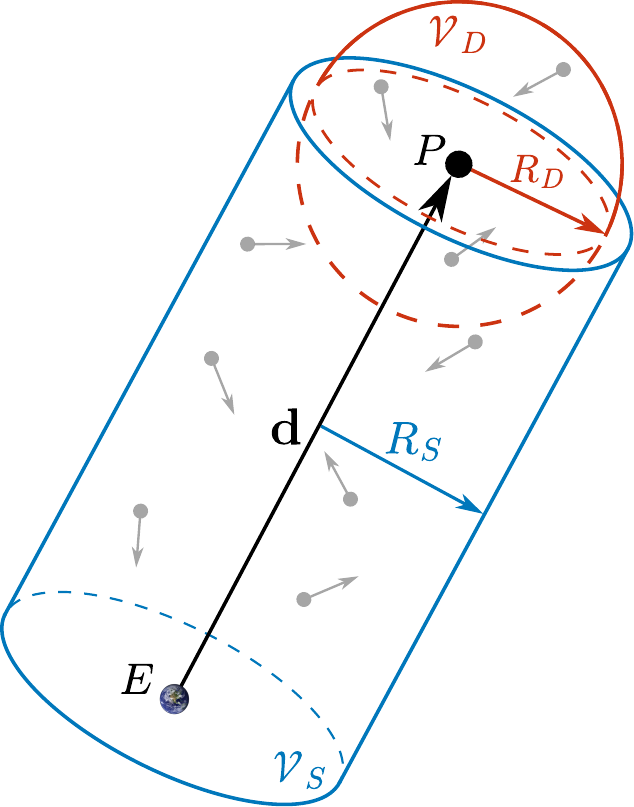}
    \caption{Illustration of the sampling region geometries relevant for the pulsar-term Doppler and Shapiro effects. The Earth location is denoted by \(E\), the pulsar by \(P\), and the line of sight by \(\mathbf d\), with the length \(d\) corresponding to the Earth--pulsar distance. The Doppler sampling region, \(\mathcal V_D\), is the sphere of radius \(R_D\) centered on the pulsar. The Shapiro sampling region, \(\mathcal V_S\), is the cylinder of radius \(R_S\) around the Earth--pulsar line segment, with length \(d\). Grey points indicate perturbing substructures with associated velocity vectors.
    }
  \label{fig:SamplingVolumes}
\end{figure}

In this section, we review the Doppler and Shapiro timing delays induced by DM substructure and describe the Monte Carlo signal model used to sample their realizations in pulsar timing data. In Sec.~\ref{sec:time_delay_review}, we summarize the timing-delay expressions used throughout the analysis, following Refs.~\cite{Dror:2019twh,Ramani:2020hdo,Lee:2020wfn,Lee:2021zqw,NANOGrav:2023hvm}. In Sec.~\ref{sec:MCSignalModel}, we define the Doppler and Shapiro sampling volumes and the corresponding population model. In Sec.~\ref{sec:limiting_cases}, we review the static, dynamic, and stochastic regimes and introduce the characteristic travel distance over the observing span. In Sec.~\ref{sec:defining_radii}, we use these considerations to define the fiducial and extended sampling regions used throughout the analysis. In total, our treatment is highly similar to that of \cite{Lee:2021zqw}.

Throughout this work, we consider the limit of pointlike substructure masses, as relevant for primordial black holes or highly compact substructures, which maximize the expected PTA sensitivity. However, non-compact substructure may also generate potentially observable signals, which can be straightforwardly incorporated in our model framework with form-factor corrections associated with the substructure density profile (see \cite{Ramani:2020hdo, Lee:2021zqw} for details). For simplicity, we consider only monochromatic substructure mass functions, but our treatment could also be applied to more realistic extended mass functions with minimal modifications \cite{Lee:2020wfn}.

\subsection{Review of substructure-induced time delays}
\label{sec:time_delay_review}

Consider a pulsar observed along the line of sight \(\hat{\mathbf d}\), directed from the Earth toward the pulsar, and let \(d\) denote the distance from the Earth to the pulsar. Let a pointlike substructure of mass \(M_\mathrm{sub}\) have an initial displacement \(\mathbf r_0\) from the pulsar at \(t=0\) and a fixed velocity \(\mathbf v\). Following \cite{Dror:2019twh,Lee:2020wfn,Lee:2021zqw,NANOGrav:2023hvm}, the Doppler timing delay may be expressed as
\begin{equation}
    r_D(t) = \frac{G M_\mathrm{sub}}{v^2} \hat{\mathbf d} \cdot \left(\sqrt{1 + x_D^2}\hat{\mathbf b}_D - \sinh^{-1}(x_D)\hat{\mathbf v} \right),
\label{eq:Doppler}
\end{equation}
where \(t_{D,0}\) is the time of closest approach, \(\mathbf b_D\) is the Doppler impact-parameter vector, \(\tau_D\) is the characteristic signal duration, and \(x_D\) is the normalized time variable, each defined by
\begin{equation}
\begin{split}
t_{D,0} &\equiv - \frac{\mathbf r_0 \cdot \mathbf v}{v^2}, \qquad 
\mathbf b_D \equiv \mathbf r_0 + \mathbf v\, t_{D,0}, \\
\tau_D &\equiv \frac{|\mathbf r_0 \times \mathbf v|}{v^2}, \qquad 
x_D \equiv \frac{t - t_{D,0}}{\tau_D}.
\end{split}
\label{eq:dopplerdefs}
\end{equation}
Similarly, the Shapiro time delay can be written as
\begin{equation}
    r_S(t) = 2G M_\mathrm{sub} \log(1 + x_S^2),
    \label{eq:Shapiro}
\end{equation}
where one introduces the associated quantities $t_{S,0}$, $\mathbf{b}_S$, $\tau_S$, and $x_S$ defined as follows using the auxiliary quantities $\mathbf r_\times \equiv \mathbf r_0 \times \hat{\mathbf d}$ and $\mathbf v_\times \equiv \mathbf v \times \hat{\mathbf d}$:
\begin{equation}
\begin{split}
t_{S,0} &\equiv -\frac{\mathbf r_\times \cdot \mathbf v_\times}{v_\times^2}, \qquad 
\mathbf b_S \equiv \hat{\mathbf d} \times (\mathbf r_\times + \mathbf v_\times t_{S,0}), \\
\tau_S &\equiv \frac{|\mathbf r_\times \times \mathbf v_\times|}{v_\times^2}, \qquad 
x_S \equiv \frac{t-t_{S,0}}{\tau_S}.
\end{split}
\label{eq:shapirodefs}
\end{equation}
Because both the Doppler and Shapiro timing delays are linear in the substructure mass, it is convenient to define the corresponding mass-normalized signals,
\begin{equation}
    s_D(t) \equiv \frac{r_D(t)}{M_\mathrm{sub}},
    \qquad
    s_S(t) \equiv \frac{r_S(t)}{M_\mathrm{sub}}.
\end{equation}
In the remainder of this work, we use \(r_D\) and \(r_S\) to denote physical timing residuals and \(s_D\) and \(s_S\) to denote the corresponding residuals per unit mass.

The two timing delays are sensitive to substructure in qualitatively different regions: the Doppler delay is largest for substructures passing near the pulsar, while the Shapiro delay is largest for substructures passing near the Earth--pulsar line of sight. This difference will play an important role in the construction of the corresponding sampling regions and in the statistics of the resulting signal realizations.

\subsection{A Monte Carlo model for substructure signals}
\label{sec:MCSignalModel}

To model the time-delay signal generated by a population of compact DM substructures, we sample substructures from a finite spatial region around the geometry relevant for each observable. As illustrated in Fig.~\ref{fig:SamplingVolumes}, for the Doppler signal we take the sampling region \(\mathcal V_D\) to be a sphere of radius \(R_D\) centered on the pulsar with volume
\begin{equation}
    V_D = \frac{4\pi}{3} R_D^3.
    \label{eq:SphereVolume}
\end{equation}
For the Shapiro signal, we take the sampling region \(\mathcal V_S\) to be a cylinder of radius \(R_S\) around the Earth--pulsar line of sight. Its volume is therefore
\begin{equation}
    V_S = \pi R_S^2 d,
    \label{eq:CylinderVolume}
\end{equation}
where \(d\) is the distance from the Earth to the pulsar. 

We characterize the substructure population by \(M_{\rm sub}\) and assume a uniform number density \(n_{\rm sub}\) within the relevant sampling region. Later in this work, we perform inference over the fractional abundance of substructures $f_\mathrm{sub}$, which is related to the number density by
\begin{equation}
    f_\mathrm{sub} =
    \frac{M_\mathrm{sub} n_\mathrm{sub}}{\rho_\mathrm{DM}} ,
\end{equation}
where \(\rho_\mathrm{DM}\) is the local dark matter density, which we take to be $\rho_\mathrm{DM} = 0.4\, \mathrm{GeV}/\mathrm{cm}^3$ throughout this work. We also assume that both the observer and the pulsar are at rest with respect
to the Galactic halo, and that the substructure velocity distribution is spatially independent and isotropic. We therefore model the velocity distribution using the Standard Halo Model,
\begin{equation}
    f_v(\mathbf v) = \frac{1}{\left(2 \pi \sigma_v^2 \right)^{3/2}}
    \exp\!\left(-\frac{|\mathbf v|^2}{2\sigma_v^2}\right),
\end{equation}
with \(\sigma_v \approx 155\,\mathrm{km/s}\) \cite{Freese:2012xd}. Including Solar motion and pulsar peculiar velocities would anisotropize and shift the relative-velocity distribution; we neglect these effects in order to isolate the dependence on signal statistics and red noise.

For either signal class \(\mathcal I \in \{D,S\}\), the expected number of substructures in the corresponding sampling region is
\begin{equation}
    \langle N_{\mathcal I} \rangle = n_{\rm sub} V_{\mathcal I}.
\end{equation}
A signal realization is then constructed by drawing the number of substructures \(N\) from a Poisson distribution with mean \(\langle N_{\mathcal I}\rangle\), sampling their initial positions uniformly within the appropriate region, and drawing their velocities from the halo distribution above. The total signal is obtained by summing the individual mass-normalized contributions and multiplying by the common substructure mass as
\begin{equation}
    r_{\mathcal I}(t) = M_\mathrm{sub} \sum_i s_{\mathcal I}^{(i)}(t).
\end{equation}
To simulate a signal realization as it would be measured in a PTA dataset, we evaluate the continuous signal at a discrete set of times \(\mathbf t = \{t_i\}\), and we denote the corresponding sampled signal vector by \(\mathbf{r}_\mathcal{I} = \{r_\mathcal{I}(t_i)\}\). 

Sampling the substructures within a finite-radius volume is sufficient because substructures far from the pulsar or far from the Earth--pulsar line segment produce only weak and slowly varying contributions for Doppler and Shapiro signals, respectively. In a full pulsar-timing analysis, such smooth trends are largely absorbed by the fitted timing model (see, \textit{e.g.}, \cite{Taylor:2021yjx, NANOGrav:2023hde}); here we capture that effect in simplified form by removing the constant, linear, and quadratic components of the signal. We implement the simplified timing-model subtraction with the operation
\begin{equation}
    \tilde{\mathbf{r}}_\mathcal{I} = \mathbf P \mathbf{r}_\mathcal{I},
\end{equation}
where \(\mathbf P\) is the residualizing projection matrix that removes the constant, linear, and quadratic components over the finite observing span.

The timing delays in Eq.~\eqref{eq:Doppler} and Eq.~\eqref{eq:Shapiro} have already had their asymptotic constant, linear, and quadratic pieces removed analytically, since these are degenerate with the secular terms absorbed by pulsar timing models. The action of the residualizing projection is to remove the finite-time constant, linear, and quadratic combinations that persist. Throughout the text, a tilde indicates that the corresponding quantity has been acted on by this projection operator. See App.~\ref{app:projection} for details.

\begin{figure*}[!htb]
  \centering
  \includegraphics[width=0.99\textwidth]{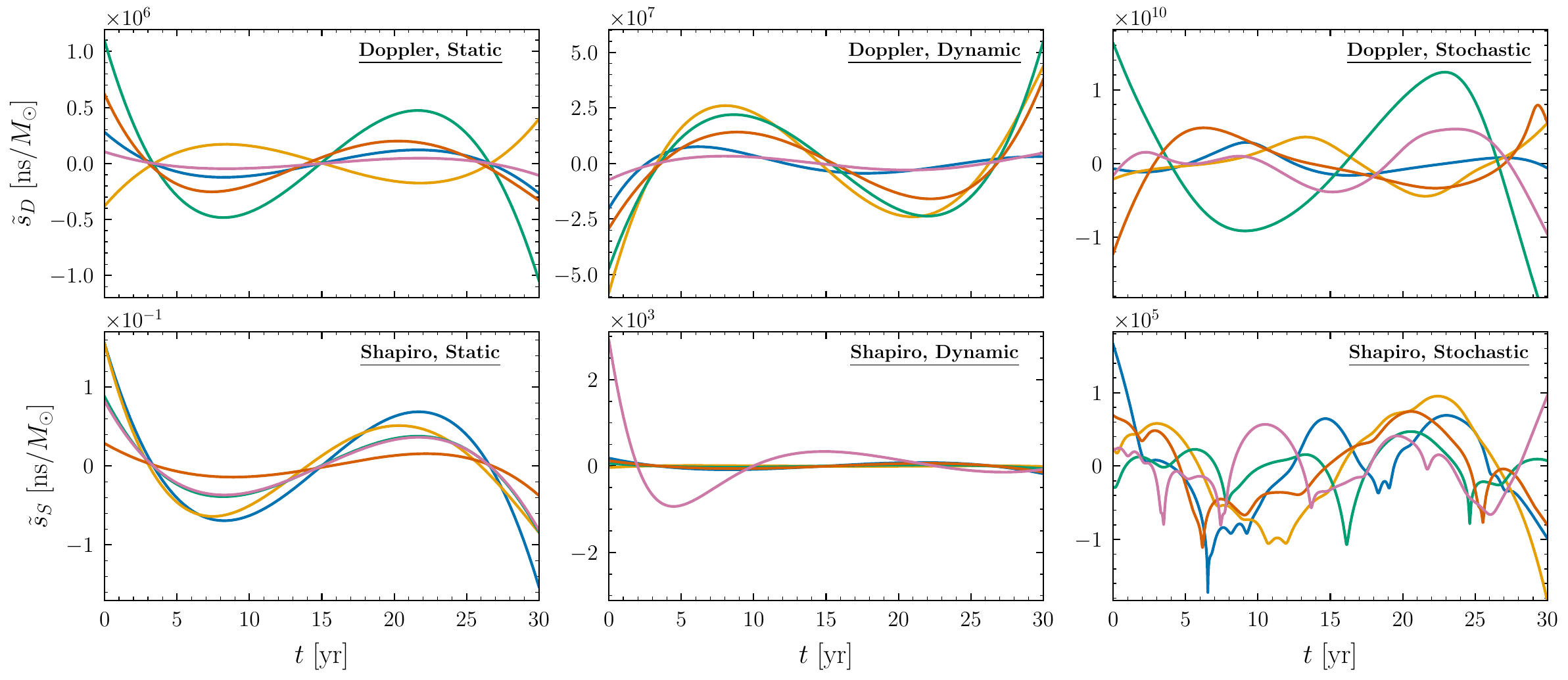}
  \caption{Example signal realizations drawn from the Monte Carlo signal model. Each panel shows five independent realizations at fixed expected number of substructures in the fiducial volume. The upper row shows Doppler signals and the lower row shows Shapiro signals, with \(\langle N_{D,S}^{\rm dyn}\rangle = 10^{-2}, 1,\) and \(10^{3}\) from left to right, corresponding to the static, dynamic, and stochastic regimes, respectively.
  }
  \label{fig:SignalExamples}
\end{figure*}

\subsection{Limiting cases and the dynamical radius}
\label{sec:limiting_cases}

Substructure-induced timing signals are often described in three regimes: \emph{static}, \emph{dynamic}, and \emph{stochastic} \cite{Dror:2019twh, Ramani:2020hdo, Lee:2021zqw}. A useful scale for organizing these regimes is the distance a typical substructure travels over the observing span, 
\begin{equation}
\begin{split}
    R_{\rm dyn} & \equiv v\, T_{\rm obs}\\
    &  \approx 7.5 \times 10^{-3}\,\mathrm{pc}
    \left(\frac{v}{250\,\mathrm{km/s}} \right)
    \left(\frac{T_{\rm obs}}{30\,\mathrm{yr}} \right).
    \label{eq:R_dyn}
\end{split}
\end{equation}
where $T_{\rm obs}$ is the total observing time over which PTA data are collected. We refer to this scale as the \emph{dynamical radius}. The terminology is chosen deliberately: a substructure within this distance can move an appreciable fraction of the relevant impact parameter during the observing span, and can therefore produce an appreciably time-dependent—\textit{i.e.}, dynamic—timing signal. However, the volume over which we sample substructures must be larger than this dynamical radius; otherwise, important dynamical events that contribute to the total time-delay signal may not be resolved.

For each signal class, we define the corresponding dynamical volume by using \(R_{\rm dyn}\) as the characteristic radius of the relevant geometry. Thus, \(\mathcal V_D^{\rm dyn}\) is the sphere of radius \(R_{\rm dyn}\) centered on the pulsar, while \(\mathcal V_S^{\rm dyn}\) is the cylinder of radius \(R_{\rm dyn}\) around the Earth--pulsar line of sight. The associated volumes are
\begin{equation}
    V_D^{\rm dyn} = \frac{4\pi}{3} R_{\rm dyn}^3,
    \qquad
    V_S^{\rm dyn} = \pi R_{\rm dyn}^2 d,
\end{equation}
corresponding to \(R_D=R_S=R_{\rm dyn}\) in Eqs.~\eqref{eq:SphereVolume} and \eqref{eq:CylinderVolume}. We denote the expected number of substructures in these volumes by
\begin{equation}
    \langle N^\mathrm{dyn}_{\mathcal I}\rangle
    =
    n_{\rm sub} V^\mathrm{dyn}_{\mathcal I},
    \qquad
    \mathcal I \in \{D,S\}.
    \label{eq:NdynDef}
\end{equation}

The limiting regimes can then be understood in terms of \(\langle N^\mathrm{dyn}_{\mathcal I}\rangle\). In the \emph{static} regime, \(\langle N^\mathrm{dyn}_{\mathcal I}\rangle \ll 1\), the timing residual is dominated by one or more distant substructures with characteristic timescales \(\tau \gtrsim T_{\rm obs}\), so the signal varies only weakly over the observing span. In the \emph{dynamic} regime, \(\langle N^\mathrm{dyn}_{\mathcal I}\rangle \sim 1\), an order-unity number of substructures can achieve closest approach during the observing window and produce signals with \(\tau \lesssim T_{\rm obs}\). In the \emph{stochastic} regime, \(\langle N^\mathrm{dyn}_{\mathcal I}\rangle \gg 1\), many substructures contribute independent time-varying signals that add incoherently, producing a highly stochastic residual. The transition between these regimes is smooth, but the limiting cases remain useful for interpreting the Monte Carlo signal realizations and for defining the dynamical-volume coordinate \(\langle N_{\mathcal I}^{\rm dyn}\rangle\).

\subsection{Defining the sampling radii}
\label{sec:defining_radii}

We now specify the finite sampling radii used in the Monte Carlo signal model. The idealized substructure population is spatially homogeneous and extends over all space, but the simulation must truncate this volume for computational tractability. We choose this truncation to satisfy two numerical requirements. First, the sampled region must extend well beyond the dynamical radius so that all dynamically relevant events are included. Second, it must contain enough substructures at the target number density to capture the statistics of the nearest objects that can contribute appreciably to the residualized signal.

We implement the first requirement by defining fiducial Doppler and Shapiro sampling radii that extend well beyond the dynamical radius,
\begin{equation}
\begin{split}
    R_D^{\rm fid} &= R_S^{\rm fid}
    = 10\,R_\mathrm{dyn} \\
    &= 7.5 \times 10^{-2}\,\mathrm{pc}
    \left(\frac{v}{250\,\mathrm{km/s}} \right)
    \left(\frac{T_{\rm obs}}{30\,\mathrm{yr}} \right).
\end{split}
\end{equation}
We take $v = 250\,\mathrm{km/s}$ as our representative velocity, while the longest observing duration considered in this work is $30\,\mathrm{yr}$. We therefore take fixed values of $R_D^\mathrm{fid} = R_S^\mathrm{fid} = 0.075~\mathrm{pc}$ throughout this work, validating this choice in detail in App.~\ref{app:validating_sampling_radii}. These radii define the fiducial sampling regions \(\mathcal V_D^{\rm fid}\) and \(\mathcal V_S^{\rm fid}\), with corresponding volumes \(V_D^{\rm fid}\) and \(V_S^{\rm fid}\). For each signal class \(\mathcal I \in \{D,S\}\), we define the fiducial-volume expectation value
\begin{equation}
    \langle N_{\mathcal I}^{\rm fid}\rangle
    \equiv n_{\rm sub} V_{\mathcal I}^{\rm fid}.
    \label{eq:NDef}
\end{equation}
Unless otherwise specified, the unadorned quantity \(\langle N\rangle\) used below as a template-bank or likelihood coordinate refers to \(\langle N_{\mathcal I}^{\rm fid}\rangle\). We use \(\langle N_{\mathcal I}^{\rm dyn}\rangle\) only when explicitly discussing the number of substructures in the dynamical volume.

The second requirement becomes important when the fiducial region is sparsely populated. In this regime, a fixed cutoff at \(R_{\mathcal I}^{\rm fid}\) can fail to include the nearest relevant objects drawn from the underlying homogeneous population. We therefore keep the target number density fixed and enlarge the Monte Carlo sampling region until it contains at least \(N_{\min}\) substructures in expectation. The radii used to generate signal realizations are
\begin{equation}
\begin{gathered}
    R_D =
    R_D^{\rm fid}
    \max\!\left[1,\,
    \frac{N_{\min}}{\langle N_D^{\rm fid}\rangle}
    \right]^{1/3},\\
    R_S =
    R_S^{\rm fid}
    \max\!\left[1,\,
    \frac{N_{\min}}{\langle N_S^{\rm fid}\rangle}
    \right]^{1/2}.
\end{gathered}
\label{eq:sampling_radii}
\end{equation}
The powers \(1/3\) and \(1/2\) reflect the volume scalings \(V_D\propto R_D^3\) and \(V_S\propto R_S^2\), respectively. When the radius is enlarged according to Eq.~\eqref{eq:sampling_radii}, the Poisson mean used to generate a Monte Carlo realization is \(n_{\rm sub}V_{\mathcal I}(R_{\mathcal I})\), with \(R_{\mathcal I}\) given by the enlarged radius. This enlargement is only a numerical device for sampling the nearest relevant contributors at fixed underlying number density; the likelihood coordinate \(\langle N\rangle\) continues to denote the fiducial-volume expectation value \(\langle N_{\mathcal I}^{\rm fid}\rangle\). The prescription therefore reduces to the fiducial cutoff when the fiducial region is sufficiently populated, and otherwise expands the sampled volume at fixed number density.

Throughout this work, we take
\begin{equation}
    N_{\min} = 10^4 .
\end{equation}
The simulation procedure of Ref.~\cite{Lee:2021zqw} used a similar strategy to model static signals, though with \(N_{\min} = 10^2\) and a fiducial sampling radius \(2R_\mathrm{dyn}\).

We validate this sampling-radius prescription in App.~\ref{app:validating_sampling_radii}. There we show that, for spatially homogeneous substructure populations with Standard Halo Model velocities, the resulting Monte Carlo signal realizations reproduce the relevant residualized signal statistics with negligible error. Figure~\ref{fig:SignalExamples} shows example realizations drawn from this model. Each panel shows five independent realizations at fixed values of the expected number of substructures in the corresponding dynamical volume, \(\langle N_{\mathcal I}^{\rm dyn}\rangle = 10^{-2}, \, 10^0, \, 10^3\), for both Doppler (top) and Shapiro signals (bottom). These examples illustrate the broad range of signal morphologies that arise as the substructure number density is varied.

\section{Observational Scenarios and the Substructure Likelihood}
\label{sec:mock_data}

In this section, we define the benchmark PTA datasets and the likelihood target used for the uncorrelated pulsar-term substructure searches. In Sec.~\ref{sec:observing_backgrounds}, we specify the observing scenarios and construct the corresponding Gaussian background model, including both idealized white-noise-only datasets and datasets with an SGWB-motivated red-noise contribution. In Sec.~\ref{sec:substructure_likelihood}, we define the projected residuals and the likelihood obtained by marginalizing over unknown substructure signal realizations. In Sec.~\ref{sec:template_bank_construction}, we describe the Monte Carlo template banks used to sample the signal distribution and train the surrogate likelihoods introduced in Sec.~\ref{sec:likelihoods}.

\subsection{Observational scenarios and backgrounds}
\label{sec:observing_backgrounds}

\begin{table}[!t]{
\renewcommand{\arraystretch}{1.2}
    \ra{1.3}
    \begin{center}
    \tabcolsep=0.15cm
    \begin{tabular}{l  c c c c c }
    \hlinewd{1pt}
     & $N_p$ & $d$ [kpc] & $T_\mathrm{obs}$ [yr] & $\Delta t$ [week] & $\sigma_{\mathrm{WN}}$ [ns]\\
    \hlinewd{0.5pt}
    SKA & 200 & 5 & 20 & 2 & 50 \\
    Optimistic & 1000 & 10 & 30 & 1 & 10 \\
    \hlinewd{1pt}
    \end{tabular}
    \end{center}}
\caption{Benchmark observing scenarios considered in this work, specified by the number of pulsars in the array (\(N_p\)), the common Earth--pulsar distance (\(d\)), the observing span (\(T_\mathrm{obs}\)), the measurement cadence (\(\Delta t\)), and the root-mean-square white-noise level (\(\sigma_{\mathrm{WN}}\)).}
\label{tab:parameters}
\end{table}

Following Refs.~\cite{Dror:2019twh,Ramani:2020hdo,Lee:2021zqw}, we consider two idealized future PTA observing scenarios, summarized in Tab.~\ref{tab:parameters}. The SKA scenario is intended to represent a future pulsar timing campaign with SKA-like capabilities~\cite{Keane:2014vja}. We model it as \(200\) pulsars at a common distance of \(5\,\mathrm{kpc}\), observed for \(20\) years with a two-week cadence and root-mean-square white-noise residuals \(\sigma_{\rm WN}=50\,\mathrm{ns}\). The Optimistic scenario represents a more ambitious long-baseline array, modeled as \(1000\) pulsars at a common distance of \(10\,\mathrm{kpc}\), observed for \(30\) years with weekly cadence and \(\sigma_{\rm WN}=10\,\mathrm{ns}\).

These benchmarks fix the pulsar number, observing grid, white-noise level, and Earth--pulsar distance entering the mock data. They are deliberately simplified: all pulsars in a given scenario share the same distance, cadence, observing span, and white-noise root-mean-square. This idealization allows us to isolate the effect of substructure statistics and red timing noise on the projected sensitivity. We also neglect intrinsic pulsar red noise, corresponding to an optimistic sample of exceptionally stable pulsars. We do, however, retain the stochastic gravitational-wave background, since its pulsar-term contribution represents an irreducible red-noise component for the uncorrelated Doppler and Shapiro searches.

For pulsar \(a\), we write the unprojected background residual as
\begin{equation}
    \mathbf r_{{\rm bkg},a}
    =
    \mathbf r_{{\rm WN},a}
    +
    \mathbf r_{{\rm GW},a}^{\rm Earth}
    +
    \mathbf r_{{\rm GW},a}^{\rm pulsar}.
    \label{eq:background_decomposition}
\end{equation}
Here \(\mathbf r_{{\rm WN},a}\) is the white measurement noise, while the SGWB contribution has been decomposed into an Earth term and a pulsar term. The Earth term is sourced by the metric perturbation at the Solar System barycenter and is correlated across the PTA. The pulsar term is sourced by the metric perturbation at the pulsar and is statistically independent between distinct pulsars. We model all background components as zero-mean Gaussian processes.

The white-noise covariance is independent between pulsars and uncorrelated between observing times,
\begin{equation}
    {\rm Cov}
    \!\left[
        \mathbf r_{{\rm WN},a},
        \mathbf r_{{\rm WN},b}
    \right]
    =
    \sigma_{\rm WN}^2\,
    \delta_{ab}\mathbf I,
    \label{eq:white_noise_covariance}
\end{equation}
where \(\mathbf I\) is the identity matrix on the observing grid.

For the SGWB contribution, we use the standard power-law red-noise parametrization employed in PTA analyses. The corresponding one-sided timing-residual power spectral density is
\begin{equation}
    P_\mathrm{red}(f; A,\gamma)
    =
    \frac{A^2}{12\pi^2}
    \left(
        \frac{f}{f_\mathrm{yr}}
    \right)^{-\gamma}
    \mathrm{yr}^3 ,
    \label{eq:red_noise_psd}
\end{equation}
where \(f_\mathrm{yr}=1/\mathrm{yr}\), \(A\) is the characteristic strain amplitude at \(f_\mathrm{yr}\), and \(\gamma\) is the power-law spectral index. On the discrete observing grid, we represent the associated single-pulsar covariance with a Fourier design matrix,
\begin{equation}
    \bm{\Sigma}_{\rm red}(A,\gamma)
    =
    \mathbf F
    \bm{\Phi}(A,\gamma)
    \mathbf F^T .
    \label{eq:red_noise_covariance_fourier}
\end{equation}
Here \(\mathbf F\) contains sine and cosine modes evaluated at the observing times, while \(\bm{\Phi}\) is the diagonal covariance of the corresponding Fourier coefficients. We construct \(\bm{\Phi}\) by evaluating \(P_{\rm red}\) on the Fourier grid, assigning equal variance to the sine and cosine coefficients at each frequency and no cross-covariance between distinct coefficients.

The SGWB Earth and pulsar terms share this temporal covariance but have different angular structure across the PTA. The Earth-term covariance is
\begin{equation}
    {\rm Cov}
    \!\left[
        \mathbf r_{{\rm GW},a}^{\rm Earth},
        \mathbf r_{{\rm GW},b}^{\rm Earth}
    \right]
    =
    \Gamma^{\rm Earth}_{ab}\,
    \bm{\Sigma}_{\rm red}(A_{\rm GW},\gamma_{\rm GW}) ,
    \label{eq:earth_term_sgwb_covariance}
\end{equation}
with
\begin{equation}
\begin{gathered}
    \Gamma^{\rm Earth}_{ab}
    =
    \frac12
    -
    \frac{x_{ab}}{4}
    +
    \frac32 x_{ab}\ln x_{ab},
    \quad
    x_{ab}
    =
    \frac{1-\hat{\mathbf d}_a\cdot\hat{\mathbf d}_b}{2},
\end{gathered}
    \label{eq:earth_term_orf}
\end{equation}
where \(A_{\rm GW}\) and \(\gamma_{\rm GW}\) are the SGWB power-law amplitude and index. The pulsar-term covariance is diagonal in pulsar space,
\begin{equation}
    {\rm Cov}
    \!\left[
        \mathbf r_{{\rm GW},a}^{\rm pulsar},
        \mathbf r_{{\rm GW},b}^{\rm pulsar}
    \right]
    =
    \Gamma^{\rm pulsar}_{ab}\,
    \bm{\Sigma}_{\rm red}(A_{\rm GW},\gamma_{\rm GW}) ,
    \label{eq:pulsar_term_sgwb_covariance}
\end{equation}
where
\begin{equation}
    \Gamma^{\rm pulsar}_{ab}
    =
    \frac12\delta_{ab}.
    \label{eq:pulsar_term_orf}
\end{equation}
The Earth and pulsar terms are statistically independent, so their cross-covariance vanishes. Their sum gives the standard Hellings--Downs covariance,
\begin{equation}
\begin{split}
    {\rm Cov}
    \!\left[
        \mathbf r_{{\rm GW},a},
        \mathbf r_{{\rm GW},b}
    \right]
    &=
    \left(
        \Gamma^{\rm Earth}_{ab}
        +
        \Gamma^{\rm pulsar}_{ab}
    \right)
    \bm{\Sigma}_{\rm red}(A_{\rm GW},\gamma_{\rm GW}) \\
    &=
    \Gamma^\mathrm{HD}_{ab}\,
    \bm{\Sigma}_{\rm red}(A_{\rm GW},\gamma_{\rm GW}) ,
\end{split}
    \label{eq:sgwb_covariance_hd}
\end{equation}
with
\begin{equation}
    \Gamma_{ab}^\mathrm{HD}
    =
    \frac12\delta_{ab}
    +
    \frac12
    -
    \frac{x_{ab}}{4}
    +
    \frac32 x_{ab}\ln x_{ab}.
    \label{eq:HD_correlation}
\end{equation}
Thus, the off-diagonal SGWB covariance is entirely due to the correlated Earth term, while the diagonal SGWB variance receives equal contributions from the Earth and pulsar terms~\cite{Hellings:1983fr,Taylor:2021yjx}.

Combining the independent white-noise and SGWB contributions gives the total unprojected background covariance. Defining
\begin{equation}
    \left(\mathbf C_{\rm bkg}\right)_{ab}
    \equiv
    {\rm Cov}
    \!\left[
        \mathbf r_{{\rm bkg},a},
        \mathbf r_{{\rm bkg},b}
    \right],
\end{equation}
we have
\begin{equation}
\begin{split}
    \left(\mathbf C_{\rm bkg}\right)_{ab}
    =
    \sigma_{\rm WN}^2\delta_{ab}\mathbf I
    +
    \Gamma^\mathrm{HD}_{ab}\,
    \bm{\Sigma}_{\rm red}(A_{\rm GW},\gamma_{\rm GW}) .
\end{split}
\label{eq:total_unprojected_background_covariance}
\end{equation}
The corresponding projected covariance on the residual space is
\begin{equation}
    \left(\tilde{\mathbf C}_{\rm bkg}\right)_{ab}
    =
    \mathbf P
    \left(\mathbf C_{\rm bkg}\right)_{ab}
    \mathbf P^T .
    \label{eq:full_projected_pta_background_covariance}
\end{equation}
Our fiducial SGWB-motivated red-noise model uses parameters motivated by the NANOGrav 15-year free-index power-law fit~\cite{NANOGrav:2023gor}. However, we will also consider systematic variations around this fiducial model, including the fixed-index SMBHB-motivated choice \(\gamma_{\rm GW}=13/3\) and changes to the overall amplitude.

\subsection{Projected residuals and the uncorrelated substructure likelihood}
\label{sec:substructure_likelihood}

We now specialize to the uncorrelated pulsar-term Doppler and Shapiro analyses. In this case, we adopt an optimistic treatment in which the SGWB Earth term is taken to be perfectly modeled and subtracted. This assumption is motivated by the coherent Hellings--Downs angular correlation of the Earth term, which distinguishes it from signals that are independent from pulsar to pulsar in a sufficiently large array. Thus, for the uncorrelated likelihood developed below, the full Hellings--Downs covariance of Eq.~\eqref{eq:total_unprojected_background_covariance} is replaced by the single-pulsar covariance that remains after subtracting the correlated Earth-term realization. The SGWB pulsar terms do not share this coherent angular structure: they arise from the metric perturbation at widely separated pulsar locations and are statistically independent between distinct pulsars. They therefore remain as an irreducible red-noise contribution after Earth-term subtraction.

For pulsar \(a\), after the SGWB Earth term is subtracted, the unprojected residual is
\begin{equation}
    \mathbf r_a
    =
    \mathbf r_{{\rm WN},a}
    +
    \mathbf r_{{\rm GW},a}^{\rm pulsar}
    +
    M_{\rm sub}\mathbf s_{\mathcal I,a}.
    \label{eq:single_pulsar_unprojected_residual}
\end{equation}
The retained background is independent between pulsars and has single-pulsar covariance
\begin{equation}
    \mathbf C_{\rm bkg}
    =
    \sigma_{\rm WN}^2\mathbf I
    +
    \frac12
    \bm{\Sigma}_{\rm red}(A_{\rm GW},\gamma_{\rm GW}) .
    \label{eq:single_pulsar_uncorrelated_background_covariance}
\end{equation}
Equivalently, the retained pulsar-term component is a common uncorrelated red-noise process with the same spectral shape but an effective single-pulsar amplitude
\(A_\mathrm{RN} = A_\mathrm{GW}/\sqrt{2}\), since the covariance scales as \(A^2\).
The white-noise-only case corresponds to \(A_\mathrm{GW}\to 0\).

The likelihood is evaluated after applying the timing-model projection,
\begin{equation}
    \tilde{\mathbf r}_a
    =
    \mathbf P\mathbf r_a,
    \quad
    \tilde{\mathbf s}_{\mathcal I,a}
    =
    \mathbf P\mathbf s_{\mathcal I,a},
    \quad
    \tilde{\mathbf C}_{\rm bkg}
    =
    \mathbf P
    \mathbf C_{\rm bkg}
    \mathbf P^T .
    \label{eq:projected_residual_signal_covariance}
\end{equation}
Because \(\mathbf P\) removes the constant, linear, and quadratic timing-model components, the projected covariance \(\tilde{\mathbf C}_{\rm bkg}\) is rank deficient when viewed as a matrix on the original space of time samples. The Gaussian density should therefore be interpreted as a density on the projected residual space, \(\mathrm{Image}(\mathbf P)\), rather than on the full unprojected data space. Equivalently, one may evaluate the likelihood in any orthonormal basis for \(\mathrm{Image}(\mathbf P)\). In practice, we use the corresponding coordinate-independent form in terms of the Moore--Penrose pseudoinverse and the pseudo-determinant. For a projected residual vector \(\tilde{\mathbf x}\), mean \(\tilde{\bm\mu}\), and projected covariance \(\tilde{\mathbf C}\), we define
\begin{equation}
    \mathcal N_{\mathbf P}
    \!\left(
        \tilde{\mathbf x};
        \tilde{\bm\mu},
        \tilde{\mathbf C}
    \right)
    =
    \frac{
        \exp\!\left[
            -\frac12
            \left(
                \tilde{\mathbf x}
                -
                \tilde{\bm\mu}
            \right)^T
            \tilde{\mathbf C}^{+}
            \left(
                \tilde{\mathbf x}
                -
                \tilde{\bm\mu}
            \right)
        \right]
    }{
        \sqrt{
            \operatorname{pdet}
            \!\left(
                2\pi \tilde{\mathbf C}
            \right)
        }
    },
    \label{eq:projected_Gaussian_density}
\end{equation}
where \(\tilde{\mathbf C}^{+}\) is the Moore--Penrose pseudoinverse, and  \(\operatorname{pdet}(\tilde{\mathbf C})\) is the product of the nonzero eigenvalues of \(\tilde{\mathbf C}\).

Conditioned on a projected signal realization, the single-pulsar likelihood is then
\begin{equation}
    p
    \!\left(
        \tilde{\mathbf r}_a
        \mid
        \tilde{\mathbf s}_{\mathcal I,a},
        M_{\rm sub}
    \right)
    =
    \mathcal N_{\mathbf P}
    \!\left(
        \tilde{\mathbf r}_a;
        M_{\rm sub}\tilde{\mathbf s}_{\mathcal I,a},
        \tilde{\mathbf C}_{\rm bkg}
    \right).
    \label{eq:single_pulsar_conditional_likelihood}
\end{equation}
Here \(\mathcal N_{\mathbf P}\) denotes the projected Gaussian density defined in Eq.~\eqref{eq:projected_Gaussian_density}. This notation makes explicit that the likelihood depends only on the components of the residuals that survive timing-model projection.

The likelihood for the projected data at fixed population parameters \(\langle N\rangle\) and \(M_{\rm sub}\), rather than at fixed substructure realization, is obtained by marginalizing this conditional probability over the projected signal distribution,
\begin{equation}
\begin{split}
    p
    \!\left(
        \tilde{\mathbf r}_a
        \mid
        \langle N\rangle,
        M_{\rm sub}
    \right)
    =
    \int
    d\tilde{\mathbf s}_{\mathcal I}\,
    &\mathcal N_{\mathbf P}
    \!\left(
        \tilde{\mathbf r}_a;
        M_{\rm sub}\tilde{\mathbf s}_{\mathcal I},
        \tilde{\mathbf C}_{\rm bkg}
    \right)\\
    &\times 
    p_{\rm sig}
    \!\left(
        \tilde{\mathbf s}_{\mathcal I}
        \mid
        \langle N\rangle
    \right).
\end{split}
\label{eq:single_pulsar_signal_marginal_likelihood}
\end{equation}
Here \(p_{\rm sig}\) is the projected signal distribution induced by the Monte Carlo signal model after applying the timing-model projection. 

Since the retained background and the pulsar-term substructure signals are independent between distinct pulsars, the joint likelihood can be evaluated directly as a product over independent single-pulsar likelihoods,
\begin{equation}
    p
    \!\left(
        \{\tilde{\mathbf r}_a\}_{a=1}^{N_p}
        \mid
        \langle N\rangle,
        M_{\rm sub}
    \right)
    =
    \prod_{a=1}^{N_p}
    p
    \!\left(
        \tilde{\mathbf r}_a
        \mid
        \langle N\rangle,
        M_{\rm sub}
    \right).
    \label{eq:uncorrelated_pta_likelihood}
\end{equation}
The remaining challenge is to approximate \(p_{\rm sig}\), which is defined implicitly by the stochastic substructure population and is not available in closed form. The next subsection describes the template banks used to sample this distribution, while Sec.~\ref{sec:likelihoods} introduces the surrogate likelihoods used to approximate Eq.~\eqref{eq:single_pulsar_signal_marginal_likelihood}.

\subsection{Signal distributions and template banks}
\label{sec:template_bank_construction}

The signal distribution \(p_{\rm sig}(\tilde{\mathbf s}_{\mathcal I}\mid \langle N\rangle)\) is sampled using template banks generated from the calibrated Monte Carlo signal model of Sec.~\ref{sec:SignalModel}. For each observing scenario and each signal class, we precompute a separate bank of projected, mass-normalized signal realizations. Each template therefore has units of timing residual per substructure mass, so changing \(M_{\rm sub}\) simply rescales the template amplitude in the likelihood.

The banks are sampled on a uniform grid in \(\log_{10}\langle N\rangle\),
\begin{equation}
    -5 \le \log_{10} \langle N\rangle \le 9.
    \label{eq:Nprior}
\end{equation}
For each decade in \(\langle N\rangle\), we use \(1600\) distinct values of \(\log_{10}\langle N\rangle\) and generate \(100\) independent signal realizations at each value, for a total of \(160{,}000\) realizations per decade. This construction provides dense coverage in the conditioning variable while also sampling the intrinsic stochastic variability of the signal distribution at fixed \(\langle N\rangle\). The density of the grid is useful for training conditional surrogate models, since it discourages the networks from memorizing isolated conditioning values and instead motivates them to learn the smooth evolution of the distribution with abundance.

These template banks provide the empirical representation of \(p_{\rm sig}\) used throughout the surrogate likelihood constructions. They therefore complete the ingredients needed to evaluate the formal likelihood target up to the approximations introduced in the next section.

\begin{figure*}[!ht]
    \centering
    \includegraphics[width=\linewidth]{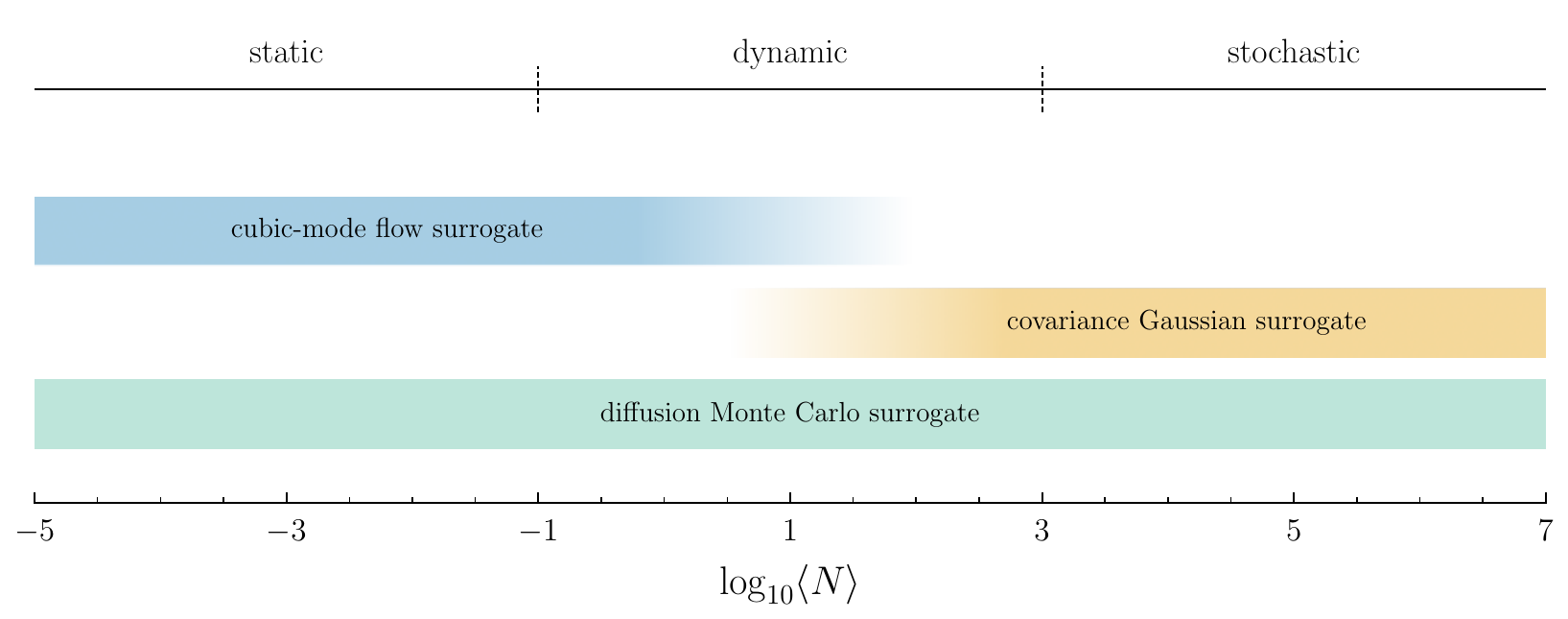}
    \caption{Schematic summary of the surrogate likelihood constructions used in this work. The horizontal axis indicates the expected number of substructures in the relevant fiducial sampling volume. At small \(\langle N\rangle\), the projected signal distribution is strongly non-Gaussian but often dominated by smooth low-order structure, motivating the cubic-mode flow surrogate. At large \(\langle N\rangle\), the signal is sourced by the superposition of many independent substructures, motivating the covariance-based surrogate when the Gaussian limit is reached. The diffusion-based Monte Carlo surrogate is designed to remain applicable across regimes by generating full projected time-domain signal realizations. The transitions between static, dynamic, and stochastic regimes are illustrative rather than sharp.
    }
    \label{fig:Roadmap}
\end{figure*}

\section{Likelihoods for Substructure Signals with Surrogate Models}
\label{sec:likelihoods}

Equation~\eqref{eq:single_pulsar_signal_marginal_likelihood} defines the uncorrelated Doppler and Shapiro likelihood as a marginalization over the projected signal distribution \(p_{\rm sig}\). This distribution is generated by the stochastic substructure population and sampled by the template banks of Sec.~\ref{sec:template_bank_construction}, but it is not available in closed form. Direct Monte Carlo evaluation of the marginalization is also computationally prohibitive across the full grid of substructure masses, abundances, observing scenarios, and noise models considered in this work. We therefore approximate the marginal likelihood using three complementary constructions, which provide computationally tractable approximations to likelihoods that are otherwise too expensive to evaluate directly. We refer to these constructions as \textit{surrogate models}.

The three surrogate models are summarized in Fig.~\ref{fig:Roadmap}. In Sec.~\ref{sec:covariance_surrogate}, we introduce a covariance-based Gaussian surrogate for the large-\(\langle N\rangle\) regime, where the signal can be approximated by its two-point statistics. In Sec.~\ref{sec:cubic_flow_surrogate}, we construct a compressed normalizing-flow surrogate based on the cubic-mode summary statistic, targeting the static, low-number-density regime where the projected signal is non-Gaussian but often smooth. Finally, in Sec.~\ref{sec:diffusion_mc_surrogate}, we introduce a diffusion-based Monte Carlo surrogate that samples the full projected signal distribution and provides our fiducial likelihood approximation for the uncorrelated pulsar-term projections.

\subsection{Covariance-based surrogate at large-\(\langle N\rangle\)} 
\label{sec:covariance_surrogate}

\begin{figure}[!t]
    \centering
    \includegraphics[width=\linewidth]{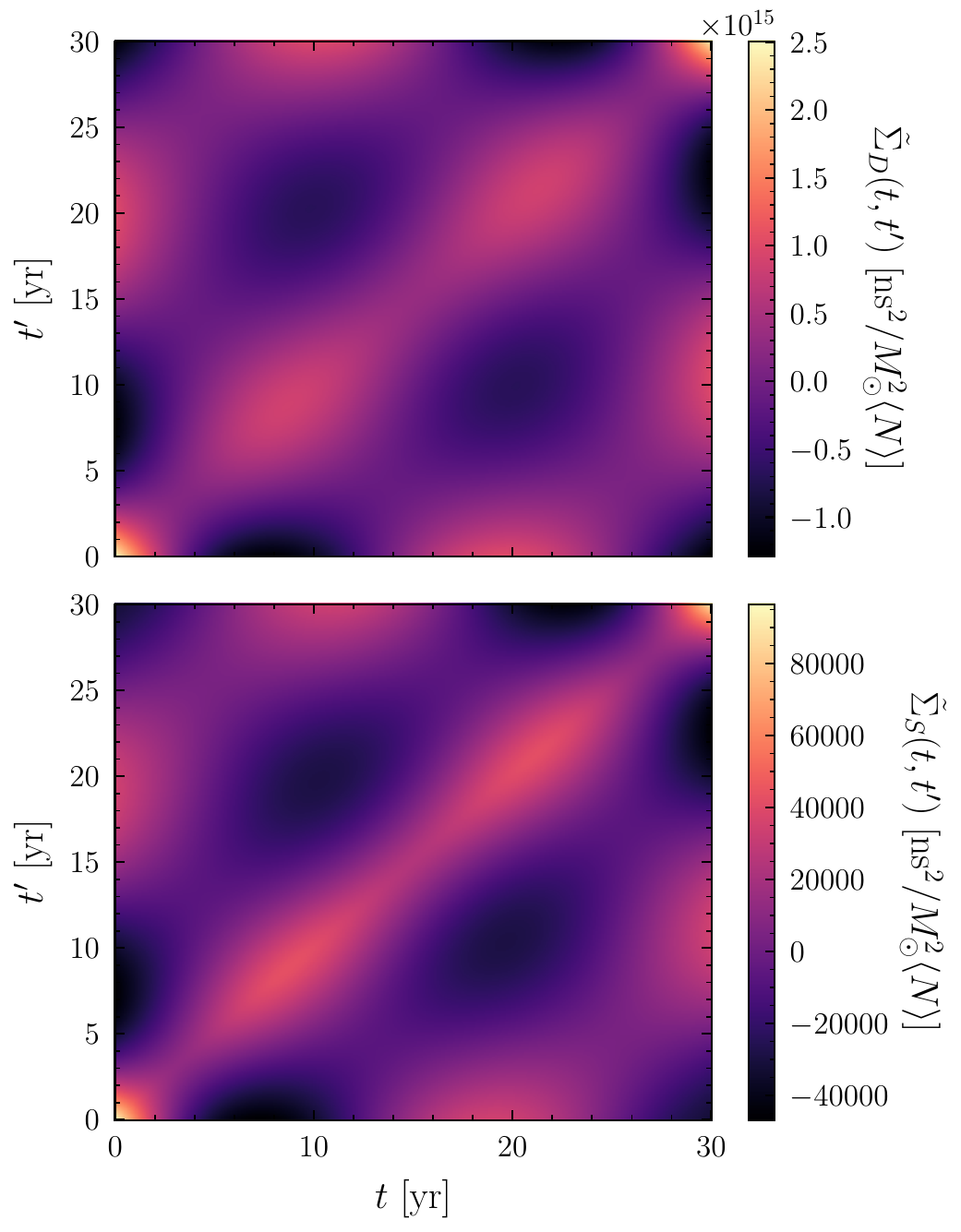}
    \caption{
    Residualized covariance matrices in the large-\(\langle N\rangle\) limit for the Doppler signal (\textit{upper}) and Shapiro signal (\textit{lower}). In both cases, the covariance depends on \(t\) and \(t'\) separately, rather than only on \(t-t'\), showing that the residualized signals are nonstationary processes. The Doppler covariance exhibits broader, more coherent structure across the observing span, indicating stronger long-timescale correlations and therefore a greater potential degeneracy with low-frequency red-noise backgrounds. The Shapiro covariance is comparatively more localized in time, reflecting the shorter-duration structure of line-of-sight encounters after timing-model projection.
    }
    \label{fig:CovarianceSummary}
\end{figure}

In the large-\(\langle N\rangle\) regime, the signal is generated by the superposition of many independent substructures. This motivates a covariance-based surrogate in which the projected signal distribution is approximated by a Gaussian whose covariance is determined by the underlying two-point statistics. If this approximation is adequate, the marginal likelihood of Eq.~\eqref{eq:single_pulsar_signal_marginal_likelihood} becomes analytically tractable and provides a simple surrogate likelihood when the Gaussian approximation is valid.

\subsubsection{Covariance of a substructure population}

For this derivation, we suppress the signal-class label and write the mass-normalized signal as
\begin{equation}
    s(t)
    =
    \sum_{i=1}^{N}
    s_i(t),
\end{equation}
where \(s_i(t)\equiv s(t\mid \mathbf r_0^{(i)},\mathbf v^{(i)})\) is the contribution from the \(i\)th substructure. The number of substructures \(N\) is Poisson distributed, and the initial positions and velocities are drawn independently from the distributions specified in Sec.~\ref{sec:SignalModel}.

The mean mass-normalized signal is therefore
\begin{equation}
\begin{split}
    \langle s(t) \rangle
    &=
    \left\langle
        \sum_{i=1}^{N} s_i(t)
    \right\rangle \\
    &=
    \langle N \rangle
    \int_{\mathcal V}
    \frac{d^3 \mathbf r_0}{V}
    \int d^3 \mathbf v\,
    f_v(\mathbf v)\,
    s(t\mid \mathbf r_0,\mathbf v),
\end{split}
    \label{eq:mean_mass_normalized_signal}
\end{equation}
where \(\mathcal V\) is the relevant sampling volume, \(V\) is its volume, and \(f_v(\mathbf v)\) is the velocity distribution defined in Sec.~\ref{sec:MCSignalModel}. Here \(\langle N\rangle\) denotes the expected number of substructures in this sampling volume.

For an infinite homogeneous population, the ensemble mean is time independent. This follows directly from Eq.~\eqref{eq:mean_mass_normalized_signal}: for each single-substructure contribution, a time shift can be absorbed into a translation of the initial position along the straight-line trajectory,
\begin{equation}
    s_i(t+\Delta t\mid \mathbf r_0,\mathbf v)
    =
    s_i(t\mid \mathbf r_0+\mathbf v\Delta t,\mathbf v).
    \label{eq:time_shift_position_translation}
\end{equation}
In the infinite-volume limit, the position integral is invariant under the change of variables
$\mathbf r_0'=\mathbf r_0+\mathbf v\Delta t$, with unit Jacobian and unchanged integration domain. Therefore,
\begin{equation}
\begin{split}
    \langle s_i(t+\Delta t) \rangle_\infty
    &=
    n_{\rm sub}
    \int d^3\mathbf r_0
    \int d^3\mathbf v\,
    f_v(\mathbf v)\,
    s_i(t+\Delta t\mid \mathbf r_0,\mathbf v) \\
    &=
    n_{\rm sub}
    \int d^3\mathbf r_0'
    \int d^3\mathbf v\,
    f_v(\mathbf v)\,
    s_i(t\mid \mathbf r_0',\mathbf v) \\
    &=
    \langle s_i(t) \rangle_\infty .
\end{split}
\end{equation}
Thus the infinite-volume mean is at most a constant in time. Since the timing-model projection removes constant modes, this mean does not contribute to the residualized covariance. The finite sampling volumes used in the Monte Carlo model break this exact translation invariance through boundary effects, but we have checked numerically that any residual projected mean is negligible compared with the signal fluctuations encoded by the covariance.

To compute the two-point function, we write
\begin{equation}
\begin{split}
    \left\langle
        s(t)s(t')
    \right\rangle
    &=
    \left\langle
        \sum_{i,j=1}^{N}
        s(t \mid \mathbf r_0^{(i)}, \mathbf v^{(i)})
        s(t' \mid \mathbf r_0^{(j)}, \mathbf v^{(j)})
    \right\rangle \\
    &=
    \langle N \rangle
    \left\langle
        s(t \mid \mathbf r_0,\mathbf v)
        s(t' \mid \mathbf r_0,\mathbf v)
    \right\rangle \\
    &\quad
    +
    \langle N \rangle^2
    \left\langle
        s(t \mid \mathbf r_0,\mathbf v)
    \right\rangle
    \left\langle
        s(t' \mid \mathbf r_0,\mathbf v)
    \right\rangle .
\end{split}
\label{eq:signal_two_point_function}
\end{equation}
Here the remaining expectation values are over a single substructure drawn uniformly from the relevant sampling volume and with velocity drawn from \(f_v(\mathbf v)\). Subtracting the product of means then gives the covariance of the unprojected mass-normalized signal,
\begin{equation}
\begin{split}
    \Sigma(t,t')
    &=
    \langle N \rangle
    \left\langle
        s(t \mid \mathbf r_0,\mathbf v)
        s(t' \mid \mathbf r_0,\mathbf v)
    \right\rangle \\
    &=
    \langle N \rangle
    \int_{\mathcal V}
    \frac{d^3\mathbf r_0}{V}
    \int d^3\mathbf v\,
    f_v(\mathbf v)\,
    s(t \mid \mathbf r_0,\mathbf v)
    s(t' \mid \mathbf r_0,\mathbf v).
\end{split}
\label{eq:mass_normalized_signal_covariance}
\end{equation}
Here the distinct-substructure contribution in Eq.~\eqref{eq:signal_two_point_function} cancels against the product of means, while the single-substructure contribution remains. The corresponding residualized covariance is obtained by applying the timing-model projection,
\begin{equation}
    \tilde{\bm\Sigma}
    =
    \mathbf P
    \bm\Sigma
    \mathbf P^T .
\label{eq:projected_signal_covariance}
\end{equation}
Inserting the Doppler or Shapiro single-substructure response, together with the corresponding sampling volume, gives \(\tilde{\bm\Sigma}_D\) or \(\tilde{\bm\Sigma}_S\), respectively. In Fig.~\ref{fig:CovarianceSummary}, we illustrate the numerically evaluated Doppler and Shapiro covariances.

We present the detailed evaluations of these two cases in App.~\ref{app:DopplerCovariance} and App.~\ref{app:ShapiroCovariance}.
Because the Gaussian surrogate model is only expected to be appropriate in the large-$\langle N \rangle$ limit, we fix the volume to be the fiducial, unexpanded one. If the expanded volume procedure were applied, then it would generate nontrivial $\langle N \rangle$-dependence in the covariance, but only for $\langle N \rangle$ where the Gaussian approximation would not be expected to be accurate.

\subsubsection{Likelihood with the Gaussian surrogate}

If the projected signal distribution is well-approximated by a Gaussian in this regime, then the convolution integral of Eq.~\eqref{eq:single_pulsar_signal_marginal_likelihood} can be performed analytically, yielding
\begin{equation}
    p(\tilde{\mathbf r} \mid \langle N\rangle, M_\mathrm{sub})
    \approx
    \mathcal N_{\mathbf P}\!\left(
        \tilde{\mathbf r};\,
        \mathbf 0,\,
        \tilde{\mathbf C}_\mathrm{bkg} 
        + 
        M_\mathrm{sub}^2 
        \tilde{\bm{\Sigma}}_{\mathcal I}(\langle N\rangle)
    \right).
    \label{eq:covariance_surrogate_likelihood}
\end{equation}
Here \(\tilde{\bm{\Sigma}}_{\mathcal I}\) denotes the projected version of the Doppler or Shapiro signal covariance, and \(\mathcal N_{\mathbf P}\) is the projected Gaussian density defined in Eq.~\eqref{eq:projected_Gaussian_density}. The usefulness of this construction therefore depends on whether the large-\(\langle N\rangle\) signal is in fact close to Gaussian.

The two signal classes behave rather differently in this regard. For the Shapiro signal, the covariance integral is finite, so the large-\(\langle N\rangle\) process is well-motivated to approach a Gaussian limit. At finite \(\langle N\rangle\), however, this convergence can be delayed, and coherent non-Gaussian features may persist even when low-dimensional Gaussianity tests are approximately satisfied. For the Doppler signal, by contrast, the covariance is dominated by rare close flybys and is logarithmically sensitive to the small-impact-parameter region. Even after imposing a physical cutoff, convergence to Gaussianity is therefore expected to be much slower than in the Shapiro case, and the Gaussian approximation need not be reliable over the range of \(\langle N\rangle\) relevant for the projected sensitivities considered here. We study these finite-\(\langle N\rangle\) departures from Gaussianity in greater detail in App.~\ref{app:NonGaussianityTests}.

\subsection{Cubic-mode flow surrogate}
\label{sec:cubic_flow_surrogate}

\begin{figure*}[!htb]
    \centering
    \includegraphics[width=0.99\linewidth]{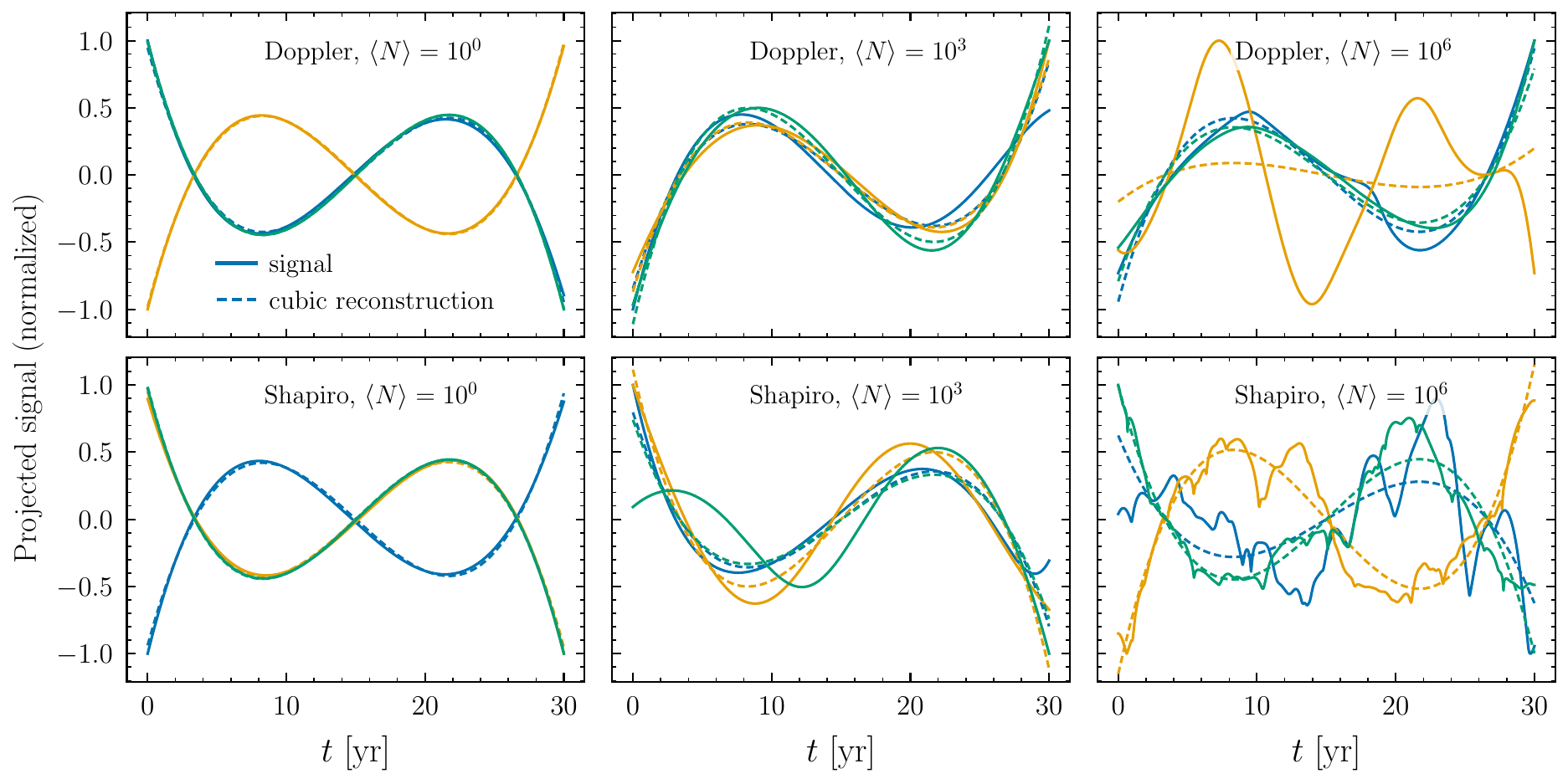}
    \caption{Illustration of the cubic-mode compression used in the flow-based surrogate likelihood. In each panel, solid curves show representative projected signal realizations, while dashed curves show their cubic reconstructions \(\tilde{s}^{(3)}_\mathcal{I}\tilde{\bm \ell}_3\). The signal class and the corresponding value of \(\langle N\rangle\) are indicated by the in-panel annotations. In each panel, the realizations are normalized independently to unit maximum absolute value in order to emphasize shape rather than overall amplitude. At small \(\langle N\rangle\), the projected signal is well-captured by the cubic mode. As \(\langle N\rangle\) increases, additional non-cubic structure becomes increasingly important, so the cubic compression becomes progressively more conservative.}
\label{fig:CubicCompression}
\end{figure*}

The covariance surrogate is appropriate when the signal is close to Gaussian, but it cannot capture the strongly non-Gaussian behavior expected when only a small number of substructures dominate the total signal. To address this regime, we construct a compressed flow surrogate likelihood based on the cubic mode, which is the leading polynomial component that survives timing-model subtraction in the static, low-number-density limit. In this construction, both the observed residuals and the simulated signal realizations are reduced to their cubic-mode amplitudes, and the likelihood is evaluated only for this one-dimensional summary statistic. We then model the abundance-dependent distribution of this cubic-mode signal amplitude with a conditional normalizing flow: an invertible neural density estimator in which the transformation, and hence the resulting probability density, depends on the conditioning variable \(\langle N\rangle\) \cite{Durkan:2019,2019arXiv191202762P}.

\subsubsection{Cubic-mode compression}

The motivation for the cubic-mode compression is physical. In the static, low-number-density regime, the signal is dominated by distant substructures whose timing residuals vary smoothly over the observing span and can therefore be expanded in low-order polynomials in time. Because the residualizing projection removes the constant, linear, and quadratic components, see App.~\ref{app:projection}, the leading surviving polynomial contribution is cubic. We therefore expect the projected signal in this regime to be well described by a \(t^3\)-like shape. At larger \(\langle N\rangle\), however, closer dynamical encounters and the superposition of many time-dependent contributions generate non-cubic structure. The cubic compression then remains a valid one-dimensional summary statistic, but it no longer captures all of the available signal information.

This distinction is important for interpreting the approximation. Previous cubic-mode estimates effectively searched for a \(t^3\)-like component in the full time-domain residuals \cite{Lee:2021zqw, NANOGrav:2023hvm}. Such an approach is accurate and well-motivated in the small-$\langle N \rangle$ limit when the signal itself is dominated by the cubic mode, but it becomes a misspecified time-domain likelihood once non-cubic signal components carry appreciable power: the data vector contains those components, while the signal model does not. In contrast, our compressed likelihood applies the same one-dimensional projection to the data, the signal model, and the background covariance. Components orthogonal to the cubic mode are therefore discarded from the likelihood, rather than retained in the data but omitted from the signal model. In this sense, the cubic compression is conservative rather than biased: it sacrifices sensitivity by ignoring signal information outside the cubic mode, but it avoids the artificial full-time-domain analysis in which a \(t^3\)-only signal model is fit to residuals that may contain substantial non-cubic substructure signal. Critically, this makes our projection-based analysis unbiased in the large-$\langle N \rangle$ limit, though at the expense of some sensitivity in the small-$\langle N \rangle$ limit.

We implement this compression using the cubic Legendre polynomial on the observing grid. Let \(\tau_i \in [-1,1]\) denote the affine-rescaled observing times introduced in App.~\ref{app:projection}, and define
\begin{equation}
    \ell_{3,i}
    =
    P_3(\tau_i),
    \qquad
    P_3(\tau)
    =
    \frac12(5\tau^3-3\tau).
    \label{eq:cubic_legendre_mode}
\end{equation}
Applying the residualizing projection and normalizing to unit norm gives the corresponding cubic mode in the projected residual space,
\begin{equation}
    \tilde{\bm \ell}_3
    =
    \frac{\mathbf P \bm \ell_3}
    {\sqrt{\bm \ell_3^T\mathbf P\bm \ell_3}} .
    \label{eq:projected_cubic_mode}
\end{equation}
Because the lower-order Legendre modes have been projected out, this is the leading polynomial template available to the residual signal in the static limit. The extent to which this cubic mode captures the projected signal is illustrated in Fig.~\ref{fig:CubicCompression}, which shows that the approximation is excellent at small \(\langle N\rangle\), while non-cubic structure becomes increasingly important as \(\langle N\rangle\) grows.

We now define a scalar-compressed data space by retaining only the amplitude along this cubic mode. For the observed projected residual, this amplitude is
\begin{equation}
    \tilde r^{(3)}
    \equiv
    \tilde{\bm \ell}_3^{\,T}\tilde{\mathbf r}.
    \label{eq:cubic_amplitude_data}
\end{equation}
For a mass-normalized projected signal realization, the corresponding cubic-mode amplitude is
\begin{equation}
    \tilde s^{(3)}
    \equiv
    \tilde{\bm \ell}_3^{\,T}\tilde{\mathbf s}.
    \label{eq:cubic_amplitude_signal}
\end{equation}
The likelihood constructed below is a likelihood for the scalar quantity \(\tilde r^{(3)}\), not a full time-domain likelihood with the signal forced to be proportional to \(\tilde{\bm \ell}_3\). Thus, the non-cubic components of both the data and the signal realizations are discarded by the compression rather than modeled inconsistently.

In this one-dimensional compressed space, the additive residual decomposition becomes
\begin{equation}
    \tilde r^{(3)}
    =
    \tilde r_{\mathrm{bkg}}^{(3)}
    +
    M_\mathrm{sub}\tilde s^{(3)},
    \label{eq:cubic_compressed_decomposition}
\end{equation}
where
\begin{equation}
    \tilde r_{\mathrm{bkg}}^{(3)}
    \equiv
    \tilde{\bm \ell}_3^{\,T}\tilde{\mathbf r}_{\mathrm{bkg}}
    \label{eq:cubic_background_amplitude}
\end{equation}
is the corresponding compressed background contribution.

\subsubsection{Likelihood with the cubic-mode flow surrogate}

Under the Gaussian background model, the compressed background contribution
\(\tilde{r}_{\mathrm{bkg}}^{(3)}\) is normally distributed with variance
\begin{equation}
    \sigma_3^2
    =
    \tilde{\bm \ell}_3^{\,T}
    \tilde{\mathbf{C}}_\mathrm{bkg}
    \tilde{\bm \ell}_3 .
    \label{eq:cubic_background_variance}
\end{equation}
Since the likelihood in this construction is defined only for the scalar cubic-mode amplitude, the relevant background model is the corresponding one-dimensional Gaussian distribution.

The resulting likelihood for the observed cubic-mode amplitude is
\begin{equation}
\begin{split}
    p(\tilde{r}^{(3)} \mid \langle N\rangle, M_\mathrm{sub})
    =
    \int d\tilde{s}^{(3)}\,
    &\mathcal N\!\left(
        \tilde{r}^{(3)};
        M_\mathrm{sub}\tilde{s}^{(3)},
        \sigma_3^2
    \right) \\
    &\times
    p_\mathrm{sig}(\tilde{s}^{(3)} \mid \langle N\rangle),
\end{split}
\label{eq:flow_compressed_likelihood}
\end{equation}
where $\mathcal{N}$ is the Gaussian probability density function and \(p_\mathrm{sig}(\tilde{s}^{(3)} \mid \langle N\rangle)\) denotes the distribution of the mass-normalized cubic-mode amplitude induced by the substructure signal model. This is the compressed analogue of Eq.~\eqref{eq:single_pulsar_signal_marginal_likelihood}: the Gaussian factor describes the background after compression to the cubic amplitude, while \(p_\mathrm{sig}\) describes the stochastic distribution of the compressed substructure signal. Because the compression onto the cubic mode is nonvanishing under the residualizing projection, all of $\tilde r^{(3)}$, $\tilde s^{(3)}$, and $\sigma_3^2$ are nonzero, and the standard Gaussian probability density function, rather than the projected Gaussian density $\mathcal{N}_\mathbf{P}$, suffices.

We estimate this one-dimensional signal distribution using the template banks introduced in Sec.~\ref{sec:template_bank_construction}. For each template-bank realization, we compute the corresponding value of \(\tilde{s}^{(3)}\), thereby obtaining Monte Carlo samples from \(p_\mathrm{sig}(\tilde{s}^{(3)} \mid \langle N\rangle)\) across the full range of \(\langle N\rangle\) considered in this work. We then train a conditional normalizing flow to represent this family of one-dimensional densities continuously as a function of the conditioning variable \(\langle N\rangle\)~\cite{Durkan:2019,2019arXiv191202762P}. Denoting the learned density by \(p_{\mathrm{flow}}\), we use
\begin{equation}
    p_\mathrm{sig}(\tilde{s}^{(3)} \mid \langle N\rangle)
    \approx
    p_{\mathrm{flow}}(\tilde{s}^{(3)} \mid \langle N\rangle)
    \label{eq:flow_surrogate_density}
\end{equation}
as the surrogate signal distribution in Eq.~\eqref{eq:flow_compressed_likelihood}. Additional details of the flow parameterization and training procedure are given in App.~\ref{app:FlowDetails}.

\subsection{Diffusion-based Monte Carlo surrogate}
\label{sec:diffusion_mc_surrogate}

The cubic-mode flow surrogate provides a simple compressed likelihood that is well motivated in the static regime, but it becomes increasingly incomplete in the dynamic regime and beyond, where non-cubic time-domain structure can carry substantial signal information.  To construct a likelihood that remains applicable across the full range of \(\langle N\rangle\), we instead learn a generative model for the full projected time-domain signal distribution and use its samples to approximate the marginalization in Eq.~\eqref{eq:single_pulsar_signal_marginal_likelihood}.

Concretely, we train a conditional diffusion model on the template banks described in Sec.~\ref{sec:template_bank_construction}, using the projected mass-normalized signal realizations \(\tilde{\mathbf{s}}\) as training data and conditioning on \(\langle N\rangle\). Diffusion models are generative models that learn to transform simple noise into samples from a target distribution through a sequence of denoising steps; in the conditional case, this denoising process also depends on an external conditioning variable, which here is the expected number of substructures \(\langle N\rangle\) \cite{ho2020denoisingdiffusionprobabilisticmodels,nichol2021improveddenoisingdiffusionprobabilistic}. Given a value of \(\langle N\rangle\), the trained model generates samples
\begin{equation}
    \tilde{\mathbf s}^{(k)}
    \sim
    p_\mathrm{diff}(\tilde{\mathbf s} \mid \langle N\rangle),
    \qquad
    k=1,\ldots,N_{\rm MC}.
    \label{eq:diffusion_signal_samples}
\end{equation}
These samples are used as a surrogate ensemble for draws from the target projected signal distribution \(p_\mathrm{sig}(\tilde{\mathbf s}\mid \langle N\rangle)\).

Under the Gaussian background model, the marginal likelihood can then be approximated by replacing the integral over \(p_\mathrm{sig}\) in Eq.~\eqref{eq:single_pulsar_signal_marginal_likelihood} with an average over diffusion-generated signal realizations,
\begin{equation}
    p(\tilde{\mathbf r} \mid \langle N\rangle, M_\mathrm{sub})
    \approx
    \frac{1}{N_{\rm MC}}
    \sum_{k=1}^{N_{\rm MC}}
    \mathcal N_{\mathbf P}\!\left(
        \tilde{\mathbf r};\,
        M_\mathrm{sub} \tilde{\mathbf s}^{(k)},
        \tilde{\mathbf C}_\mathrm{bkg}
    \right).
\label{eq:diffusion_mc_likelihood}
\end{equation}
This expression is the direct Monte Carlo analogue of Eq.~\eqref{eq:single_pulsar_signal_marginal_likelihood}: each generated signal realization defines a conditional Gaussian likelihood for the projected residuals, and the marginal likelihood is obtained by averaging these likelihoods over the generated signal ensemble.

In practice, for each value of \(\langle N\rangle\) at which we evaluate the likelihood, we draw \(N_{\rm MC}=128{,}000\) independent diffusion samples and evaluate the Monte Carlo average in Eq.~\eqref{eq:diffusion_mc_likelihood}. This choice is fixed by the convergence tests in App.~\ref{app:Convergence}, where we verify that the remaining Monte Carlo error is small compared with the features relevant for the projected sensitivity curves.

Our use of a diffusion model, rather than direct samples from the generative model, to perform this marginalization is motivated purely by computational cost. At large $\langle N \rangle$, generating signal realizations is extraordinarily costly, while generating samples from the diffusion model is orders of magnitude faster. In this manner, the use of the diffusion model allows us to build up a much larger template bank for likelihood marginalization, leading to better converged results, than would be possible with direct generative model samples.

In contrast to the cubic-mode flow surrogate in Sec.~\ref{sec:cubic_flow_surrogate}, this approach retains the full projected time-domain structure of the signal and is therefore sensitive to non-cubic and strongly non-Gaussian features that would be discarded by a low-dimensional compression. Although computationally more expensive, it provides a more faithful approximation of the target likelihood for the noise-only projected sensitivities considered here. Because the Monte Carlo likelihood remains costly to evaluate, we restrict its use in this work to projected sensitivities rather than fully calibrated inference on high-signal datasets. Additional details of the diffusion model and its training procedure are given in App.~\ref{app:DiffusionDetails}. Because this diffusion-based approach preserves all time information, we find that this analysis is marginally more sensitive in the static limit, where the signal is well-described by its cubic component, as compared to our projected cubic mode analysis.

\section{Projected sensitivities}
\label{sec:projections}

In this section, we use the surrogate likelihoods developed in Sec.~\ref{sec:likelihoods} to project the sensitivity of future pulsar timing observations to DM substructure. We focus on noise-only synthetic datasets generated using the benchmark observing scenarios and background models defined in Sec.~\ref{sec:mock_data}, and report the median expected \(95\%\) credible upper limit on the fractional abundance of substructure, \(f_\mathrm{sub}\), at fixed substructure mass \(M_\mathrm{sub}\). In Sec.~\ref{sec:Results}, we describe the null analyses used to compute projected limits. In Sec.~\ref{sec:likelihood_method_comparison}, we compare the covariance, cubic-flow, and diffusion-based surrogate likelihoods. In Sec.~\ref{sec:previous_projections}, we present our fiducial projections and compare them with previous white-noise-only estimates~\cite{Dror:2019twh,Ramani:2020hdo,Lee:2020wfn,Lee:2021zqw}. Finally, in Sec.~\ref{sec:noise_variations}, we study how the projected reach changes under systematic variations of the SGWB red-noise model.

Throughout this section, our goal is to understand precisely how the projected sensitivities are degraded when red noise is included in the timing model.
As discussed in Sec.~\ref{sec:substructure_likelihood}, the SGWB leads to an irreducible pulsar-term red noise contribution.
For some pulsars this will be the dominant pulsar-term red noise contribution, while for others additional intrinsic red-noise processes associated with pulsar spin noise or other pulsar-specific instabilities could enhance the red noise contribution~\cite{NANOGrav:2023hde}.
Since our goal here is to understand how red-noise processes degrade the sensitivity, it is easiest to quantify this effect when only a few red-noise parameters are included.
Therefore we omit the pulsar-specific red noise contributions in our timing model, which typically add $2 \times N_p$ parameters to the timing model~\cite{NANOGrav:2023gor} associated with the amplitude and index of the red-noise spectrum for each pulsar, and focus on understanding how the sensitivity changes as a function of the SGWB parameters, $A_\text{GW}, \gamma_\text{GW}$.
Unless otherwise specified, the SGWB background uses the fiducial NANOGrav-motivated parameters \(A_\mathrm{GW}=6.4\times 10^{-15}\) and \(\gamma_\mathrm{GW}=3.2\)~\cite{NANOGrav:2023gor}.

\subsection{Null analyses and projected limits}
\label{sec:Results}

\begin{figure*}[!htb]
    \centering
    \includegraphics[width=0.99\linewidth]{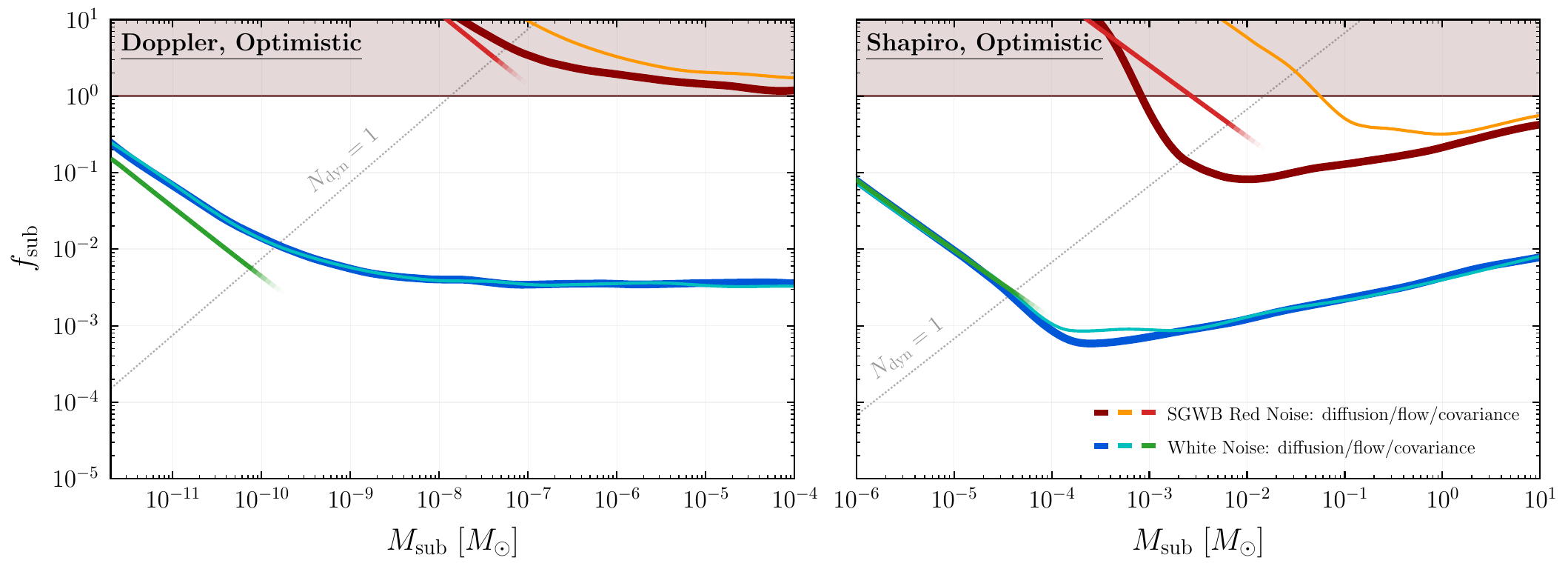}
    \caption{
    Projected sensitivities in the Optimistic observing scenario using the three surrogate likelihoods developed in this work. The left panel shows Doppler signals and the right panel shows Shapiro signals. Curves are shown both with white noise alone and with the additional uncorrelated pulsar-term SGWB red-noise contribution. The thick curves show the diffusion-based likelihood, the thinner curves show the cubic-mode flow surrogate, and the faded curves show the covariance-based surrogate in the regime where the large-\(\langle N\rangle\) approximation is expected to apply. The dotted grey line labeled \(N_\mathrm{dyn}=1\) indicates where the expected number of substructures in the dynamical volume is unity; regions above this line approach the large-\(\langle N\rangle\) limit, while regions below it approach the small-\(\langle N\rangle\) limit. See Eq.~\eqref{eq:NdynDef} for the definition of the expected dynamic number, as compared to the expected number $\langle N \rangle$ in the fiducial sampling volume as defined in Eq.~\eqref{eq:NDef}. Across the surrogate constructions, inclusion of the SGWB red-noise contribution substantially degrades the projected sensitivity, especially for Doppler signals. In the case of the Doppler signal, the parametric non-Gaussianity of the signal, even deep in the large $\langle N \rangle$ limit, results in systematic disagreement between the fiducial diffusion model sensitivity projections and those developed under the covariance model.
    }
    \label{fig:AnalysisVariations}
\end{figure*}

To compute projected limits, we generate \(25\) independent noise-only PTA datasets for each observing and noise scenario. At fixed substructure mass \(M_\mathrm{sub}\), we place a log-uniform prior on the fractional abundance of substructure over
\begin{equation}
    10^{-4} \leq f_\mathrm{sub} \leq 10^{2}.
    \label{eq:prior}
\end{equation}
The \(95^\mathrm{th}\) percentile of this log-uniform prior is \(f_\mathrm{sub}\simeq 50\), and therefore our limits for \(f_\mathrm{sub}\leq 1\) are insensitive to the prior range in Eq.~\eqref{eq:prior}.

For a given signal class, the fractional abundance determines the expected number of substructures in the fiducial sampling volume according to
\begin{equation}
    \langle N \rangle
    =
    \frac{
        f_\mathrm{sub}\,\rho_\mathrm{DM}\,V^\mathrm{fid}
    }{
        M_\mathrm{sub}
    },
    \label{eq:number_fraction}
\end{equation}
where we take \(\rho_\mathrm{DM}=0.4\,\mathrm{GeV}/\mathrm{cm}^3\). The likelihood at fixed \(M_\mathrm{sub}\) and \(f_\mathrm{sub}\) is therefore evaluated as
\begin{equation}
    p(\{\tilde{\mathbf r}_i\}\mid M_\mathrm{sub},f_\mathrm{sub})
    =
    p\!\left(
        \{\tilde{\mathbf r}_i\}
        \mid
        M_\mathrm{sub},
        \langle N\rangle
    \right),
\end{equation}
with $\langle N \rangle$ related to $f_\mathrm{sub}$ determined by Eq.~\eqref{eq:number_fraction}, using one of the surrogate likelihoods developed in Sec.~\ref{sec:likelihoods}. For the substructure masses considered in this work, the prior interval for the fractional abundance in Eq.~\eqref{eq:prior} maps entirely into the training range in $\langle N \rangle$ defined in Eq.~\eqref{eq:Nprior}. 

We then compute the marginal posterior for \(f_\mathrm{sub}\) at fixed \(M_\mathrm{sub}\) and extract the 95\% credible upper limit. Repeating this procedure for each of the \(25\) noise-only datasets, we report the median \(95\%\) credible upper limit as the expected sensitivity.

\subsection{Comparison of likelihood approximation}
\label{sec:likelihood_method_comparison}

We first compare the three surrogate likelihoods developed in Sec.~\ref{sec:likelihoods}: the covariance-based likelihood of Sec.~\ref{sec:covariance_surrogate}, the cubic-flow likelihood of Sec.~\ref{sec:cubic_flow_surrogate}, and the diffusion-based likelihood of Sec.~\ref{sec:diffusion_mc_surrogate}. The comparison is shown in Fig.~\ref{fig:AnalysisVariations} for both Doppler and Shapiro signals in the Optimistic observing scenario, with and without the uncorrelated pulsar-term SGWB red-noise contribution.

The three methods behave as expected from the discussion in Sec.~\ref{sec:likelihoods}. In the small-\(\langle N\rangle\) regime, to the right of the dotted \(N_\text{dyn}=1\) line in Fig.~\ref{fig:AnalysisVariations}, the cubic-flow likelihood is close to the diffusion-based likelihood for both Doppler and Shapiro signals because the projected signal is dominated by the leading cubic mode that survives timing-model subtraction. At larger \(\langle N\rangle\), the cubic-flow likelihood becomes conservative: the cubic compression discards non-cubic signal information rather than fitting a misspecified cubic-only model to the full time-domain residuals. For the Shapiro signal, the covariance-based likelihood agrees with the diffusion-based likelihood in the large-\(\langle N\rangle\) regime, to the left of the dotted line, where the signal is well described by a Gaussian process. These limiting agreements validate the diffusion-based likelihood in the regimes where simpler and more interpretable approximations are expected to be accurate.

The Doppler signal provides a useful contrast. Because its covariance is sensitive to rare close passages, the Doppler signal approaches Gaussianity slowly, as discussed in App.~\ref{app:DopplerCovariance}. As a result, the covariance-based likelihood is not expected to reproduce the diffusion-based result over the mass range considered here, even when \(\langle N\rangle\) is large. This behavior is visible in Fig.~\ref{fig:AnalysisVariations} from the different large-\(\langle N\rangle\) mass scalings: the covariance approximation gives \(f_\text{sub}\propto M_\text{sub}^{-1}\), while the diffusion approximation gives a shallower scaling, approximately \(f_\text{sub}\propto M_\text{sub}^{-3/4}\).

Up to prior-volume effects, the diffusion-based likelihood gives the following scaling laws in the small- and large-\(\langle N\rangle\) limits over the masses considered in this work. For the Doppler signal,
\begin{align}
    f_\text{sub} \propto \begin{cases}
        M_\text{sub}^{-3/4} & \langle N \rangle \gg 1 \,, \\[1ex]
        M_\text{sub}^0 & \langle N \rangle \ll 1 \,,
    \end{cases}
\end{align}
while for the Shapiro signal,
\begin{align}
    f_\text{sub} \propto \begin{cases}
        M_\text{sub}^{-1} & \langle N \rangle \gg 1 \,, \\[1ex]
        M_\text{sub}^{1/3} & \langle N \rangle \ll 1 \,.
    \end{cases}
\end{align}
The Shapiro scalings and the Doppler small-\(\langle N\rangle\) scaling are consistent with previous estimates~\cite{Schutz:2016khr, Dror:2019twh, Ramani:2020hdo}. The Doppler large-\(\langle N\rangle\) scaling differs from the covariance-based expectation used in previous estimates~\cite{Ramani:2020hdo}, reflecting the non-Gaussian structure that remains important over the mass range shown. At still smaller masses, we expect the Doppler diffusion result to asymptote slowly toward the Gaussian scaling \(f_\text{sub}\propto M_\text{sub}^{-1}\), consistent with the slow approach to Gaussianity discussed in App.~\ref{app:DopplerCovariance}.

The diffusion-based likelihood is therefore most valuable in the intermediate regime, where neither limiting approximation is fully reliable. For the Shapiro signal, this occurs when the signal is not yet sufficiently Gaussian for the covariance surrogate to apply, but substantial information is already present outside the cubic mode. For the Doppler signal, the diffusion-based likelihood remains important even deeper into the large-\(\langle N\rangle\) regime because the signal retains appreciable non-Gaussian structure.

\subsection{Comparison with previous projections}
\label{sec:previous_projections}

\begin{figure*}[!htb]
    \centering
    \includegraphics[width=0.99\linewidth]{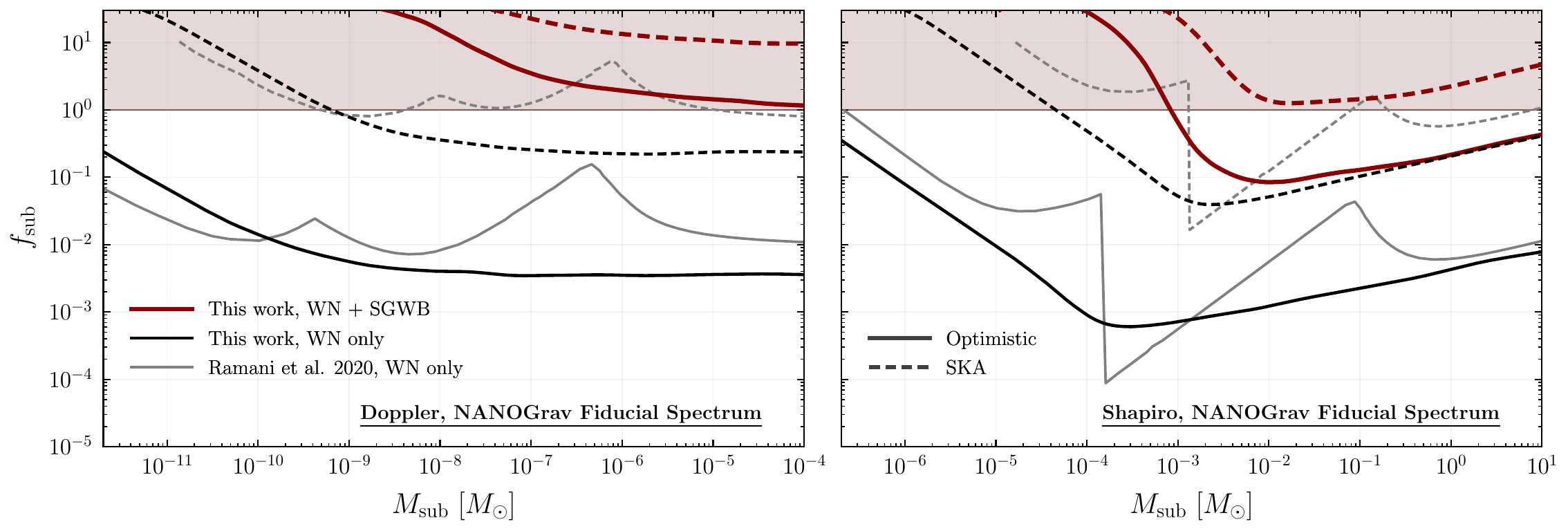}
    \caption{
    Fiducial projected sensitivities obtained with the diffusion-based likelihood. The left panel shows Doppler signals and the right panel shows Shapiro signals. Red curves show the projected sensitivities obtained in this work after including both white noise and the uncorrelated pulsar-term SGWB red-noise contribution, assuming the NANOGrav 15-year free-index power-law fit with \(A_\mathrm{GW}=6.4\times 10^{-15}\) and \(\gamma_\mathrm{GW}=3.2\). Black curves show the corresponding white-noise-only projections from this work. Grey curves show the previous white-noise-only projections of Ref.~\cite{Ramani:2020hdo}. Solid and dashed curves correspond to the Optimistic and SKA observing scenarios, respectively. The comparison between the black and grey curves shows that our white-noise projections broadly reproduce previous estimates, while the red curves illustrate how the SGWB red-noise contribution substantially weakens the projected sensitivity, especially for Doppler signals.
    }
    \label{fig:FiducialProjections}
\end{figure*}

\begin{figure*}[!htb]
    \centering
    \includegraphics[width=\linewidth]{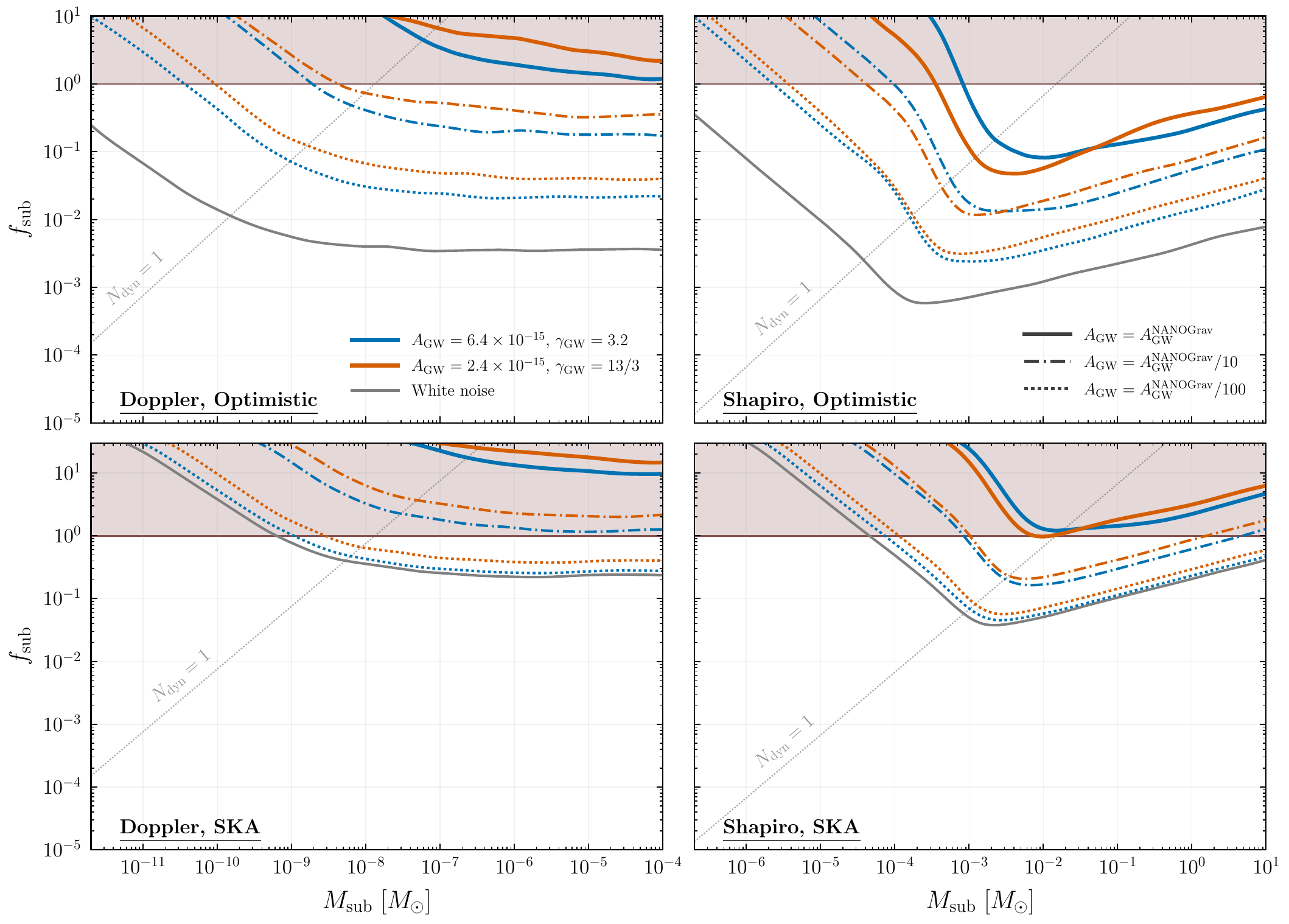}
    \caption{
    Projected sensitivities to substructure signals under systematic variations of the SGWB noise model. The top panels show our Optimistic observing scenario, while the bottom panels show the SKA observing scenario. The left panels show Doppler signals, and the right panels show Shapiro signals. Blue curves use the free-index SGWB power-law fit from the NANOGrav 15-year analysis, with $\gamma_\mathrm{GW}=3.2$, while orange curves use the fixed-index SMBHB-motivated value, $\gamma_\mathrm{GW}=13/3$. For each spectral index, solid, dot-dashed, and dotted curves show projections for $A_\mathrm{GW}=A_\mathrm{GW}^{\mathrm{NANOGrav}}$, $A_\mathrm{GW}^{\mathrm{NANOGrav}}/10$, and $A_\mathrm{GW}^{\mathrm{NANOGrav}}/100$, respectively. For $\gamma_\mathrm{GW}=3.2$, the NANOGrav-normalized amplitude is $A_\mathrm{GW}^{\mathrm{NANOGrav}}=6.4\times10^{-15}$; for $\gamma_\mathrm{GW}=13/3$, it is $A_\mathrm{GW}^{\mathrm{NANOGrav}}=2.4\times10^{-15}$. The gray curves show the corresponding white-noise-only projections. The dotted gray diagonal indicates the approximate transition where the expected number of dynamically relevant substructures is unity, and the shaded region marks $f_\mathrm{sub}>1$. 
    }
    \label{fig:NoiseVariations}
\end{figure*}

Our fiducial diffusion-based sensitivities are compared with previous white-noise-only projections in Fig.~\ref{fig:FiducialProjections}. The red curves show the projected sensitivity obtained in this work using the NANOGrav-fiducial SGWB, the black curves show the projected sensitivity obtained in this work assuming only white noise, while the grey curves show the white-noise-only estimates of Ref.~\cite{Ramani:2020hdo}. The optimistic PTA observing scenario limits are shown as solid lines, while those assuming the SKA scenario are shown as dashed lines.

Overall we find relatively good agreement in the sensitivity derived here compared to previous works when only white noise is included.
The limits derived here scale identically with respect to $M_\text{sub}$ to those in Ref.~\cite{Ramani:2020hdo} for both small and large $M_\text{sub}$.
The main discrepancies occur at intermediate $M_\text{sub}$.
In this regime the previous, analytic results rely on using approximate signal shapes for the closest substructure, and are cut-off arbitrarily at large $M_\text{sub}$ due to the expectation that these signals would not persist when a substructure is not expected to pass within the dynamical radius~\cite{Dror:2019twh}.
The fact that a more complete analysis can improve the sensitivity by actually modeling the signal in this regime is then not surprising.

The effect of red noise is most severe for Doppler signals. In both the SKA and Optimistic scenarios, the Doppler limits remain above \(f_\mathrm{sub}=1\) across the mass range shown once the fiducial SGWB red-noise contribution is included. This represents a substantial weakening relative to the corresponding white-noise-only estimates.
The Shapiro channel is slightly more robust to red noise. The SGWB still weakens the projected sensitivity relative to the white-noise-only estimates, but it does not eliminate the reach entirely. In the Optimistic scenario, Shapiro timing retains sensitivity to \(f_\mathrm{sub}<1\) over part of the mass range shown. In the SKA scenario, the reach is more marginal, but Shapiro remains more promising than Doppler, with sensitivity near fractional abundances of order unity in the presence of the NANOGrav-fiducial SGWB red-noise background.

Although these projected sensitivities are weaker than existing compact-lensing constraints for sufficiently compact pointlike substructures, extended profiles provide a possible target for future Shapiro timing searches because their compact-lensing signatures are suppressed~\cite{Croon:2020wpr}. Assessing this case quantitatively would require replacing the point-mass response used here with an extended-profile timing response and retraining the corresponding surrogate likelihoods.

\subsection{Systematic Variations of the SGWB}
\label{sec:noise_variations}

While the NANOGrav-motivated SGWB model provides a useful fiducial benchmark, the relevant quantity for the uncorrelated pulsar-term forecasts is the effective red-noise floor that remains in the timing residuals after all deterministic modeling and subtraction. If the observed nanohertz excess is truly a stochastic process with an amplitude comparable to current PTA-inferred values, then its impact on the uncorrelated Doppler and Shapiro searches is largely independent of whether it is interpreted specifically as a gravitational-wave background or as another source of stochastic red noise with the same single-pulsar covariance. Intrinsic pulsar red noise, for example, would also degrade sensitivity to the uncorrelated pulsar-term Doppler and Shapiro signals. Improved prospects would therefore require the effective stochastic red-noise floor to be lower than the fiducial model considered here, which could be the case if the real $A_\text{GWB}, \gamma_\text{GWB}$ are at the edge of the currently preferred posterior~\cite{NANOGrav:2023gor}.

We therefore consider systematic variations of the SGWB power-law model as a proxy for varying this effective red-noise floor. The first spectrum is the NANOGrav 15-year free-index fit \cite{NANOGrav:2023gor} used in our fiducial projections,
\begin{equation}
    A_\mathrm{GW}=6.4\times10^{-15},
    \qquad
    \gamma_\mathrm{GW}=3.2 .
\end{equation}
The second is the fixed-index fit with \(\gamma_\mathrm{GW}=13/3\), motivated by the canonical SMBHB spectrum, for which the corresponding NANOGrav-preferred amplitude is \(A_\mathrm{GW}=2.4\times10^{-15}\) \cite{NANOGrav:2023gor}. Similar values for $A_\mathrm{GW}$ under $\gamma = 13/3$ were obtained by other pulsar timing array collaborations \cite{EPTA:2023fyk, Reardon:2023gzh, Xu:2023wog, Miles:2024seg}. In general, the projected sensitivity to substructure is set by the total red noise power, which is jointly determined by $A_\mathrm{GW}$ and $\gamma_\mathrm{GW}$. It is therefore critical that variations in $\gamma_\mathrm{GW}$ be accompanied by their associated inferred value of $A_\mathrm{GW}$ in order to accurately represent the impact of the existing statistical uncertainty in the SGWB as it propagates into uncertainty in the projected reach. For each spectral index, we also consider amplitudes reduced by factors of \(10\) and \(100\). These variations should therefore be interpreted as alternate effective red-noise budgets in which the stochastic component relevant for the substructure search is lower than in the fiducial NANOGrav-motivated model. Such a reduction would require future analyses to revise the inferred SGWB strain downward, for example because systematic effects or other modeled low-frequency components are identified and mitigated.

The projected sensitivities for the Optimistic observing scenario with varied SGWB assumptions are shown in the top panels of Fig.~\ref{fig:NoiseVariations}. Our conclusions regarding the degradation of Doppler and Shapiro sensitivity are relatively insensitive to the choice of \(\gamma_\mathrm{GW}\) given a self-consistently estimated $A_\mathrm{GW}$ for that index. Although the NANOGrav fixed-index analysis prefers a lower value of \(A_\mathrm{GW}\), taking \(\gamma_\mathrm{GW}=13/3\) instead of \(3.2\) makes the background spectrum steeper, increasing the relative amount of low-frequency noise power while decreasing the high-frequency power. The net effect is modest: the steeper spectrum slightly weakens the Shapiro reach in the static regime, improves it in the dynamic and stochastic regimes, and weakens the Doppler reach across the full range of masses shown.

Artificially varying the SGWB amplitude at fixed index has a much larger effect.\footnote{By artificial, we mean that the variations do not correspond to parameter uncertainties derived in the standard NANOGrav analysis. Instead these variations should be understood as illustrative ones which clarify the parametric dependence of the estimated sensitivity on the SGWB amplitude.} For the Doppler signal, the fiducial NANOGrav-motivated amplitude keeps the projected sensitivity above \(f_\mathrm{sub}=1\) across the full mass range shown. Reducing \(A_\mathrm{GW}\) by one decade improves the reach, but still leaves the projection far from the white-noise-only limit. A two-decade reduction is needed before the Doppler sensitivity begins to recover the qualitative behavior of the white-noise projection. The Shapiro signal is less severely affected: even at the fiducial SGWB amplitude, the Optimistic scenario retains sensitivity to \(f_\mathrm{sub}<1\) over a finite range of substructure masses. Reducing \(A_\mathrm{GW}\) by one or two decades steadily improves the reach and moves the projection toward the white-noise-only result, especially near the dynamic regime where the strongest limits are obtained. Nevertheless, the Shapiro projections also remain visibly degraded relative to the white-noise limit. Thus, in the Optimistic scenario, Shapiro timing is the more robust pulsar-term search, while Doppler timing requires an effective stochastic red-noise floor substantially below the fiducial SGWB value to recover competitive sensitivity.

The analogous projections for the SKA observing scenario are shown in the bottom panels of Fig.~\ref{fig:NoiseVariations}. The same qualitative conclusions hold, but with the region of best reach shifted toward larger substructure masses and with weaker limits overall relative to the Optimistic case. For Doppler timing, at the fiducial SGWB amplitude and for either power-law index, the projected limits remain above \(f_\mathrm{sub}=1\) over the full mass range. Even a one-decade reduction in \(A_\mathrm{GW}\) is insufficient to recover sensitivity below unity. In contrast to the Optimistic scenario, a two-decade reduction brings the projection much closer to the white-noise result, due to the larger white-noise amplitude assumed in this scenario. The Shapiro channel again fares better: the fiducial SGWB case is close to \(f_\mathrm{sub}=1\) near its optimal mass range, while reducing \(A_\mathrm{GW}\) by one or two decades produces sub-unity sensitivity over an increasingly broad interval. Thus, for the SKA scenario as well, Shapiro timing is the more robust pulsar-term channel, whereas Doppler timing requires a substantially reduced effective stochastic background to become competitive.

Taken together, these variations show that the pulsar-term projections are highly sensitive to the residual stochastic red-noise amplitude, not only to the white-noise timing precision. To place these variations in context, it is useful to compare them with the range of theoretically motivated \(A_\mathrm{GW}\) values. While we do not attempt a comprehensive survey here, App.~A of Ref.~\cite{NANOGrav:2023hfp} compiles SMBHB population-model predictions in terms of \(68\%\) containment intervals for \(A_{\rm yr}\), which is identical to our definition of \(A_\mathrm{GW}\). Across the models listed there, the lowest lower edge of the interval is \(A_\mathrm{GW}=6.1\times 10^{-17}\), comparable to our two-decade-suppressed case, while the highest upper edge is \(A_\mathrm{GW}=1.5\times 10^{-14}\). Thus, although the fiducial NANOGrav-motivated amplitude is the appropriate benchmark for current PTA evidence, the suppressed-amplitude cases considered here are also useful reference points for the lower end of published SMBHB population predictions.

It is important, however, not to overstate the plausibility of these suppressed-amplitude scenarios. They are not nearby points in the current NANOGrav posterior. The NANOGrav-inferred SGWB amplitude is measured with uncertainties of order tens of percent at fixed \(\gamma_\mathrm{GW}\)~\cite{NANOGrav:2023gor}, whereas the variations considered here reduce \(A_\mathrm{GW}\) by factors of \(10\) and \(100\). Realizing the corresponding sensitivities would therefore require a substantial change in the interpretation of the present nanohertz signal. Any low-frequency power currently contributing to the inferred SGWB amplitude would have to be attributed to components that can be modeled or removed, with the residual stochastic component suppressed to the \(\sim 10\%\) or \(\sim 1\%\) amplitude level. At the same time, the true SMBHB background would have to lie near the low-amplitude end of published population-model predictions. The suppressed-amplitude curves should therefore be read as highly optimistic lower-background benchmarks, not as likely outcomes of the current NANOGrav measurement.

\section{Projecting sensitivity to the correlated Doppler signal}
\label{sec:correlated_doppler}

Thus far, we have focused on substructure signals that appear as independent realizations in different pulsars: the pulsar-acceleration Doppler signal and the line-of-sight Shapiro signal. A complementary channel is provided by the correlated Doppler signal induced by acceleration of the Solar System barycenter, often referred to as the Earth term. In this case, the same local acceleration field is observed across the PTA, with each pulsar responding through projection along its line-of-sight. The signal therefore produces a coherent angular pattern across the array.

A full time-domain correlated analysis is substantially more computationally demanding than the independent-pulsar analyses considered above. In the independent-pulsar searches, the pulsar-term signal realizations are statistically independent across pulsars, so the likelihood factorizes over pulsars once the common Earth-term background has been subtracted. For the correlated Doppler signal, this factorization is lost: the signal is described by a single stochastic Solar-System acceleration field whose projections determine the joint response of all pulsars. A fully general surrogate likelihood would therefore need to model the joint time-domain signal distribution across the entire PTA, including any non-Gaussian correlations induced by the common substructure realization.

Even before modeling this joint non-Gaussian signal distribution, the correlated Gaussian background sets a significant computational floor. The SGWB contribution must be represented by the full cross-pulsar covariance, including the Hellings--Downs-correlated Earth term together with the uncorrelated pulsar terms. The resulting background covariance lives in the residual space of all pulsars and observing times, and must be evaluated repeatedly over the grid of substructure masses, abundances, noise realizations, and SGWB noise models considered here. This cost is especially severe for the small projected white-noise levels in our Optimistic observing scenario. In standard PTA analyses, the correlated red-noise covariance can often be represented accurately using a truncated set of low-frequency Fourier modes \cite{NANOGrav:2023gor}. Here, however, the timing-model projection and very small white-noise levels leave many temporal eigenmodes with non-negligible SGWB power. A truncation to only the lowest few frequencies is therefore not sufficient to represent the correlated background covariance at the precision required for these projected sensitivities. This motivates a compressed approximation to the correlated Doppler likelihood.

The approximation we use is to model only the cubic component of the correlated Doppler signal, analogous to the cubic-mode compression introduced in Sec.~\ref{sec:cubic_flow_surrogate}. This choice is motivated by two observations. First, the independent-pulsar Doppler analyses show that the cubic-mode flow gives conservative sensitivities that remain comparable to the diffusion-based likelihood in both white-noise and red-noise-dominated scenarios; see Fig.~\ref{fig:AnalysisVariations}. This indicates that, for the noise-only sensitivity projections considered here, the leading cubic component captures much of the Doppler information relevant for setting limits. Second, unlike an independent scalar compression applied separately to each pulsar, the correlated cubic compression does not discard the angular structure of the Earth-term signal.

To model the cubic component while incorporating this angular structure, we use a cubic-\textit{vector} flow model. In this construction, the latent signal is represented by a three-dimensional cubic vector, whose projections onto the pulsar line-of-sight directions determine the coherent response across the PTA. This projection structure follows directly from Eq.~\eqref{eq:Doppler}.

The remainder of this section constructs and applies this correlated-search likelihood. In Sec.~\ref{sec:correlated_doppler_flow}, we define the cubic-vector flow model for the correlated Doppler signal. In Sec.~\ref{sec:correlated_doppler_noise}, we construct the compressed cross-pulsar noise covariance, retaining the full Hellings--Downs angular structure of the SGWB. In Sec.~\ref{sec:correlated_doppler_likelihood}, we combine the signal and noise models into the correlated-Doppler likelihood and describe the corresponding null analyses. Finally, in Sec.~\ref{sec:correlated_doppler_results}, we present the projected sensitivities and compare the correlated Earth-term reach with the independent pulsar-term Doppler search.

For notational simplicity, in this section we suppress labels indicating that the signal type is Doppler, since it always is. We retain tildes for quantities whose definitions explicitly depend on the timing-model projection.

\begin{figure}[!tb]
  \centering
  \includegraphics[width=0.4\textwidth]{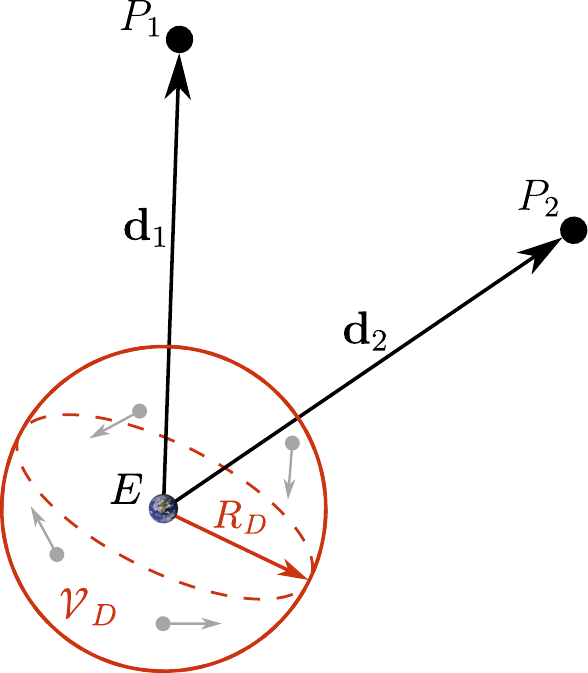}
  \caption{As in Fig.~\ref{fig:SamplingVolumes}, but illustrating the sampling volume for the correlated Doppler signal sourced by acceleration of the Solar System barycenter.}
  \label{fig:EarthSampling}
\end{figure}

\subsection{Cubic-vector flow model}
\label{sec:correlated_doppler_flow}

The correlated Doppler signal is determined by a single Solar-System acceleration response vector, projected along each pulsar line-of-sight. This projection structure follows directly from Eq.~\eqref{eq:Doppler}, where the Doppler residual is proportional to the dot product between \(\hat{\mathbf d}\) and the substructure-induced vector response. The signal variable for the correlated Earth term search is therefore a three-dimensional cubic vector rather than an independent scalar cubic amplitude for each pulsar.

We generate training samples for this vector using the same substructure Monte Carlo signal model described in Sec.~\ref{sec:SignalModel}, with the Doppler sampling volume now centered on the Earth rather than on an individual pulsar; see Fig.~\ref{fig:EarthSampling}. In this section, \(\langle N\rangle\) denotes the fiducial-volume expectation value for this Earth-centered Doppler sampling region. For each value of \(\langle N\rangle\), we draw Doppler realizations, compute the three Cartesian components of the mass-normalized vector response, apply the timing-model projection, and project each component onto the normalized residual-space cubic mode \(\tilde{\bm \ell}_3\). This defines
\begin{equation}
    \mathbf{s}^{(3)}
    \equiv
    \left(
        \tilde{\bm \ell}_3^{\,T}\tilde{\mathbf s}_{x},
        \tilde{\bm \ell}_3^{\,T}\tilde{\mathbf s}_{y},
        \tilde{\bm \ell}_3^{\,T}\tilde{\mathbf s}_{z}
    \right),
    \label{eq:correlated_doppler_cubic_vector}
\end{equation}
where \(\tilde{\mathbf s}_{x}\), \(\tilde{\mathbf s}_{y}\), and \(\tilde{\mathbf s}_{z}\) are the residualized, mass-normalized Doppler responses in the Cartesian directions. For a PTA with pulsar line-of-sight unit vectors \(\{\hat{\mathbf d}_a\}\), the correlated Doppler contribution to the compressed residual of pulsar \(a\) is
\begin{equation}
    \mathbf{r}_a^{(3)}
    =
    M_\mathrm{sub}\,
    \hat{\mathbf d}_a \cdot \mathbf s^{(3)} \, .
    \label{eq:correlated_doppler_projected_response}
\end{equation}
This expression makes explicit that the same latent vector \(\mathbf s^{(3)}\) determines the coherent signal pattern across the entire array.

Because the substructure population is isotropic, the distribution of \(\mathbf s^{(3)}\) is rotationally invariant. Defining
\begin{equation}
    s^{(3)} \equiv |\mathbf{s}^{(3)}| ,
\end{equation}
the direction \(\hat{\mathbf{s}}^{(3)} = \mathbf{s}^{(3)} / s^{(3)}\) is uniformly distributed on the sphere, and
\begin{equation}
    p_{\rm sig}(\mathbf{s}^{(3)} \mid \langle N \rangle)
    =
    \frac{
        p_s(s^{(3)} \mid \langle N \rangle)
    }{
        4\pi (s^{(3)})^2
    } .
    \label{eq:isotropic_cubic_vector_density}
\end{equation}
We therefore train a conditional normalizing flow only for the radial distribution \(p_s(s^{(3)} \mid \langle N \rangle)\), using \(\log_{10}\langle N \rangle\) as the conditioning variable. We follow the flow-modeling procedure described in App.~\ref{app:FlowDetails}, with the modification that the learned variable distribution is of the positive-definite radius \(s^{(3)}\), rather than a signed cubic amplitude.

\subsection{Compressed noise covariance for the correlated search}
\label{sec:correlated_doppler_noise}

For the correlated Doppler search, we use the full cross-pulsar background covariance rather than the Earth-term-subtracted covariance used for the uncorrelated pulsar-term searches (discussed in detail in Sec.~\ref{sec:observing_backgrounds}). After timing-model projection, the relevant time-domain covariance is the block covariance \((\tilde{\mathbf C}_{\rm bkg})_{ab}\) defined in Eq.~\eqref{eq:full_projected_pta_background_covariance}. This covariance includes independent white measurement noise and, when included, the full SGWB contribution, with both the Hellings--Downs-correlated Earth term and the uncorrelated pulsar terms.

As in Sec.~\ref{sec:cubic_flow_surrogate}, the correlated Doppler likelihood is evaluated after compressing each pulsar's projected residual onto the cubic mode. The corresponding covariance of the compressed amplitudes is therefore
\begin{equation}
    (\mathbf C^{(3)})_{ab}
    =
    \tilde{\bm \ell}_3^{\,T}
    (\tilde{\mathbf C}_{{\rm bkg}})_{ab}
    \tilde{\bm \ell}_3 ,
    \label{eq:cubic_hd_covariance}
\end{equation}
where \(\tilde{\bm \ell}_3\) is the normalized cubic mode in the residualized time-domain space. This compressed covariance \(\mathbf C^{(3)}\) is the background covariance used in the correlated-Doppler likelihood.

\subsection{Likelihood and null analyses}
\label{sec:correlated_doppler_likelihood}

\begin{figure*}[!htb]
    \centering
    \includegraphics[width=.99\linewidth]{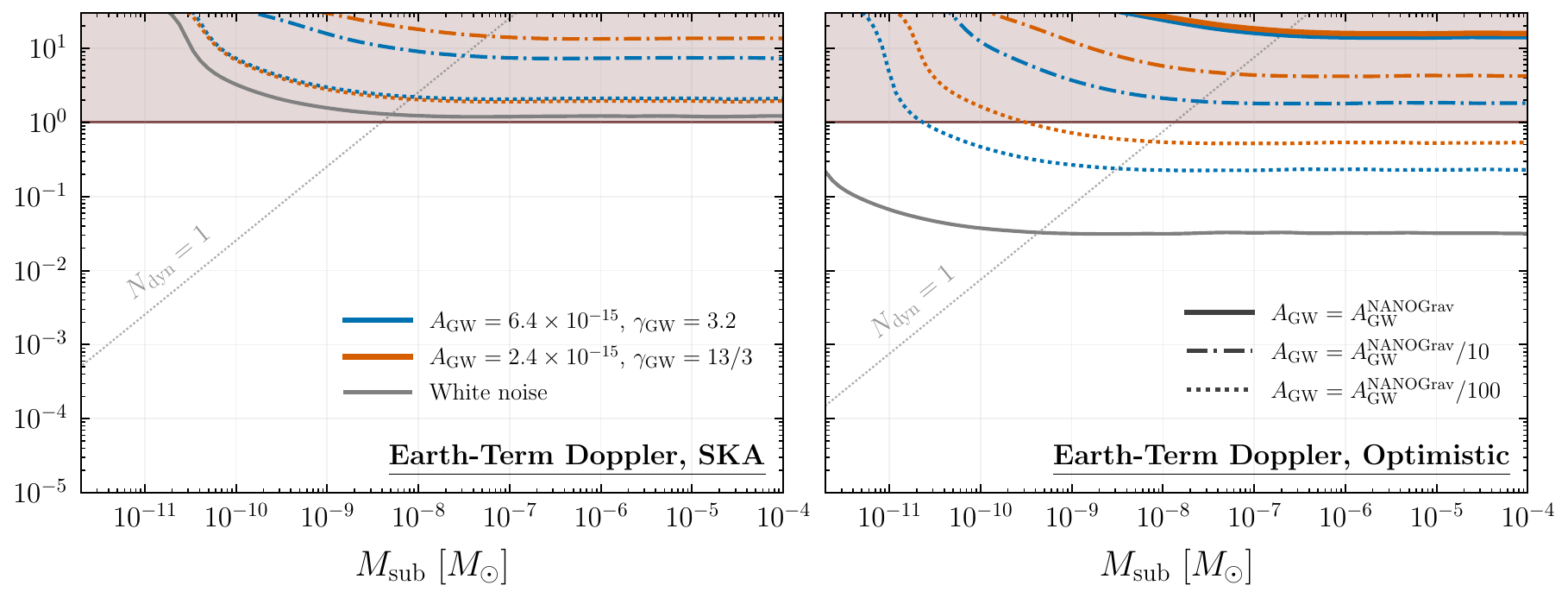}
    \caption{(\textit{Left panel}) Projected sensitivity to substructure through the correlated Doppler signal in the SKA observing scenario. (\textit{Right panel}) As in the left panel, but for the Optimistic observing scenario. We show the same SGWB noise variations considered in Sec.~\ref{sec:noise_variations}.}
    \label{fig:EarthTermVariations}
\end{figure*}

The compressed data vector, which follows from Eq.~\eqref{eq:correlated_doppler_projected_response}, is
\begin{equation}
    \mathbf r^{(3)}
    \equiv
    \left\{
        \tilde{\bm \ell}_3^{\,T}\tilde{\mathbf r}_a
    \right\}_{a=1}^{N_p}.
\end{equation}
Combining the cubic-vector signal model with the compressed noise covariance, the likelihood is obtained by marginalizing over the latent vector \(\mathbf s^{(3)}\):
\begin{widetext}
\begin{equation}
\begin{split}
    p(\mathbf r^{(3)} \mid M_\mathrm{sub}, \langle N\rangle)
    &=
    \int d^3 \mathbf s^{(3)}\,
    \mathcal N
    \!\left(
        \mathbf r^{(3)};\,
        \left\{
            M_\mathrm{sub}\,\hat{\mathbf d}_a \cdot \mathbf s^{(3)}
        \right\}_{a=1}^{N_p},
        \mathbf C^{(3)}
    \right)
    p_{\rm sig}(\mathbf s^{(3)} \mid \langle N\rangle) \\
    &=
    \int d s^{(3)}\,
    p_s(s^{(3)}\mid \langle N \rangle)
    \int \frac{d\Omega_{\hat{\mathbf s}^{(3)}}}{4\pi}\,
    \mathcal N
    \!\left(
        \mathbf r^{(3)};\,
        \left\{
            M_\mathrm{sub}\,s^{(3)}
            \hat{\mathbf d}_a \cdot \hat{\mathbf s}^{(3)}
        \right\}_{a=1}^{N_p},
        \mathbf C^{(3)}
    \right) .
\end{split}
\label{eq:correlated_doppler_likelihood}
\end{equation}
\end{widetext}
The second line uses Eq.~\eqref{eq:isotropic_cubic_vector_density} to express the latent-vector integral in terms of the radial variable \(s^{(3)}\) and the angular direction \(\hat{\mathbf s}^{(3)}\). As in the case of our pulsar-term cubic-mode analysis, use of the standard but now multivariate Gaussian probability density function $\mathcal{N}$ suffices.

We evaluate Eq.~\eqref{eq:correlated_doppler_likelihood} numerically, using the conditional normalizing flow for \(p_s(s^{(3)} \mid \langle N \rangle)\) and a fixed angular grid for \(\hat{\mathbf s}^{(3)}\). For each radial and angular grid point, we evaluate the multivariate Gaussian likelihood and integrate over both grids.
We then repeat the null-analysis procedure used for the independent-pulsar forecasts in Sec.~\ref{sec:projections}. For each observing scenario and noise model, we generate noise-only PTA datasets with covariance \(\mathbf C^{(3)}\). In each null realization, we randomly draw the pulsar sky locations isotropically, which fixes the $\{\hat{\bf d}\}$ which enter the correlated Doppler signal and the Hellings--Downs covariance. At fixed substructure mass \(M_\mathrm{sub}\), we use the same log-uniform prior on \(f_\mathrm{sub}\) as in Sec.~\ref{sec:Results}. For each value of \(M_\mathrm{sub}\), we compute the posterior over \(f_\mathrm{sub}\), extract the 95\% credible upper limit, and report the median over noise-only realizations.

\subsection{Projected sensitivities}
\label{sec:correlated_doppler_results}

The projected sensitivities to the correlated Earth-term Doppler signal are shown in Fig.~\ref{fig:EarthTermVariations} for the SKA and Optimistic observing scenarios. In the SKA scenario, the projected limits remain above \(f_\mathrm{sub}=1\) for all noise models considered, including the white-noise-only case. In the Optimistic scenario, sensitivity to fractional abundances below unity is achieved only when the SGWB amplitude is reduced by a factor of \(100\) relative to the NANOGrav-motivated value.

The comparison with the uncorrelated pulsar-term search reflects two competing effects. At high masses where sensitivities are realized in the static limit, the number density is small and the Doppler signal distribution becomes heavy-tailed, with the largest contributions coming from rare nearby substructures. The pulsar-term search obtains an independent draw from this distribution for each pulsar, increasing the chance that at least one target samples a large fluctuation. This sampling advantage allows the uncorrelated pulsar-term search to outperform the correlated Earth-term search in the rare-event regime.

At lower masses where sensitivities are derived for number densities which correspond to the dynamic or even stochastic limit, the situation is reversed. The larger substructure number density makes the Doppler signal less dependent on rare nearby objects, so the advantage of multiple independent signal draws is reduced. The Earth-term search additionally combines many pulsar responses while using its dipolar angular structure to distinguish it from the Hellings--Downs angular correlations of the SGWB. This extends the correlated Doppler reach to lower masses than are accessible through the uncorrelated pulsar-term search.

A less compressed time-domain analysis could in principle improve the projected correlated-Doppler reach. However, the independent-pulsar Doppler comparison in Fig.~\ref{fig:AnalysisVariations} suggests that the leading cubic component captures the signal information most relevant for the noise-only limits considered here, even in red-noise-dominated scenarios. We therefore do not expect the qualitative conclusion to be changed by replacing the cubic-vector surrogate with a more computationally intensive correlated analysis. We conclude that the correlated Earth-term Doppler channel does not substantially improve the prospects for detecting DM substructure with future PTA datasets unless the effective stochastic SGWB amplitude is much lower than the current PTA-inferred value.

\section{Discussion}
\label{sec:Discussion}

Pulsar timing provides a conceptually clean gravitational probe of DM substructure via induced Doppler and Shapiro delays~\cite{Siegel:2007fz, Seto:2007kj, Clark:2015sha, Baghram:2011is, Kashiyama_2012, Kashiyama:2018gsh, Dror:2019twh,Ramani:2020hdo,Lee:2020wfn, Lee:2021zqw, NANOGrav:2023hvm}.
Most of the previous analyses of these signals assumed an optimistic white-noise-dominated noise budget since, prior to the first evidence for the SGWB reported by NANOGrav in Ref.~\cite{NANOGrav:2023gor}, it was not clear precisely how large the irreducible SGWB would be.
However, given the first evidence for the SGWB, it is important to understand how this background will affect pulsar timing searches for DM substructure.
Our results show that Doppler and Shapiro searches are strongly limited by the red-noise environment implied by a NANOGrav-like nanohertz SGWB \cite{NANOGrav:2023gor}. We have considered both the uncorrelated pulsar-term Doppler and Shapiro signals and the correlated Doppler signal induced by acceleration of the Solar System barycenter. Although these channels have distinct multi-pulsar structures, each is substantially degraded once an SGWB-motivated red-noise background is included. Even under deliberately optimistic assumptions, including idealized observing scenarios, no intrinsic pulsar red noise, and estimated SGWB parameters, the resulting red-noise background remains a major limitation for PTA substructure searches.

For the benchmark scenarios considered here, the red-noise degradation is severe. Neither the pulsar-term nor correlated Earth-term Doppler searches reach fractional abundances \(f_\mathrm{sub}\leq 1\) once the NANOGrav-motivated SGWB is included in either the SKA or Optimistic observing scenario. 
The Shapiro channel therefore appears to be the most promising of the channels considered here. It can retain sensitivity to $f_\text{sub} \leq 1$ even in the presence of an SGWB with NANOGrav-motivated parameters, although this reach is marginal in the SKA scenario and stronger only in the Optimistic scenario. This channel may also provide a useful target for diffuse substructures whose compact-lensing signatures are suppressed \cite{Dror:2019twh,Ramani:2020hdo,Croon:2020wpr}. Nevertheless, the Shapiro reach is still significantly weakened by the SGWB-motivated red-noise floor relative to the white-noise-only case.

There are clear directions in which the analysis could be made more robust, for example, by including intrinsic pulsar red-noise processes \cite{Taylor:2021yjx, NANOGrav:2023hde} or anisotropy or inhomogeneity in the local substructure position and velocity distributions. For the searches considered here, such extensions are unlikely to qualitatively improve the sensitivity under a NANOGrav-like red-noise background. 
We therefore conclude that, while detection of DM substructure via PTA datasets remains a tantalizing prospect, it is unlikely to be realized in the near future if the SGWB remains close to current PTA-inferred amplitudes.

At the same time, the framework developed here provides a likelihood-level treatment of substructure signals across the static, dynamic, and stochastic regimes. If the effective stochastic red-noise floor relevant for PTA substructure searches proves substantially lower than current PTA-inferred amplitudes, this framework provides the statistical foundation for future analyses. More broadly, replacing the response model and retraining the signal surrogate would enable forecasts and searches for extended profiles, mass spectra, or other substructure populations.

\begin{acknowledgments}
We thank T.~Bert\'olez-Mart\'inez, D.~Hooper, Y.~Kahn and N.~Rodd for helpful conversations. The work of JWF and FV was supported by the Office of the Vice Chancellor for Research at the University of Wisconsin-Madison, with funding from the Wisconsin Alumni Research Foundation. This research used resources of the National Energy Research Scientific Computing Center (NERSC), a U.S. Department of Energy Office of Science User Facility located at Lawrence Berkeley National Laboratory, operated under Contract No. DE-AC02-05CH11231 using NERSC award HEP-ERCAP0023978. This work made use of computing resources provided by the Center for High Throughput Computing at the University of Wisconsin, Madison \cite{https://doi.org/10.21231/gnt1-hw21}.
\end{acknowledgments}

\appendix
\section{Details of timing-model projection}
\label{app:projection}

To remove low-order timing-model structure from the data, we construct a projection operator that eliminates the constant, linear, and quadratic components of a discretely sampled time series. Let \(\mathbf d\) denote a vector of timing data evaluated at sampling times \(\{t_i\}\). We first affine-rescale the times to a dimensionless coordinate \(\tau \in [-1,1]\) in order to improve numerical conditioning. On this grid, we evaluate the first three Legendre polynomials,
\begin{equation}
    P_0(\tau)=1,
    \quad
    P_1(\tau)=\tau,
    \quad
    P_2(\tau)=\frac12(3\tau^2-1).
\end{equation}
These are assembled as columns of a design matrix \(\mathbf X\), whose column space spans the quadratic subspace to be removed.

We then perform a reduced QR factorization,
\begin{equation}
    \mathbf X = \mathbf Q \mathbf R,
\end{equation}
so that the columns of \(\mathbf Q\) form an orthonormal basis for this subspace. The residualizing projection matrix is
\begin{equation}
    \mathbf P \equiv \mathbf I - \mathbf Q \mathbf Q^{\mathsf T}.
\end{equation}
By construction, \(\mathbf P\) is symmetric and idempotent, and therefore acts as an orthogonal projector onto the complement of the quadratic timing-model subspace.

Applied to a data vector \(\mathbf r\), the projected vector
\begin{equation}
    \tilde{\mathbf r} = \mathbf P \mathbf r
\end{equation}
is the timing residual after removal of its best-fit constant, linear, and quadratic components in the least-squares sense. Throughout this work, we apply the same projection operator to data, signal realizations, and background realizations before constructing likelihoods or summary statistics.

\section{Validating the sampling radii}
\label{app:validating_sampling_radii}

The calibrated Monte Carlo signal model developed in Sec.~\ref{sec:SignalModel} is used throughout our subsequent statistical and machine-learning analyses, so it is important to verify that the finite sampling prescription does not distort the signal realizations in any phenomenologically relevant way. In this appendix, we validate both our fiducial sampling radii and our low-density volume-expansion prescription with a suite of Monte Carlo tests. In all cases, we take the observing duration to be
\begin{equation*}
    T_{\rm obs} = 30\,\mathrm{yr},
\end{equation*}
corresponding to the longest baseline considered in this work and therefore providing the most stringent test of the Monte Carlo signal model.

Our validation proceeds in three steps. First, we test whether substructures outside the fiducial sampling region make an appreciable contribution in the high-density regime. Next, we test whether our adaptive expansion prescription with \(N_{\min}=10^4\) is sufficient in the low-density regime. Finally, we perform a more quantitative distributional test by checking whether the times of closest approach remain consistent with the uniform statistics expected from an underlying homogeneous population. Taken together, these tests show that truncating the sampling region at finite radius does not materially affect the timing residuals or the passage statistics relevant for PTA observations.

\subsection{Substructure beyond the sampling region}
\label{sec:larger_radius_examples}

\begin{figure*}[!htb]
  \centering
  \includegraphics[width=0.99\textwidth]{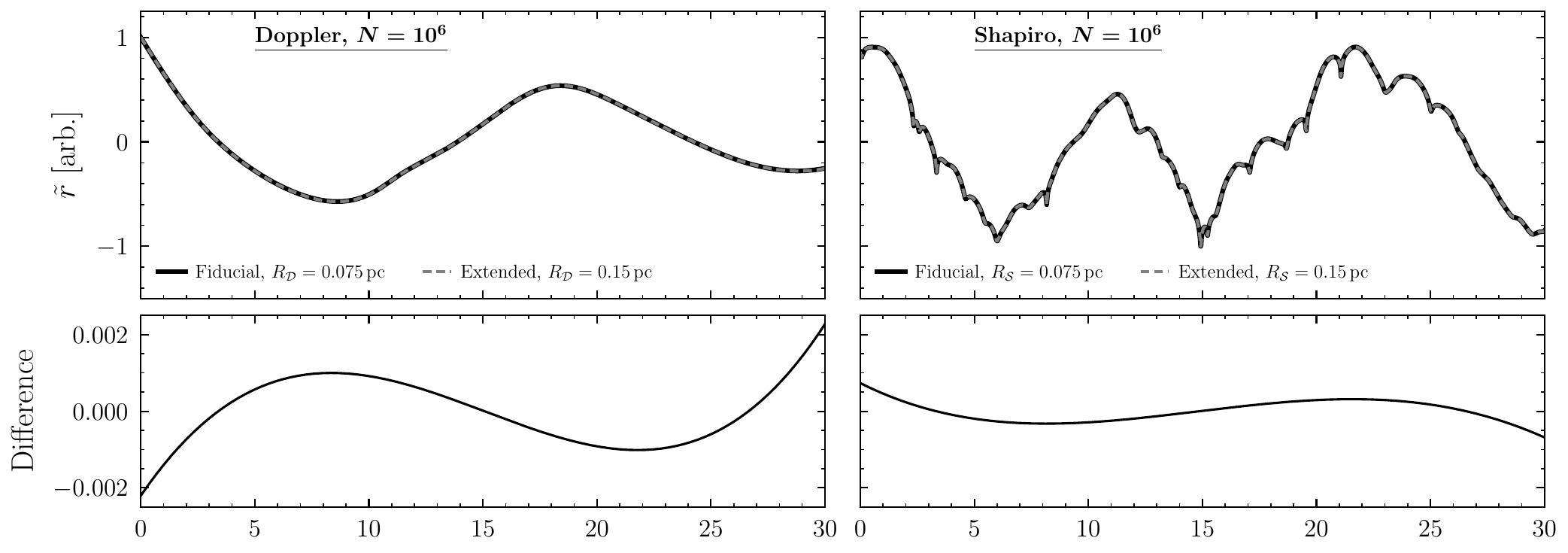}
  \caption{(\textit{Upper left}) Example Doppler signal generated at the fixed substructure number density for which the expected number of substructures in the fiducial Doppler volume is \(\langle N_D^{\rm fid}\rangle = 10^{6}\). The solid black curve is obtained using the fiducial sampling radius \(R_D^{\rm fid}=0.075\,\mathrm{pc}\), while the dashed gray curve is obtained using the extended sampling radius \(R_D=0.15\,\mathrm{pc}\). (\textit{Upper right}) As in the upper left, but for the Shapiro signal, with \(\langle N_S^{\rm fid}\rangle = 10^{6}\), fiducial radius \(R_S^{\rm fid}=0.075\,\mathrm{pc}\), and extended radius \(R_S=0.15\,\mathrm{pc}\). (\textit{Lower left}) Difference between the fiducial and extended Doppler signals shown in the upper left. (\textit{Lower right}) Difference between the fiducial and extended Shapiro signals shown in the upper right. In both cases, the difference is sub-percent throughout, demonstrating that substructure outside the fiducial sampling region makes a negligible contribution to the total timing residual. The signals are normalized such that the fiducial curve has maximum absolute magnitude \(1\).}
  \label{fig:extended_sampling}
\end{figure*}

We first consider a high-density scenario in which the expected number of substructures in the corresponding fiducial sampling volume is \(10^6\). In this regime, no adaptive volume expansion is invoked in our standard procedure, so this test directly probes whether the fiducial radius itself is sufficiently large.

For the Doppler signal, we compare the fiducial choice \(R_D^{\rm fid}=0.075\,\mathrm{pc}\) against an extended sampling radius \(R_D=0.15\,\mathrm{pc}\). At fixed number density, this increases the sampled volume by a factor of eight. For the Shapiro signal, we similarly compare \(R_S^{\rm fid}=0.075\,\mathrm{pc}\) against \(R_S=0.15\,\mathrm{pc}\), which increases the sampled cylindrical volume by a factor of four.

For each signal class, we first generate the total timing residual using all substructures in the enlarged sampling region. We then recompute the residual after discarding the substructures whose initial positions lie outside the corresponding fiducial sampling region. The difference between these two constructions isolates the contribution of substructure beyond the fiducial radius.

As shown in Fig.~\ref{fig:extended_sampling}, the contribution from substructures outside the fiducial sampling region is sub-percent for both Doppler and Shapiro signals. We therefore conclude that, in the high-density regime, extending the sampling region beyond our fiducial choice produces only negligible corrections to the residualized timing signal.

\subsection{Examining the adaptive volume expansion}
\label{sec:volumetric_expansion}

\begin{figure*}[!htb]
  \centering
  \includegraphics[width=0.99\textwidth]{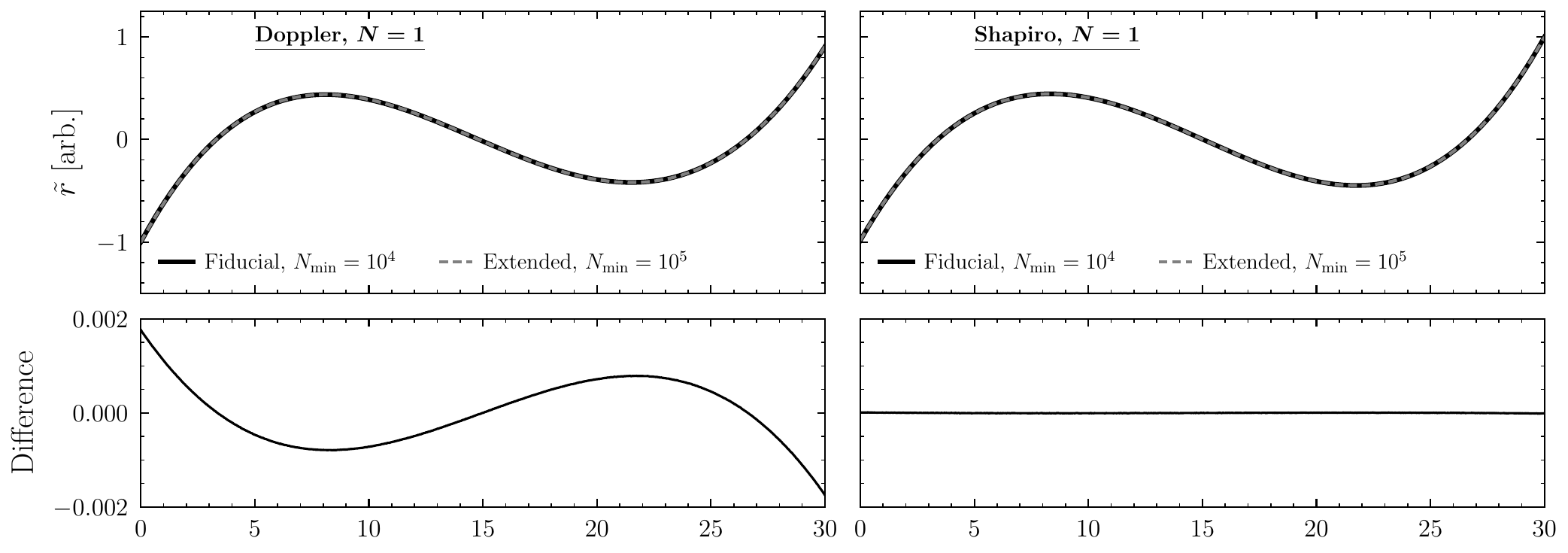}
  \caption{(\textit{Upper left}) Example Doppler signal generated at the fixed substructure number density for which the expected number of substructures in the fiducial Doppler volume is \(\langle N_D^{\rm fid}\rangle = 1\). The solid black curve is obtained using the fiducial adaptive expansion with \(N_{\min}=10^{4}\), while the dashed gray curve is obtained using an extended adaptive expansion with \(N_{\min}=10^{5}\). (\textit{Upper right}) As in the upper left, but for the Shapiro signal, with \(\langle N_S^{\rm fid}\rangle = 1\). (\textit{Lower left}) Difference between the fiducial and extended Doppler signals shown in the upper left. (\textit{Lower right}) Difference between the fiducial and extended Shapiro signals shown in the upper right. In both cases, increasing the adaptive expansion from \(N_{\min}=10^{4}\) to \(N_{\min}=10^{5}\) produces a negligible change in the total timing residual. As in Fig.~\ref{fig:extended_sampling}, signals are normalized such that the fiducial curve has maximum absolute magnitude \(1\).}
  \label{fig:small_sampling}
\end{figure*}

We now turn to the low-density regime, in which the expected number of substructures in the corresponding fiducial sampling volume is unity. In our standard prescription, this requires expanding the sampling region until the expected number of sampled substructures reaches $N_{\min} = 10^4$. The purpose of this test is to determine whether that choice is already sufficient, or whether more aggressive expansion materially changes the realized signal.

To test this, we compare against a larger sampling region chosen so that the expected number of sampled substructures is instead \(10^5\) at the same underlying number density. For both Doppler and Shapiro signals, we first generate the total timing residual using the larger expanded region. We then recompute the residual using only those sampled substructures that would have been included in the smaller expanded region corresponding to \(N_{\min}=10^4\).

This comparison directly tests the sufficiency of our fiducial adaptive prescription. As shown in Fig.~\ref{fig:small_sampling}, increasing the expected number of sampled substructures from \(10^4\) to \(10^5\) changes the total timing residual only at a negligible sub-percent level. We therefore conclude that our fiducial choice \(N_{\min}=10^4\) is adequate to capture the signal contributions relevant for later analyses.

\subsection{The distribution of times of closest approach}
\label{sec:closest_approach_distribution}

\begin{figure*}[!ht]
    \centering
    \includegraphics[width=\linewidth]{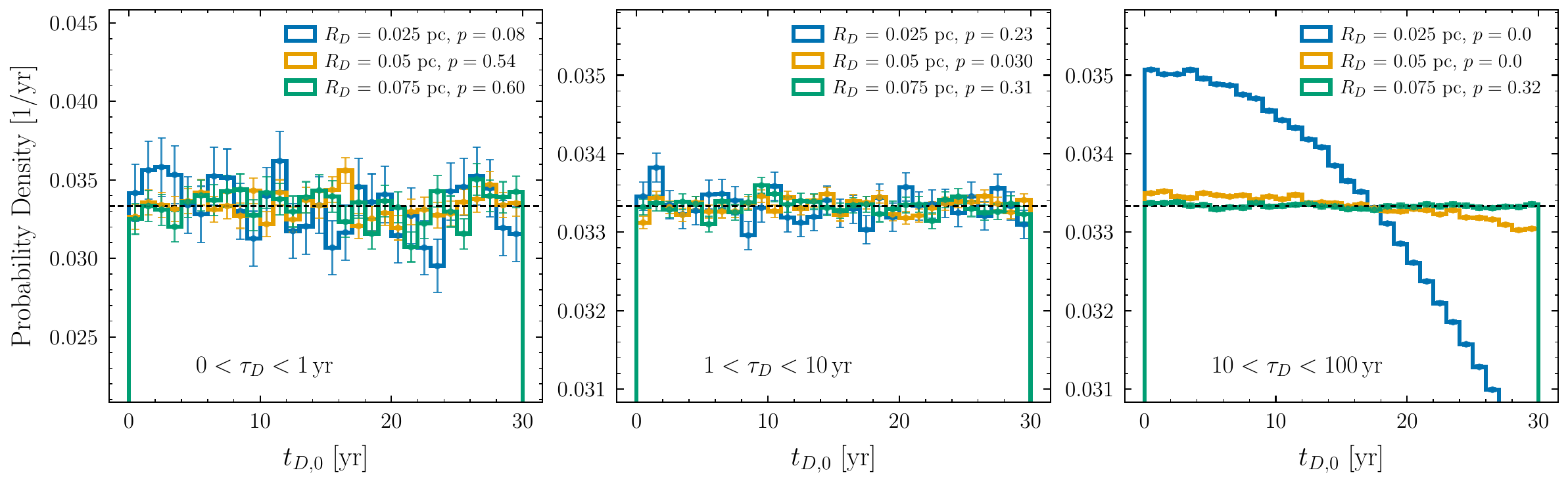}
    \caption{(\textit{Left}) The normalized distribution of \(t_{D,0}\) restricted to substructures that contribute signals with characteristic timescale \(\tau_D < 1\,\mathrm{yr}\) and whose closest approach occurs during the observing interval. Blue, orange, and green curves show Monte Carlo results for \(R_D = 0.025\,\mathrm{pc}\),  \(R_D = 0.05\,\mathrm{pc}\), and \(R_D^{\rm fid}=0.075\,\mathrm{pc}\), respectively. The caption also reports the \(p\)-value obtained by comparing the Monte Carlo distribution to the expected uniform distribution with a Kolmogorov--Smirnov test. All three radii are sufficient in this short-timescale regime. (\textit{Center}) As in the left panel, but for \(1\, \mathrm{yr}  \leq \tau_D < 10\,\mathrm{yr}\). (\textit{Right}) As in the left panel, but for \(10\,\mathrm{yr} \leq \tau_D < 100\,\mathrm{yr}\). The choices \(R_D = 0.025\,\mathrm{pc}\) and \(R_D = 0.05\,\mathrm{pc}\) are no longer compatible with the expected uniform distribution, while \(R_D^{\rm fid}=0.075\,\mathrm{pc}\) remains compatible. We therefore adopt \(R_D^{\rm fid}=0.075\,\mathrm{pc}\) as our fiducial Doppler sampling radius.}
    \label{fig:DopplerValidation}
\end{figure*}

\begin{figure*}[!ht]
    \centering
    \includegraphics[width=\linewidth]{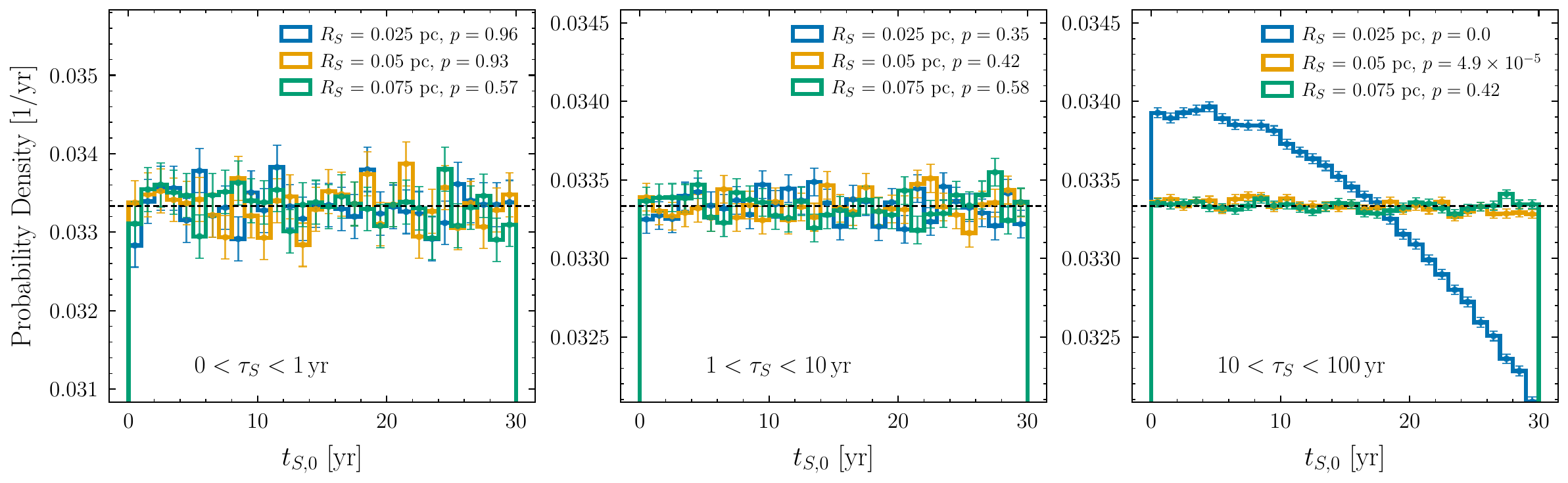}
    \caption{As in Fig.~\ref{fig:DopplerValidation}, but for the Shapiro signal. Based on the rightmost panel, we adopt \(R_S^{\rm fid}=0.075\,\mathrm{pc}\) as our fiducial Shapiro sampling radius.}
    \label{fig:ShapiroValidation}
\end{figure*}

We now perform a more quantitative test of the Monte Carlo signal model by examining the distribution of closest-approach times, \(t_{D,0}\) for Doppler signals and \(t_{S,0}\) for Shapiro signals, as defined in Eqs.~\eqref{eq:dopplerdefs} and~\eqref{eq:shapirodefs}. Under our physical assumptions, the substructure population is spatially homogeneous and its velocity distribution is independent of position. It follows that the time of closest approach, \(t_0\), should be uniformly distributed and statistically independent of the characteristic timescale \(\tau\).

Because our Monte Carlo signal model excludes substructures whose initial positions lie outside the finite sampling region, exact spatial homogeneity is broken. As a result, the induced distribution of \(t_0\) need not remain exactly uniform, and it need not remain exactly independent of \(\tau\). However, this matters only if the violation occurs for substructures whose closest approach falls within the observing window, since those are the passages that can make an observable contribution to the data.

To test this directly, we perform Monte Carlo simulations of substructures drawn from the signal model and compute their associated values of \(t_0\) and \(\tau\). We consider three timescale ranges independently:
\begin{equation*}
\begin{split}
    &\text{(i)}\ \tau < 1\,\mathrm{yr}, \\
    &\text{(ii)}\ 1\,\mathrm{yr} \leq  \tau < 10\,\mathrm{yr}, \\
    &\text{(iii)}\ 10\,\mathrm{yr} \leq \tau < 100\,\mathrm{yr}.
\end{split}
\end{equation*}
Within each range, we examine the distribution of \(t_0\) over the interval \(0 < t_0 < 30\,\mathrm{yr}\), corresponding to substructures whose closest approach occurs during the observing period. The resulting Doppler and Shapiro distributions are shown in Fig.~\ref{fig:DopplerValidation} and Fig.~\ref{fig:ShapiroValidation}, respectively.

We quantify agreement with the expected uniform distribution using the one-sample Kolmogorov--Smirnov test \cite{Conover1999}. For our fiducial choices \(R_D^{\rm fid} = R_S^{\rm fid} = 0.075\,\mathrm{pc}\), the resulting closest-approach-time distributions are compatible with uniformity at \(p>0.01\) in all three timescale ranges. By contrast, radii of \(0.025\,\mathrm{pc}\) and \(0.05\,\mathrm{pc}\) are adequate for shorter-timescale events, but fail for the \(\tau < 100\,\mathrm{yr}\) sample. This provides a quantitative justification for our adoption of \(R_D^{\rm fid}=R_S^{\rm fid}=0.075\,\mathrm{pc}\) as the fiducial sampling radii.

Taken together, the tests in this section show that our finite sampling prescription captures the timing residuals and passage statistics relevant for PTA observations while remaining computationally tractable. This provides the calibration needed to use the resulting Monte Carlo signal model confidently in the inference and machine-learning analyses that follow.

\section{Doppler covariance at large-\(\langle N\rangle\)}
\label{app:DopplerCovariance}

In this appendix, we derive the covariance of the mass-normalized Doppler timing signal in the large-\(\langle N\rangle\) limit. As discussed in Sec.~\ref{sec:covariance_surrogate}, the covariance of a signal generated by a Poisson population of independent substructures may be written as
\begin{equation}
    \Sigma_D(t,t')
    =
    \langle N \rangle
    \big\langle
    s_D(t \mid \mathbf r_0,\mathbf v)\,
    s_D(t' \mid \mathbf r_0,\mathbf v)
    \big\rangle,
\end{equation}
where the average is taken over the one-substructure position and velocity distributions. Our task here is therefore to evaluate this one-substructure correlator for the mass-normalized Doppler signal.

It is convenient to work in a coordinate system adapted to the substructure velocity and the pulsar line-of-sight. Let the velocity vector \(\mathbf v\) have magnitude \(v\), polar angle \(\theta_v\) with respect to the line-of-sight \(\hat{\mathbf d}\), and azimuthal angle \(\phi_v\), so that
\begin{equation}
    \hat{\mathbf d}\cdot \hat{\mathbf v} = \cos\theta_v.
\end{equation}
We then define an orthonormal basis by
\begin{equation}
\begin{gathered}
    \hat{\mathbf e}_3 = \hat{\mathbf v}, \\
    \hat{\mathbf e}_1
    =
    \frac{
        \hat{\mathbf d}
        -
        (\hat{\mathbf d}\cdot \hat{\mathbf e}_3)\hat{\mathbf e}_3
    }{
        \left|
        \hat{\mathbf d}
        -
        (\hat{\mathbf d}\cdot \hat{\mathbf e}_3)\hat{\mathbf e}_3
        \right|
    }, \\
    \hat{\mathbf e}_2 = \hat{\mathbf e}_3 \times \hat{\mathbf e}_1.
\end{gathered}
\end{equation}
In this basis, the initial substructure position may be parameterized by
\begin{equation}
    \mathbf r_0
    =
    r_0
    \left(
        \sin\theta\cos\phi\,\hat{\mathbf e}_1
        +
        \sin\theta\sin\phi\,\hat{\mathbf e}_2
        +
        \cos\theta\,\hat{\mathbf e}_3
    \right).
\end{equation}
This choice makes the relevant quantities quite simple:
\begin{equation}
\begin{gathered}
    t_{D,0} = -\frac{r_0}{v}\cos\theta, \qquad b \equiv |\mathbf b_D| = r_0\sin\theta, \\
    \tau_D = \frac{r_0}{v}\sin\theta = \frac{b}{v}, \qquad \hat{\mathbf b}_D \cdot \hat{\mathbf d} = \sin\theta_v \cos\phi,\\
    x_D(t) = \frac{t-t_{D,0}}{\tau_D} = \frac{v t + r_0\cos\theta}{r_0\sin\theta}.
\end{gathered}
\end{equation}
Here \(b\) is the Doppler impact parameter. Using Eq.~\eqref{eq:Doppler}, the mass-normalized Doppler signal is
\begin{align}
    s_D(t) & = \frac{G}{v^2} \bigg[\sqrt{1+x_D^2(t)}\,\sin\theta_v\cos\phi \nonumber \\
    & \qquad \qquad \qquad \qquad - \sinh^{-1}\!\big(x_D(t)\big)\cos\theta_v \bigg].
\end{align}
Inserting this into the general covariance formula gives
\begin{equation}
\begin{split}
    \frac{\Sigma_D(t,t')}{\langle N \rangle G^2}
    &=
    \int d^3\mathbf v\, f_v(\mathbf v)
    \int_{\mathcal V_D}\frac{d^3\mathbf r}{V_D}\,
    s_D(t)\,s_D(t').
\end{split}
\end{equation}
It is useful to define the normalized speed distribution
\begin{equation}
    f_s(v)
    =
    \sqrt{\frac{2}{\pi}}
    \frac{v^2}{\sigma_v^3}
    e^{-v^2/2\sigma_v^2},
\end{equation}
which is related to the isotropic velocity distribution by $f_v(\mathbf v) = f_s(v)/4\pi v^2$. The angular integrals over \(\theta_v\), \(\phi_v\), and \(\phi\) may then be performed analytically, yielding
\begin{widetext}
\begin{equation}
\begin{split}
    \Sigma_D(t,t')
    &=
    \frac{2\pi \langle N \rangle G^2}{3V_D}
    \int_0^\infty \frac{dv\,f_s(v)}{v^4}
    \int_0^{R_D} dr\, r^2
    \int_0^\pi d\theta\,\sin\theta \\
    &\qquad \times
    \left[
        \sqrt{1+x_D^2(t)}\,\sqrt{1+x_D^2(t')}
        +
        \sinh^{-1}\!\big(x_D(t)\big)\,
        \sinh^{-1}\!\big(x_D(t')\big)
    \right].
\end{split}
\end{equation}
\end{widetext}
A further simplification is obtained by changing variables from \((r,\theta)\) to \((b,z)\), where
\begin{equation}
    b = r_0\sin\theta,
    \qquad
    z = r_0\cos\theta.
\end{equation}
Here \(z\) is the initial displacement along the velocity direction, measured relative to the plane containing the point of closest approach. In these variables,
\begin{equation}
    z_\pm = \pm \sqrt{R_D^2-b^2},
    \qquad
    x_D(t) = \frac{v t + z}{b}.
\end{equation}
The covariance therefore takes the form
\begin{widetext}
\begin{equation}
\begin{split}
    \Sigma_D(t,t')
    &=
    \frac{2\pi \langle N \rangle G^2}{3V_D}
    \int_0^\infty \frac{dv\,f_s(v)}{v^4}
    \int_0^{R_D} db\, b
    \int_{z_-}^{z_+} dz
    \left[
        \sqrt{1+x_D^2(t)}\,\sqrt{1+x_D^2(t')}
        +
        \sinh^{-1}\!\big(x_D(t)\big)\,
        \sinh^{-1}\!\big(x_D(t')\big)
    \right].
\end{split}
\label{eq:DopplerCovariance}
\end{equation}
\end{widetext}
Equation~\eqref{eq:DopplerCovariance} is the expression used in our numerical evaluation of the mass-normalized Doppler covariance.

The most important feature of Eq.~\eqref{eq:DopplerCovariance} is its behavior at small impact parameter. In the limit \(b\to 0\), we observe that $x_D^2(t) = (vt+z)^2/b^2 \gg 1$, so $\sqrt{1+x_D^2(t)} \sim |vt+z|/b$. Furthermore, since the square-root term in Eq.~\eqref{eq:DopplerCovariance} dominates, the integrand behaves parametrically as $db/b$:
\begin{widetext}
\begin{equation}
\begin{split}
    \Sigma_D(t,t')
    \gtrsim
    \frac{2\pi \langle N \rangle G^2}{3V_D}
    \int_0^\epsilon \frac{db}{b}
    \int_0^\infty \frac{dv\,f_s(v)}{v^4}
    \int_{z_-}^{z_+} dz\,
    |v t + z|\,|v t' + z|.
\end{split}
\end{equation}
\end{widetext}
The covariance is therefore logarithmically divergent at small \(b\). Moreover, if closest approach occurs during the observing interval, the resulting signal is not smoothly captured by the low-order polynomial trends removed by the timing-model projection, so this small-\(b\) divergence persists in the residualized covariance as well.

By contrast, the apparent singular behavior at small velocity is removed by the projection. Expanding the signal at fixed \(b\) and small \(v\), the unprojected mass-normalized Doppler signal may be written as
\begin{equation}
    s_D(t) = \frac{1}{v^2} \sum_{n=0}^{\infty} c_n \left(\frac{v t}{b}\right)^n,
\end{equation}
where the $c_n$ are $v$-independent constants. After projection, the constant, linear, and quadratic pieces are removed, leaving
\begin{equation}
    \tilde s_D(t) = \frac{1}{v^2} \sum_{n=3}^{\infty} c_n \left(\frac{v t}{b}\right)^n.
\end{equation}
As a result, the residualized covariance is suppressed rather than enhanced in the \(v\to 0\) limit. The observationally relevant divergence is therefore the small-impact-parameter divergence, not a small-velocity one.

This behavior has an important statistical consequence. The mass-normalized Doppler signal is dominated by rare close passages, so in the ideal point-mass model the variance is formally divergent and the large-\(\langle N\rangle\) process is not naturally well-described by a Gaussian distribution. In practice, any physical or analysis cutoff \(b_{\min}\) renders the covariance finite, but the disproportionate influence of the closest encounters implies that convergence toward Gaussianity remains very slow even after such a cutoff is imposed. We examine this directly in App.~\ref{app:NonGaussianityTests}.

In this work, we adopt a physical small-impact-parameter cutoff of
\begin{equation*}
    b_{\min} = 10^{-8}\,\mathrm{pc},
\end{equation*}
At smaller impact parameters, the straight-line trajectory approximation implicit in Eq.~\eqref{eq:Doppler} is expected to fail \cite{Jennings:2019qqz, Lee:2021zqw}. In our generative-model pipeline, this cutoff is implemented by rejection sampling: any substructure with impact parameter below \(10^{-8}\,\mathrm{pc}\) is resampled. In practice, the probability that a given substructure realizes an impact factor below this threshold is vanishingly small, and the impact of this cutoff is negligible.

For numerical purposes, it is simplest to evaluate the unprojected covariance with a small but nonzero infrared cutoff in the velocity integral and then apply the projection \(\bm{\Sigma}\to \tilde{\bm{\Sigma}}\) afterward. The resulting Doppler covariance remains strongly sensitive to the treatment of the small-\(b\) region, reflecting the fact that slow events with closest approach near the observing interval can make an outsized contribution to the residualized signal. 

The physical Doppler covariance for a common-mass population is obtained by multiplying the result in Eq.~\eqref{eq:DopplerCovariance} by \(M_{\rm sub}^2\).

\section{Shapiro covariance at large-\(\langle N\rangle\)}
\label{app:ShapiroCovariance}

In this appendix, we derive the covariance of the mass-normalized Shapiro timing signal in the large-\(\langle N\rangle\) limit. As discussed in Sec.~\ref{sec:covariance_surrogate}, the covariance of a signal generated by a Poisson population of independent substructures may be written as
\begin{equation}
    \Sigma_S(t,t')
    =
    \langle N \rangle
    \big\langle
    s_S(t \mid \mathbf r_0,\mathbf v)\,
    s_S(t' \mid \mathbf r_0,\mathbf v)
    \big\rangle,
\end{equation}
where the average is taken over the one-substructure position and velocity distributions. Our task here is therefore to evaluate this one-substructure correlator for the mass-normalized Shapiro signal.

It is convenient to work in a coordinate system adapted to the Earth--pulsar line-of-sight. We define orthonormal basis vectors \(\hat{\mathbf e}_i\) such that \(\hat{\mathbf e}_3\) is aligned with the line-of-sight \(\hat{\mathbf d}\), while \(\hat{\mathbf e}_1\) and \(\hat{\mathbf e}_2\) span the plane orthogonal to \(\hat{\mathbf d}\). We then parameterize the initial substructure position \(\mathbf r_0\) in cylindrical coordinates \((b,\phi,z)\), and the velocity \(\mathbf v\) in terms of transverse and longitudinal components \((v_\perp,\phi_v,v_\parallel)\), so that
\begin{equation}
\begin{split}
    \mathbf v
    &=
    v_\perp \cos\phi_v\,\hat{\mathbf e}_1
    +
    v_\perp \sin\phi_v\,\hat{\mathbf e}_2
    +
    v_\parallel \hat{\mathbf e}_3, \\
    \mathbf r_0
    &=
    b \cos(\phi+\phi_v)\,\hat{\mathbf e}_1
    +
    b \sin(\phi+\phi_v)\,\hat{\mathbf e}_2
    +
    z\,\hat{\mathbf e}_3.
\end{split}
\end{equation}
With this choice of coordinates, \(b\) is the impact parameter relative to the line-of-sight, \(\phi\) is the angle between the transverse velocity and transverse displacement, and \(z\) is the displacement along the line-of-sight.

Using the definitions introduced in Sec.~\ref{sec:SignalModel}, the time of closest approach and characteristic timescale are
\begin{equation}
    t_{S,0} = -\frac{b\cos\phi}{v_\perp},
    \qquad
    \tau_S = \frac{b|\sin\phi|}{v_\perp}.
\end{equation}
It follows that
\begin{equation}
    x_S(t)
    =
    \frac{t-t_{S,0}}{\tau_S}
    =
    \frac{t v_\perp + b\cos\phi}{b|\sin\phi|},
\end{equation}
and the one-substructure mass-normalized Shapiro signal is therefore
\begin{equation}
    s_S(t)
    =
    2G \log\!\left[1+x_S^2(t)\right].
\end{equation}
Substituting this into the general covariance formula gives
\begin{align}
    \frac{\Sigma_S(t,t')}{\langle N \rangle (2G)^2}
    & =
    \int d^3\mathbf v\, f_v(\mathbf v) \\
    & \qquad \times \int_{\mathcal V_S}\frac{d^3\mathbf r}{V_S}\,
    \log\!\left[1+x_S^2(t)\right]
    \log\!\left[1+x_S^2(t')\right], \nonumber
\end{align}
where \(V_S\) is the Shapiro sampling volume.

To proceed, it is useful to factor the isotropic velocity distribution into transverse and longitudinal parts. We define
\begin{equation}
    f_\perp(v_\perp)
    =
    \frac{v_\perp}{\sigma_v^2}
    e^{-v_\perp^2/2\sigma_v^2},
    \qquad
    f_\parallel(v_\parallel)
    =
    \frac{1}{\sqrt{2\pi\sigma_v^2}}
    e^{-v_\parallel^2/2\sigma_v^2},
\end{equation}
so that
\begin{equation}
    f_v(\mathbf v)
    =
    \frac{1}{2\pi v_\perp}
    f_\perp(v_\perp)\,f_\parallel(v_\parallel).
\end{equation}
In these variables, the integrals over \(v_\parallel\), \(\phi_v\), and \(z\) are trivial. After performing them, the covariance reduces to
\begin{widetext}
\begin{equation}
    \Sigma_S(t,t')
    =
    \frac{(2G)^2\langle N\rangle}{\pi R_S^2}
    \int_0^\infty dv_\perp\, f_\perp(v_\perp)
    \int_0^{2\pi} d\phi
    \int_0^{R_S} db\, b\,
    \log\!\left[1+x_S^2(t)\right]
    \log\!\left[1+x_S^2(t')\right].
    \label{eq:ShapiroCovariance}
\end{equation}
\end{widetext}
Equation~\eqref{eq:ShapiroCovariance} is the expression used in our numerical evaluation of the mass-normalized Shapiro covariance.

Unlike in the Doppler case, the Shapiro covariance does not diverge at small impact parameter. In the limit \(b\to 0\), we observe that the dominant piece in $x_S(t)$ goes like $b^{-1}$, so that
\begin{equation}
    \log\!\left[1+x_S^2(t)\right]
    \sim
    \log\!\left(\frac{1}{b^2}\right),
\end{equation}
and similarly for \(t'\). The integrand in Eq.~\eqref{eq:ShapiroCovariance} therefore scales as
\begin{equation}
    b\,
    \log\!\left(\frac{1}{b^2}\right)
    \log\!\left(\frac{1}{b^2}\right),
\end{equation}
which is integrable as \(b\to 0\). The covariance is thus finite without the need for an additional small-impact-parameter regulator.

This finiteness has an important consequence for the statistics of the large-\(\langle N\rangle\) signal. Because the contribution of each substructure is independent and the covariance remains finite, the mass-normalized Shapiro signal is well-motivated to approach a Gaussian process in the large-\(\langle N\rangle\) limit. In the main text, we therefore use the covariance of Eq.~\eqref{eq:ShapiroCovariance}, together with the timing-model projection, to construct the corresponding Gaussian surrogate likelihood. The quality of this approximation at finite \(\langle N\rangle\) is examined separately in App.~\ref{app:NonGaussianityTests}.

As in the case of our Doppler covariance calculation, the physical Shapiro covariance for a common-mass population is obtained by multiplying the result above by \(M_{\rm sub}^2\).

\section{Tests of Gaussianity at finite \(\langle N\rangle\)}
\label{app:NonGaussianityTests}

\begin{figure*}[!htb]
    \centering
    \includegraphics[width=\linewidth]{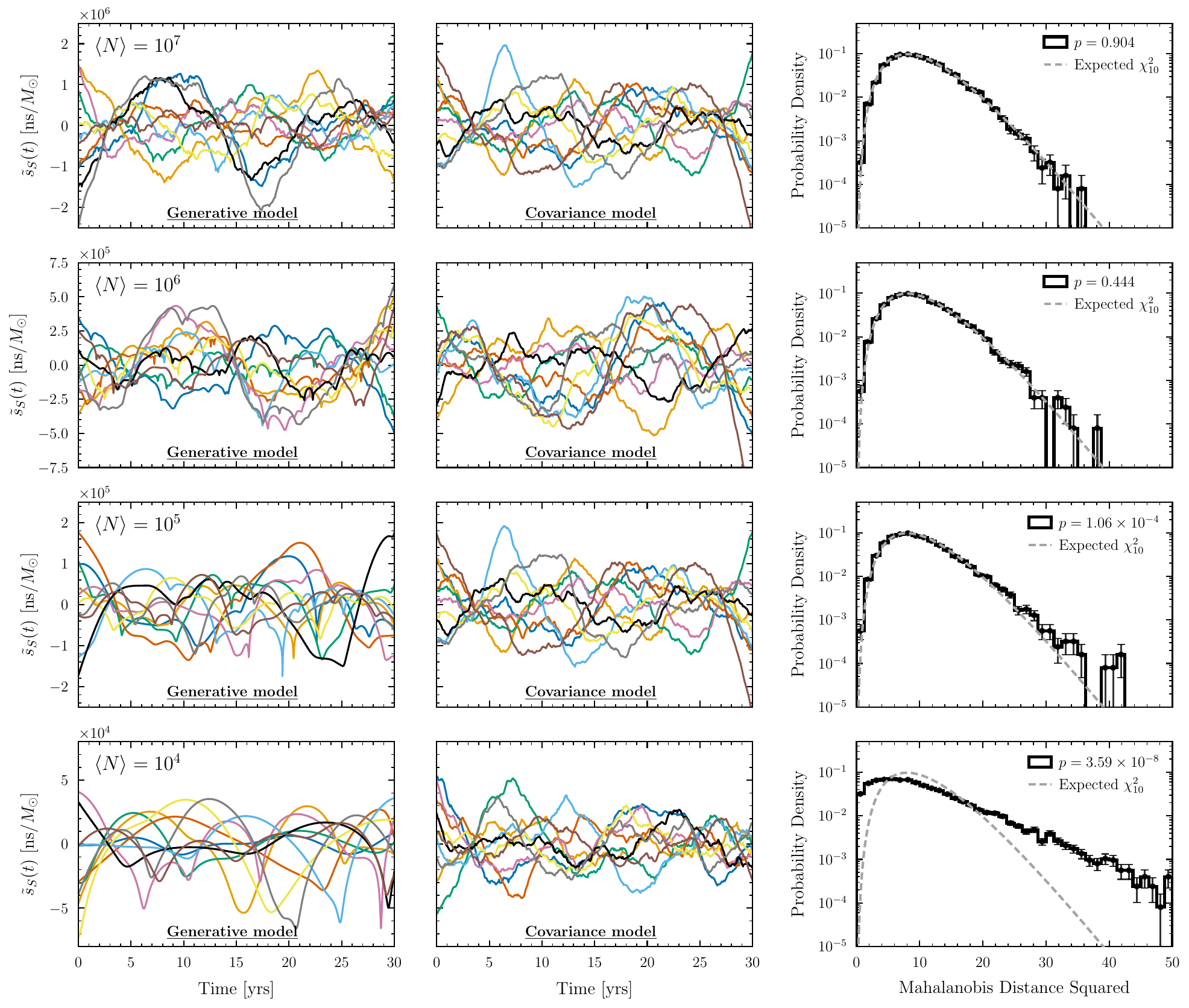}
    \caption{In each row, the left column shows 10 samples from the full Monte Carlo signal model, the middle column shows 10 samples from the covariance model, and the right column compares the empirical distribution of squared Mahalanobis distances, computed from full generative-model samples using the 10 leading eigenmodes, with the $\chi^2$ distribution with 10 degrees of freedom expected under Gaussian statistics. From top to bottom, the rows correspond to $\langle N \rangle = 10^7$, $10^6$, $10^5$, and $10^4$. The rightmost panels also report the $p$-value for consistency with the $\chi^2_{10}$ expectation. Gaussianity is disfavored for $\langle N \rangle \leq 10^5$, and the sharp coherent features visible in the generative-model samples but absent in the covariance-model samples indicate that appreciable non-Gaussianity may persist even at the largest $\langle N \rangle$ considered here.}
    \label{fig:ShapiroNonGaussianity}
\end{figure*}

In this appendix, we examine the agreement of a Gaussian covariance model with Doppler and Shapiro signals. We begin with an inspection of the Shapiro case, which we do expect to realize Gaussianity in the large $\langle N_S \rangle$ limit. As one specific test of Gaussianity, at fixed $\langle N_S \rangle$, for a 30-year observing period, we generate $10^4$ residualized signal realizations. We then use the expected covariance calculated in App.~\ref{app:ShapiroCovariance} to evaluate the squared Mahalanobis distance restricted to the subspace spanned by the 10 leading eigenmodes of the covariance. Under the assumption that the samples were realized from the expected covariance, this squared Mahalanobis distance should follow a $\chi^2$-distribution with 10 degrees of freedom \cite{Anderson2003}. The one-sample Kolmogorov-Smirnov test then allows us to test the compatibility of the Monte Carlo signal model samples with Gaussianity by comparing the distribution of the squared Mahalanobis distance with this expectation \cite{Conover1999}. The results of these Monte Carlo tests for $\langle N_S \rangle = 10^4, \, 10^5, \, 10^6,$  and $10^7$ are shown in Fig.~\ref{fig:ShapiroNonGaussianity}. Only by $\langle N_S \rangle = 10^6$ do we observe consistency at the level of $p \geq 0.05$. For lower $\langle N_S \rangle$, this test shows that Gaussianity is not a good description. 

Moreover, in Fig.~\ref{fig:ShapiroNonGaussianity}, we also compare samples drawn directly from our Monte Carlo signal model with samples drawn directly from the covariance. A visual comparison reveals that, although the Mahalanobis distance test indicates some level of compatibility with Gaussianity at $\langle N_S \rangle = 10^6$, sharp coherent features appear in the full Monte Carlo signal model samples but not in the covariance-model samples. This suggests that non-Gaussian features not captured by our simple Mahalanobis distance may persist even to higher $\langle N_S \rangle$, spoiling Gaussianity over nearly the full number range considered in this work.

\begin{figure*}[!htb]
    \centering
    \includegraphics[width=\linewidth]{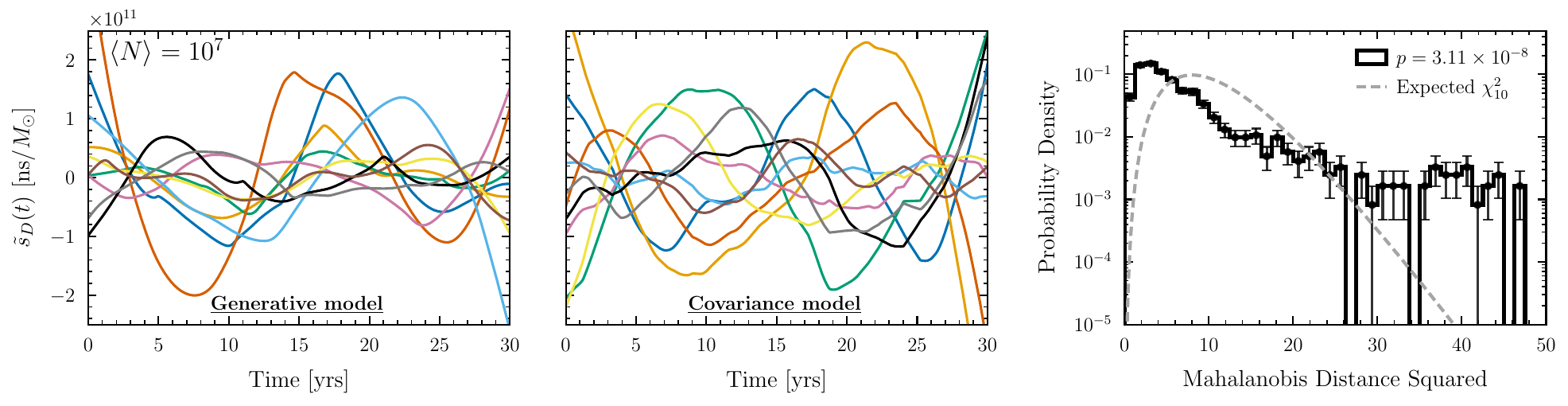}
    \caption{As in Fig.~\ref{fig:ShapiroNonGaussianity}, but for the Doppler signal. The disagreement with the Gaussian expectation is severe.}
    \label{fig:DopplerNonGaussianity}
\end{figure*}

We perform a similar test for the Doppler signal, for which convergence to Gaussianity is expected to be slow because the covariance is dominated by rare close passages. Here, we consider only $\langle N_D \rangle = 10^7$, which already suffices to show significant non-Gaussianity. Our results are summarized in Fig.~\ref{fig:DopplerNonGaussianity}.

\section{Details of the normalizing flow model}
\label{app:FlowDetails}

In this appendix, we summarize the normalizing flow used to model the distribution of the cubic-mode signal amplitude introduced in the main text. Concretely, the quantity modeled here is the mass-normalized cubic-mode amplitude \(s^{(3)}_\mathcal{I}\), obtained by projecting a signal realization onto the cubic Legendre mode defined in Sec.~\ref{sec:cubic_flow_surrogate}. We model this quantity conditionally on \(\log_{10}\langle N\rangle\).

Rather than modeling \(s^{(3)}_\mathcal{I}\) directly, we first remove its leading scaling with the expected number of substructures. For the signal populations considered here, the signed cubic-mode amplitude is symmetric under \(s^{(3)}_\mathcal I\to -s^{(3)}_\mathcal I\), so it is sufficient to model the distribution of its magnitude. We therefore define the transformed variable
\begin{equation}
    x_\mathcal{I} \equiv \log_{10}|s^{(3)}_\mathcal{I}| - \alpha \log_{10}\langle N \rangle,
\end{equation}
and train the flow to model the conditional density of \(x_\mathcal I\). The parameter \(\alpha\) is chosen to match the asymptotic scaling of the cubic-mode amplitude in the low-\(\langle N\rangle\) regime. In practice, we take
\begin{equation}
    \alpha =
    \begin{cases}
        3/2, & \text{Doppler}, \\
        1, & \text{Shapiro}.
    \end{cases}
\end{equation}
This transformation reduces the variation of the target distribution across \(\langle N \rangle\), allowing the conditional model to learn a smoother family of one-dimensional densities.

With these definitions, the flow models the one-dimensional conditional density
\begin{equation}
    p_X(x_\mathcal{I} \mid \log_{10}\langle N \rangle).
\end{equation}
The corresponding signed density for the cubic-mode amplitude \(s^{(3)}_\mathcal{I}\) is then obtained by a change of variables,
\begin{equation}
    p(s^{(3)}_\mathcal{I} \mid \log_{10}\langle N \rangle)
    =
    \frac{
        p_X(x_\mathcal{I} \mid \log_{10}\langle N \rangle)
    }{
        2 \ln(10)\,|s^{(3)}_\mathcal{I}|
    },
    \label{eq:s3_density_from_flow}
\end{equation}
where \(x_\mathcal{I} = \log_{10}|s^{(3)}_\mathcal{I}| - \alpha \log_{10}\langle N \rangle\). The factor of \(2\) accounts for the two possible signs of \(s^{(3)}_\mathcal{I}\), while the factor of \(\ln(10)\,|s^{(3)}_\mathcal{I}|\) is the Jacobian associated with the logarithmic transformation.

We implement \(p_X(x \mid \log_{10}\langle N\rangle)\) using a conditional neural spline flow \cite{Durkan:2019}. In practice, we use a one-dimensional conditional model with \(16\) spline bins, \(4\) flow transforms, and hidden-layer widths \((64,64,64)\). The spline bound is chosen to be the smallest integer value for which the full training sample lies within the spline domain, so that the nonlinear transformation acts across the entire range populated by the data.

The training data are obtained directly from the template banks described in Sec.~\ref{sec:template_bank_construction}. For each template-bank realization, we compute the corresponding cubic-mode amplitude \(s^{(3)}_\mathcal{I}\) and pair it with the associated value of \(\log_{10}\langle N\rangle\). We then form the transformed target \(x_\mathcal{I}\) defined above and train the conditional flow by maximizing the log-likelihood of \(x_\mathcal{I}\) given \(\log_{10}\langle N\rangle\). We reserve \(10\%\) of the available samples for validation and train on the remaining \(90\%\).

Optimization is performed with AdamW using learning rate \(3\times10^{-5}\), weight decay \(10^{-4}\), batch size \(4096\), and \(100\) training epochs. We employ gradient clipping with maximum norm \(1\) together with a learning-rate schedule consisting of a \(10\%\) linear warmup followed by cosine decay to a minimum learning rate of \(10^{-7}\). At the end of each epoch, we evaluate the validation loss and retain the checkpoint with the best validation performance as the inference model used in later analyses.

\section{Details of the diffusion-based surrogate}
\label{app:DiffusionDetails}

In this appendix, we summarize the generative diffusion model used to construct the Monte Carlo likelihood of Sec.~\ref{sec:diffusion_mc_surrogate}. The key idea is to factor each projected mass-normalized signal realization into an overall amplitude scale and a normalized waveform shape. The former is modeled with a one-dimensional conditional flow, while the latter is modeled with a conditional diffusion network. This separation is useful because the overall normalization and the waveform morphology vary differently with \(\langle N\rangle\), and are therefore more naturally learned by different surrogate models.

\subsection{Scale model}

For each projected mass-normalized signal realization \(\mathbf s\), we define a robust amplitude scale
\begin{equation}
    A \equiv q_{95}[\mathbf s] - q_{05}[\mathbf s]
\end{equation}
where \(q_{95}\) and \(q_{05}\) denote the 95th and 5th percentiles of the realization. We then define
\begin{equation}
    u \equiv \log_{10} A.
\end{equation}
The role of the scale model is to capture the distribution of this overall normalization as a function of \(\log_{10}\langle N\rangle\).

Rather than modeling \(u\) directly, we first remove its leading \(\log_{10}\langle N\rangle\)-dependence by defining
\begin{equation}
    x \equiv \frac{
        u - \mu(\log_{10}\langle N\rangle)
    }{
        \sigma(\log_{10}\langle N\rangle)
    },
\end{equation}
where \(\mu\) and \(\sigma\) are smooth functions fit on the training split alone. In practice, these functions are represented by monotone PCHIP interpolants on coarsened knot grids, with \(\mu\) constrained to be nondecreasing and \(\sigma\) constrained to be nonincreasing. This removes the dominant conditioner dependence before the residual distribution is passed to the flow.

We model the residual variable \(x\) with a one-dimensional conditional neural spline flow \cite{Durkan:2019}. The corresponding density for \(u=\log_{10}A\) is then recovered as
\begin{equation}
    p(u \mid \log_{10}\langle N\rangle)
    =
    \frac{
        p_X(x \mid \log_{10}\langle N\rangle)
    }{
        \sigma(\log_{10}\langle N\rangle)
    }.
\end{equation}
In practice, we use a one-dimensional conditional model with \(16\) spline bins, \(4\) flow transforms, and hidden-layer widths \((64,64,64)\). The spline bound is chosen to be the smallest integer value for which the full training sample lies within the spline domain.

The scale flow is trained with AdamW using learning rate \(3\times10^{-5}\), weight decay \(10^{-4}\), batch size \(4096\), and \(100\) training epochs. We use a \(10\%\) validation split, gradient clipping with maximum norm \(1\), and a learning-rate schedule consisting of a \(10\%\) linear warmup followed by cosine decay to a minimum learning rate of \(10^{-7}\). The checkpoint with the best validation loss is used in later analyses.

\subsection{Diffusion model for normalized waveform shapes}

Given the scale \(A\), we define the normalized waveform shape
\begin{equation}
    \hat{\mathbf s} \equiv \frac{\mathbf s}{A}.
\end{equation}
The purpose of the diffusion model is to learn the conditional distribution of \(\hat{\mathbf s}\), that is, the waveform morphology after the overall normalization has been removed.

Training is carried out entirely in the projected residual subspace. Let \(\mathbf Q\) denote a matrix whose columns span the timing-model directions removed by the residualizing projection. During both training and sampling, waveform quantities are projected according to
\begin{equation}
    \hat{\mathbf s}
    \;\mapsto\;
    \hat{\mathbf s} - (\hat{\mathbf s}\mathbf Q)\mathbf Q^T,
\end{equation}
so that the generative model remains confined to the residual subspace. The normalized waveforms are not mean-subtracted; instead, they are standardized only by dividing by a single global training-set standard deviation. By contrast, \(u=\log_{10}A\) and \(\log_{10}\langle N\rangle\) are standardized by subtracting their training-set means and dividing by their training-set standard deviations.

The diffusion model itself is a conditional one-dimensional U-Net trained with a standard noise-prediction objective. We use a cosine diffusion schedule with \(256\) diffusion steps. If \(\hat{\mathbf s}_0\) denotes a projected, standardized waveform realization and \(t\) a randomly drawn diffusion step, the noisy state is constructed as
\begin{equation}
    \hat{\mathbf s}_t
    =
    \sqrt{\bar\alpha_t}\,\hat{\mathbf s}_0
    +
    \sqrt{1-\bar\alpha_t}\,\bm\epsilon,
\end{equation}
where \(\bm\epsilon\) is Gaussian noise projected back into the residual subspace. The network is then trained to predict \(\bm\epsilon\) using a mean-squared-error loss.

Our denoising network is a conditional 1D U-Net with base channel count \(48\), channel multipliers \((1,2,4,8,8)\), and embedding dimension \(256\). In addition to the noisy waveform, the network receives sinusoidal timestep embeddings and a conditioning vector built from \((u,\log_{10}\langle N\rangle)\). We also append four fixed positional channels on the observing grid,
\begin{equation}
    P_3(\tau),\qquad P_4(\tau),\qquad \sin(\pi\tau),\qquad \cos(\pi\tau),
\end{equation}
where \(\tau \in [-1,1]\) is the affine-rescaled time coordinate and \(P_\ell\) denotes the Legendre polynomial of degree \(\ell\). The network is built from residual blocks throughout, with self-attention in the deepest levels and bottleneck.

Optimization of the diffusion model is performed with AdamW using learning rate \(10^{-4}\), weight decay \(10^{-4}\), batch size \(64\), and \(100\) training epochs. We use gradient clipping with maximum norm \(1\), together with a learning-rate schedule consisting of a linear warmup for \(5500\) steps followed by cosine decay to a minimum learning rate of \(3\times10^{-6}\). We maintain an exponential moving average of the network parameters with decay \(0.999\), and use the EMA model for validation and checkpoint selection.

\subsection{Sampling procedure}

At inference time, the scale and shape models are combined hierarchically. For a chosen value of \(\log_{10}\langle N\rangle\), we first draw a logarithmic scale
\begin{equation}
    u \sim p(u \mid \log_{10}\langle N\rangle)
\end{equation}
from the scale flow, and then set
\begin{equation}
    A = 10^u.
\end{equation}
Conditioned on the pair \((u,\log_{10}\langle N\rangle)\), we then run the reverse diffusion chain, starting from a Gaussian draw in the projected residual subspace, to obtain a sample of the normalized waveform shape \(\hat{\mathbf s}\).

The final mass-normalized signal realization is reconstructed as
\begin{equation}
    \mathbf s = A\,\hat{\mathbf s},
\end{equation}
followed by a final projection back into the residual subspace. Repeating this procedure yields the Monte Carlo ensemble used in Eq.~\eqref{eq:diffusion_mc_likelihood}.

\section{Convergence of Diffusion Projections}
\label{app:Convergence}

\begin{figure}[!htb]
    \centering
    \includegraphics[width=.99\linewidth]{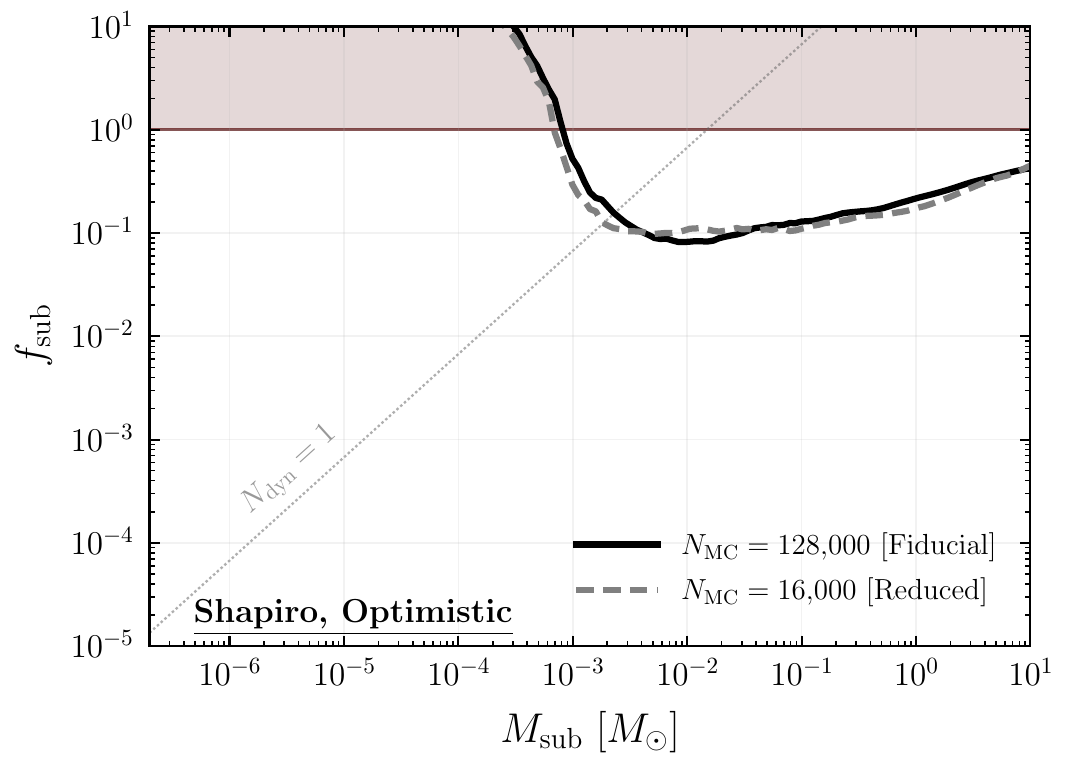}
    \caption{A comparison of our projected sensitivities to the Shapiro signal in the Optimistic observing scenario with the NANOGrav-fiducial SGWB spectrum, computed using $N_\mathrm{MC}=128{,}000$ (solid black) and $N_\mathrm{MC} = 16{,}000$ (dashed grey). Good agreement is observed, indicating that with $N_\mathrm{MC} = 128{,}000$, our null sensitivity projections are likely sufficiently converged.}
    \label{fig:MCValidation}
\end{figure}

Direct Monte Carlo marginalization of the stochastic signal process becomes increasingly demanding at large signal-to-noise. To see this, consider the exact process-marginalized likelihood
\begin{equation}
\begin{split}
    p(\tilde{\mathbf r} \mid \langle N\rangle, M_\mathrm{sub})
    =
    \int d\tilde{\mathbf s} \,&
    \mathcal N\!\left(
        \tilde{\mathbf r};
        M_\mathrm{sub}\tilde{\mathbf s}_\mathcal{I},
        \tilde{\mathbf C}_\mathrm{bkg}\right)\\
        &\times p_\mathrm{sig}(\tilde{\mathbf s}_\mathcal{I} \mid \langle N\rangle ).
\end{split}
\end{equation}
Suppose the true signal were $M_\mathrm{sub} \tilde{\mathbf{s}}_I^\mathrm{true}$, and it is large compared to characteristic variations associated with background noise parametrized by $\tilde{\mathbf C}_\mathrm{bkg}$. In this large-signal limit, the Gaussian probability density function is well approximated as a Dirac delta function, so that
\begin{equation}
p(\tilde{\mathbf r} \mid \langle N\rangle, M_\mathrm{sub}) \approx p_\mathrm{sig}(\tilde{\mathbf{s}}_\mathcal{I}^\mathrm{true} \mid \langle N\rangle). 
\end{equation}
Accurately approximating this process-marginalized likelihood via Monte Carlo sampling from the process itself might then take an exponentially large number of samples from the generative model.

We do not claim that our choice of \(N_{\rm MC}=128{,}000\) is sufficient for a fully calibrated analysis at arbitrarily large signal strength. Our goal here is instead to develop null projected sensitivities, for which it suffices to demonstrate that the resulting limits are stable with respect to the number of Monte Carlo signal samples.

To do so, we reduce our sample size from \(N_{\rm MC}=128{,}000\) to \(N_{\rm MC}=16{,}000\) and repeat our procedure to compute projected sensitivities for the Shapiro signal in the Optimistic observing scenario under the fiducial NANOGrav SGWB model. The results of this test are demonstrated in Fig.~\ref{fig:MCValidation}, validating the convergence of our projected sensitivities presented in the main text.

\bibliography{main}
\end{document}